\documentclass{aa}  
\usepackage{graphicx}
\usepackage[english]{babel}
\usepackage{amsmath}
\usepackage{natbib}
\usepackage[colorlinks=true, allcolors=blue]{hyperref}
\usepackage{xcolor}
\usepackage{lscape}
\usepackage{pdflscape}
\usepackage[utf8]{inputenc}
\DeclareUnicodeCharacter{2212}{-}
\usepackage{txfonts}
\begin{document}

   \title{Wavelength-resolved reverberation mapping of intermediate redshift quasars HE 0413-4031 and HE 0435-4312: Dissecting Mg II, optical Fe II, and UV Fe II emission regions \thanks{based on observations made with the Southern African Large Telescope (SALT)}}
   \titlerunning{ RM of Intermediate redshift quasars}
   \authorrunning{R. Prince et al.}

   \author{Raj Prince\inst{1}
          \and 
        Michal Zaja\v{c}ek\inst{2}
          \and
          Swayamtrupta Panda\inst{3}
          \and
          Krzysztof Hryniewicz\inst{4}
          \and 
          Vikram Kumar Jaiswal \inst{1}
          \and
          Bo\. zena Czerny\inst{1}
          \and 
          Piotr Trzcionkowski\inst{5}
          \and
          Mateusz Bronikowski\inst{6}
          \and
          Mateusz Ra\l owski\inst{7}
          \and
          Catalina Sobrino Figaredo\inst{8}
          \and
          Mary Loli Martinez–Aldama\inst{9}
          \and
          Marzena {\'S}niegowska\inst{10,11}
          \and
          Justyna {\'S}redzi\' nska\inst{12}
          \and
          Maciej Bilicki\inst{1}
          \and
          Mohammad-Hassan Naddaf \inst{1}
          \and
          Ashwani Pandey\inst{1}
          \and
          Martin Haas\inst{13}
          \and
          Marek Jacek Sarna\inst{10}
          \and
          Grzegorz Pietrzy\' nski\inst{10}
          \and
          Vladimir Karas\inst{14}
          \and
              Aleksandra Olejak\inst{10}
    \and 
    Robert Przy{\l}uski\inst{15}
    \and
    Ramotholo R. Sefako\inst{16}
    \and
    Anja Genade\inst{5, 16}
    \and
    Hannah L. Worters\inst{16} 
    \and
    Szymon Koz\l owski\inst{5}
    \and
    Andrzej Udalski\inst{5}
          }
   \institute{Center for Theoretical Physics, Polish Academy of Sciences, Al. Lotnik\' ow 32/46, 02-668 Warsaw, Poland\\
              \email{raj@cft.edu.pl}
              \and
                      Department of Theoretical Physics and Astrophysics, Faculty of Science, Masaryk University, Kotlářská 2, 611 37 Brno, Czech Republic
         \and
             Laboratório Nacional de Astrofísica. Rua dos Estados Unidos, 154,  37504-364 Itajubá, MG, Brazil
             \and 
        National Centre for Nuclear Research, ul. Pasteura 7, 02-093 Warsaw, Poland
        \and
            Astronomical Observatory, University of
            Warsaw, Al. Ujazdowskie 4, 00-478 Warsaw, Poland
        \and
        Centre for Astrophysics and Cosmology, University of Nova Gorica, Vipavska 11c, 5270 Ajdov\v{s}\v{c}ina, Slovenia
        \and
        Astronomical Observatory of the Jagiellonian University, Faculty of Physics, Astronomy and Applied Computer Science, ul. Orla 171, 30-244 Cracow, Poland
        \and
        University of Haifa, Abba Khoushy Ave 199, Haifa, 3498838, Israel
        \and
        Instituto de Física y Astronomía, Facultad de Ciencias, Universidad de Valparaíso, Gran Bretaña 1111, Valparaíso, Chile 
        \and
        Nicolaus Copernicus Astronomical Center, Polish Academy of Sciences, ul. Bartycka 18, 00-716 Warsaw, Poland
        \and
        School of Physics and Astronomy, Tel Aviv University, Tel Aviv 69978, Israel
        \and
        Copernicus Science Centre,  Wybrzeże Kościuszkowskie 20, 00-390 Warsaw, Poland
        \and
        Astronomisches Institut Ruhr-Universit\" at Bochum, Universit\" atsstra{\ss}e 150, 44801 Bochum, Germany
        \and
        Astronomical Institute, Academy of Sciences, Bo\v{c}n{\'i} II 1401, CZ-14131 Prague, Czech Republic
        \and 
         Space Research Center, Polish Academy of Sciences, Bartycka 18A, 00-716 Warszawa
        \and
        South African Astronomical Observatory, PO Box 9, Observatory, 7935 Cape Town, South Africa
        \and
        University of Cape Town, Rondebosch, Cape Town, 7700
            }         

 
  \abstract
  {We present the wavelength-resolved reverberation mapping (RM) of combined Mg II and UV Fe II broad-line emissions for two intermediate redshifts (z $\sim$ 1), luminous quasars - HE 0413-4031 and HE 0435-4312, monitored by the Southern African Large Telescope (SALT) and 1-m class telescopes between 2012 - 2022.}
  {Through the wavelength-resolved technique, we aim to disentangle the Mg II and Fe II emission regions and to build a radius-luminosity relation for UV Fe II emission, which has so far remained unconstrained.}
  {Several time delay methodologies have been applied to constrain the time delays for total Mg II and Fe II emissions. In addition, wavelength-resolved RM is performed to quantify the inflow or outflow of broad-line region (BLR) gas around the supermassive black hole and to disentangle the emission/emitting regions from lines produced in proximity to each other. }
  {The mean total FeII time delay is nearly equal to the mean total MgII time delay for HE 0435-4312 suggesting a co-spatiality of their emission regions. However, in HE 0413-4031, the mean FeII time delay is found to be longer than the mean MgII time delay, suggesting that FeII emission is produced at longer distances from the black hole. The UV FeII R-L relation is updated with these two quasars (total four) and compared with the optical FeII relation (20 sources), which suggests that the optical FeII emission region is located further than the UV FeII region by a factor of 1.7-1.9 i.e. $R_{\rm FeII-opt}\sim (1.7-1.9) R_{\rm FeII-UV}$. }
  {Wavelength-resolved reverberation is an efficient way to constrain the geometry and structure of the BLR. We detected a weak pattern in the time delay vs. wavelength relation, suggesting that the MgII broad line originates a bit closer to the SMBH than the UV FeII pseudo continuum, however, the difference is not very significant. Comparison of MgII, UV, and optical FeII R-L relations suggests that the difference may be larger for lower-luminosity sources, possibly with the MgII emission originating further from the SMBH. In the future, more RM data will be acquired to put better constraints on these trends, in particular the UV FeII R-L relation. }



\maketitle
%
\section{Introduction}
Reverberation mapping is a powerful technique used to map the inner structure of active galactic nuclei (AGN) based on the information on light travel time between the central disk and the clouds located at the sub-parsec/parsec scale \citep[see][for a review]{Cackett2021}. The clouds are predominantly in Keplerian orbits around the supermassive black hole (SMBH) with velocities of the order of $\sim 1000$-10000 km/s. These clouds are the sources of the broad optical and UV emission lines thus forming so-called broad-line regions (BLR). The structure and the formation of these clouds are still under discussion \citep[see e.g.][for a review]{krolik_book1999,netzer2015ara}. The region is basically unresolved apart from interferometric measurements for three AGN done with the GRAVITY instrument (3C273, \citealt{GRAVITY2018}; IRAS 09149-6206, \citealt{2019Msngr.178...20A}; and NGC 3783, \citealt{GRAVITY2020, GRAVITY2021}) which supported the picture of the flattened rotating distribution of BLR clouds. 

The reverberation technique provides an indirect measurement of the extent of the BLR and an insight into the kinematics and the geometry of the line-emitting material around the SMBH. The reverberation requires intense monitoring of the photometric (continuum UV-optical) and spectroscopic (line emission) variations of an AGN. The method was proposed by \citet{Blandford1982} and \citet{Peterson1993}. Initially, intense monitoring of many AGN has been done mostly concentrating on the H$\beta$ delay measurements \citep[e.g.][]{Kaspi2000, Peterson2004, Bentz2013,dupu2015, Du2018, Grier_2017, 2018NatAs...2...63M,fonseca2020,li_SS_2021,2023MNRAS.520.2009M} but recently also intermediate-redshift RM measurement of quasars based on Mg II line was done \citep{Shen2016,  Lira2018, Homayouni2020, Czerny2019, Zajacek2020, Yu2021, zajacek2021, prince_CTS_2022} as well as RM of high-redshift quasars based on CIV line \citep{peterson2005,kaspi2007, Lira2018,derosa2018,hoormann2019,grier2019,shen2019,kaspi2021}. 

These studies consistently showed that there is a significant correlation between the measured time delay and the monochromatic continuum luminosity, the so-called radius-luminosity (R-L) relation for all three emission lines (H${\beta}$, Mg II, and C IV), based on the monitoring of a large number of sources. The R-L relation enables us to estimate the virial mass of the black hole, from single-epoch spectroscopy, following the virial theorem. A more exciting application of the R-L relation is to infer luminosity distances using the measured time delays, from which one can obtain monochromatic luminosities using the R-L relation,  and the observed flux densities \citep{haas2011, watson2011, Czerny2013}. To avoid the circularity problem, one can determine simultaneously R-L relation parameters and cosmological parameters by maximizing the likelihood function that depends on the observed time delays as well as theoretical time delays calculated based on the assumed cosmological model. So far based on the current samples of RM quasars it has been shown that the R-L relation parameters are independent of the assumed cosmological model, hence the RM quasars are standardizable, however, the cosmological constraints are rather weak in comparison with better-established cosmological probes \citet{martinez2019, Czerny2021, zajacek2021, khadka2021, Khadka2022, Cao2022}.  

From the list of sources that have been monitored by the Southern Large African Telescope (SALT), the reverberation study was performed using the luminous quasars CTS C30.10, HE 0413-4031, and HE 0435-4312 \citep{Czerny2019,Zajacek2020, zajacek2021,prince_CTS_2022} using almost 10 years of observations.
The wavelength-resolved reverberation is a more advanced technique that allows tracing separately the wings of the lines from the core as well as the kinematics of the BLR medium when multiple lines contribute to a given wavelength range. Such studies were done for more than 35 AGNs by various authors (\citealt{2002A&A...386L..19K, 2003A&A...407..461K}; \citealt{Bentz2010}; \citealt{Denney2010}; \citealt{Grier2012}; \citealt{Du2018}; \citealt{derosa2018}; \citealt{Xiao_2018}; \citealt{Zhang2019}; \citealt{Hu_2020}; \citealt{U2021}) for low-redshift sources. The monitoring of bright, more distant quasars takes more time to get suitable data and so far only one intermediate-redshift quasar CTS C30.10 was monitored for 12 years and the wavelength-resolved analysis was presented in \citet{prince_CTS_2022}. In \citet{prince_CTS_2022}, for the first time, we provided the R-L relation for the UV pseudo continuum FeII emission using the low-luminosity and the low-redshift AGN NGC 5548 and the intermediate-redshift, luminous quasar CTS C30.10. In the present paper, we extend the previous results by adding another two intermediate-redshift quasars HE0413-4031 and HE 0435-4312 monitored for over 12 years with the Southern Large African Telescope (SALT). Our study is more focused on building the standard R-L relation for UV FeII emission and the comparison with the optical FeII R-L relation.

The paper is organized as follows. In Section~\ref{sec_obs}, we provide details about the observational data and the photometric and spectral analyses. In Section~\ref{sec_time_delay}, all the methods applied for the time-delay measurement are introduced, which is followed by results and discussion presented in Sections~\ref{sec_results} \& \ref{sec_discussion}, respectively. We summarize the main results in Section~\ref{sec_conclusions}. 

\section{Observational data}
\label{sec_obs}

The two quasars, HE 0413-4031 and HE 0435-4312 were discovered in the Hamburg-ESO slitless survey \citep{wisotzki2000}. These relatively bright quasars, located at the intermediate redshift (z $\sim$ 1), were selected for long-term monitoring with the Southern Large African Telescope (SALT). For these two sources, as well as for the third quasar, CTS C30.10, the time delay of the Mg II line with respect to the continuum was already determined \citep{Czerny2019, Zajacek2020, zajacek2021}, and the wavelength-resolved delays for CTS C30.10 were obtained \citep{prince_CTS_2022}. The monitoring was extended further with the aim to perform wavelength-resolved time delays for the other two objects. The basic properties of the two sources are listed in Table~\ref{tab:basic}. 

   \begin{table*}
       \centering
      \caption[]{Basic properties of the analyzed quasars HE 0413-4031 and HE 0435-4312. From the left to the right column, we list the source designation, redshift, right ascension, declination, visual magnitude, luminosity at 3000\,\AA, SMBH mass (from MgII), and the Eddington ratio estimated assuming the radiative efficiency of $10\%$.}
         \label{tab:basic}
      \begin{tabular}{lccccccc}
            \hline
            \noalign{\smallskip}
            Source      &  redshift & RA & Dec & V [mag] & $\log{(L_{3000}\,[{\rm erg\,s^{-1}}])}$ & $\log{(M_{BH}\,[M_{\odot}])}$ & $\lambda_{\rm Edd}$ \\
            \noalign{\smallskip}
            \hline
            \noalign{\smallskip}
            HE 0413-4031 & 1.3764$^a$ & 04h15m14s & -40d23m41s  & 16.5$^b$ & 46.74 & 8.87$^a$ & 1.66\\
            HE 0435-4312 & 1.2231$^c$  &  04h37m11.8s$^*$ &-43d06m04s & 17.1$^d$ & 46.36$^c$ & 9.34$^e$ & 0.28\\
            \noalign{\smallskip}
            \hline
         \end{tabular}
         \\
         $^a$ \citet{Zajacek2020}, $^b$ \citet{veron2010}, $^c$ \citet{zajacek2021}, $^d$ NED, $^e$ \citet{sredzinska2017}
   \end{table*}

\subsection{Spectroscopic monitoring with SALT}

The monitoring of the two quasars was done in the years 2012 - 2022. The spectroscopic measurements were performed with the Robert Stobbie Spectrograph (RSS; \citealt{burgh2003,kobul2003,smith_salt2006}) in a long-slit spectroscopy mode. The same setup was used over the years for better accuracy of the results.  All the observations were made in a service mode. We collected 31 spectra for HE 0413-4031, and 28 spectra for HE 0435-4312. We used the same methodology of data reduction over the years, as described in more detail in earlier papers \citep{sredzinska2017, Zajacek2020, zajacek2021}.  Every observing block consisted of two $\sim 10 $ min exposures, and RSS PG1300 grating was always used. 
The basic reduction of the raw spectra was done with the standard SALT pipeline \citep{2010SPIE.7737E..25C}. Next, the spectra were fitted and calibrated using the photometric data for that purpose. Later, in the fitting process, the vignetting effect in spectra was corrected using the corresponding calibration stars for each of the objects \citep[see][]{sredzinska2017}, and the spectra were corrected for the Galactic reddening, although the effect is rather small (e.g. HE 0435-4312: A$_V$=0.045, HE 0413-4031: A$_V$=0.034).

\subsubsection{Spectral decomposition method}
\label{sect:decompo}

\begin{figure*}
    \includegraphics[width=0.48\textwidth]{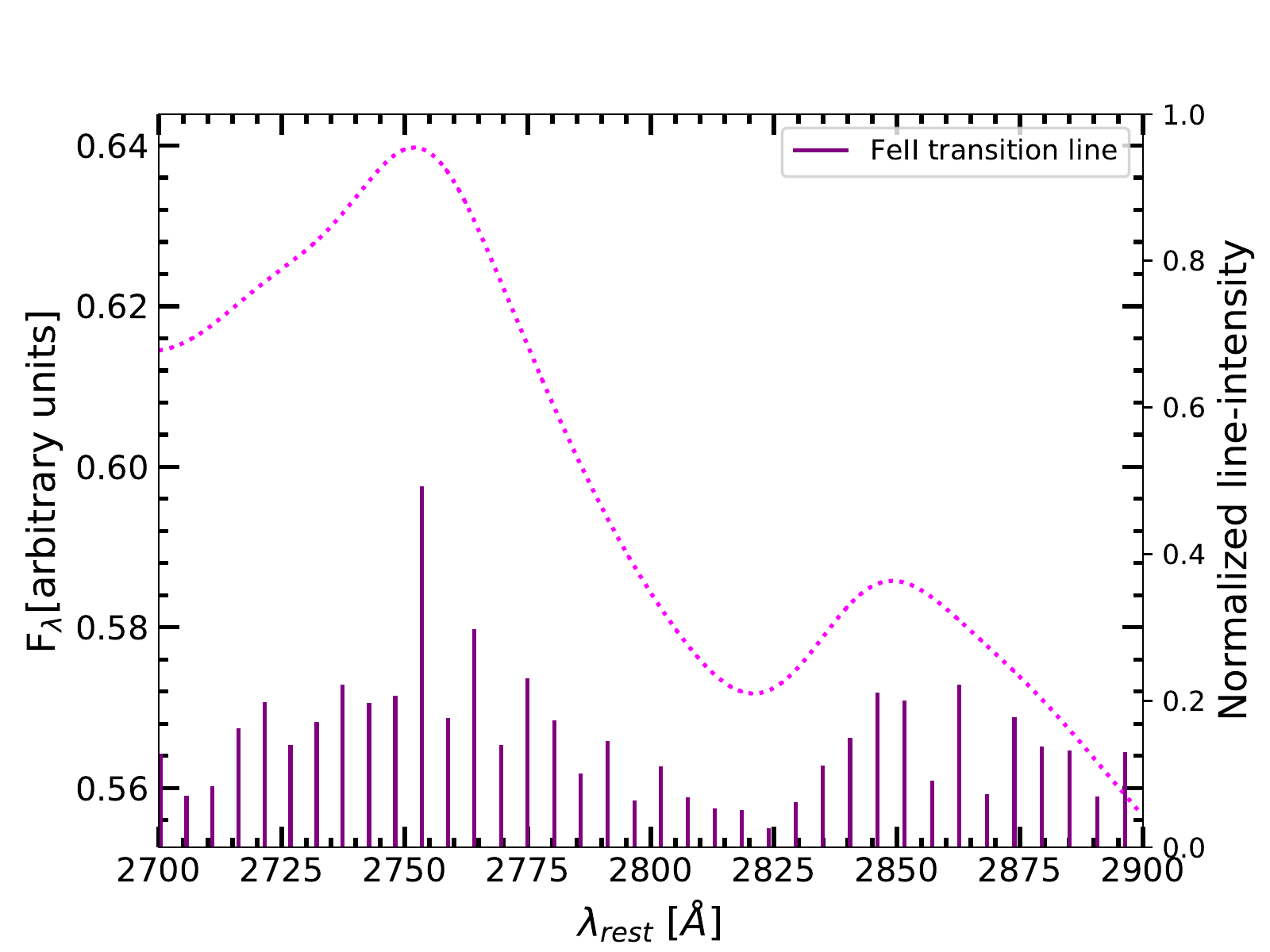}
 \includegraphics[width=0.48\textwidth]{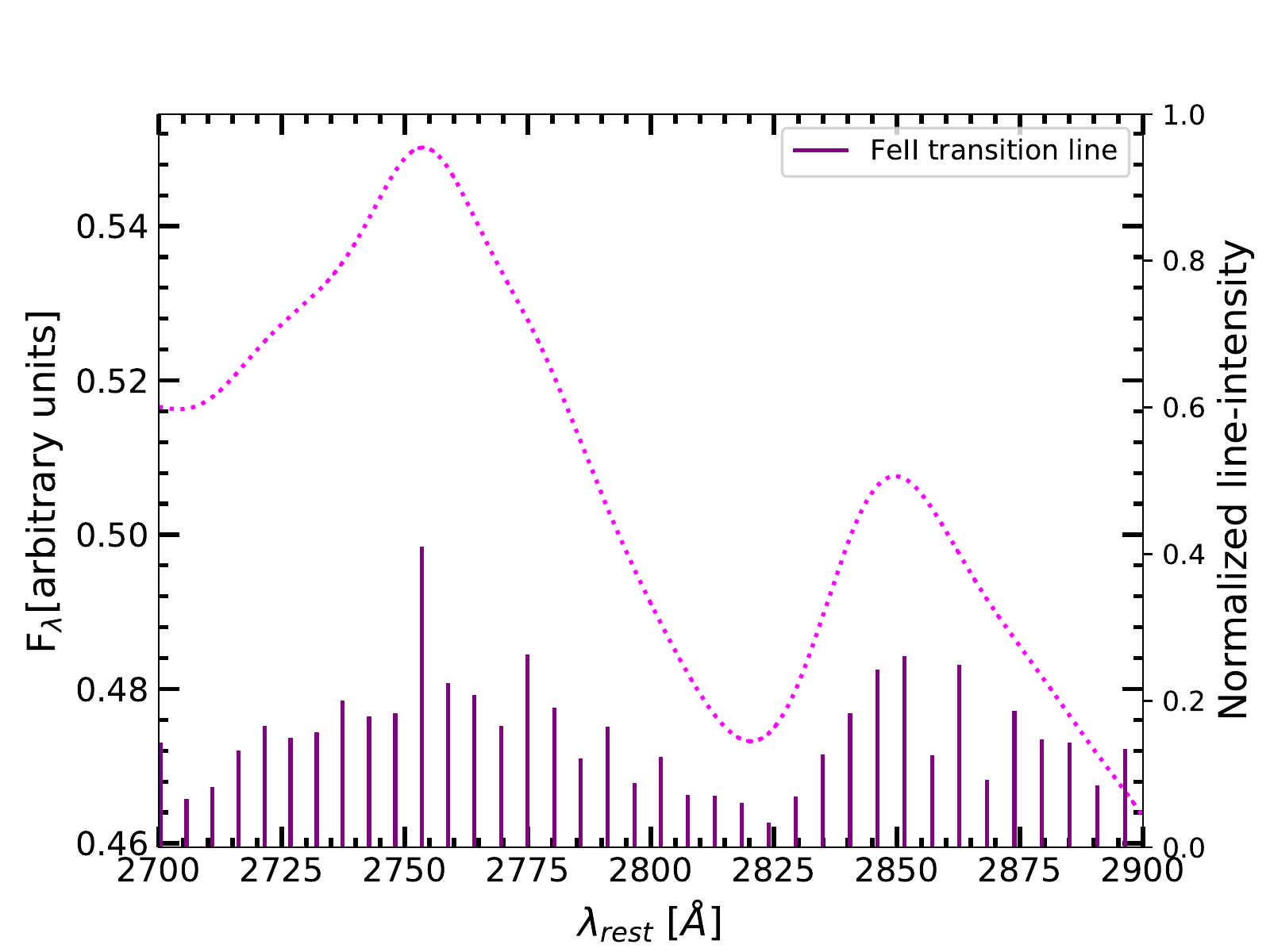}
 \caption{Pseudo-continuum UV FeII emission profile (dotted magenta) obtained by smearing the FeII transition lines (purple) by a velocity profile of 2820 km/s for HE 0413-4031 (left panel) and 2350 km/s for HE 0435-4312 (right panel). Transition lines are theoretical predictions taken from FeII templates \citep{Bruhweiler2008} \texttt{d12-m20-20-5.dat} for HE 0413-4031 and \texttt{d11-m20-20.5-735.dat} for HE 0435-4312.}
 \label{fig:FeII-trans}
\end{figure*}

\begin{figure*}
\includegraphics[width=0.49\textwidth]{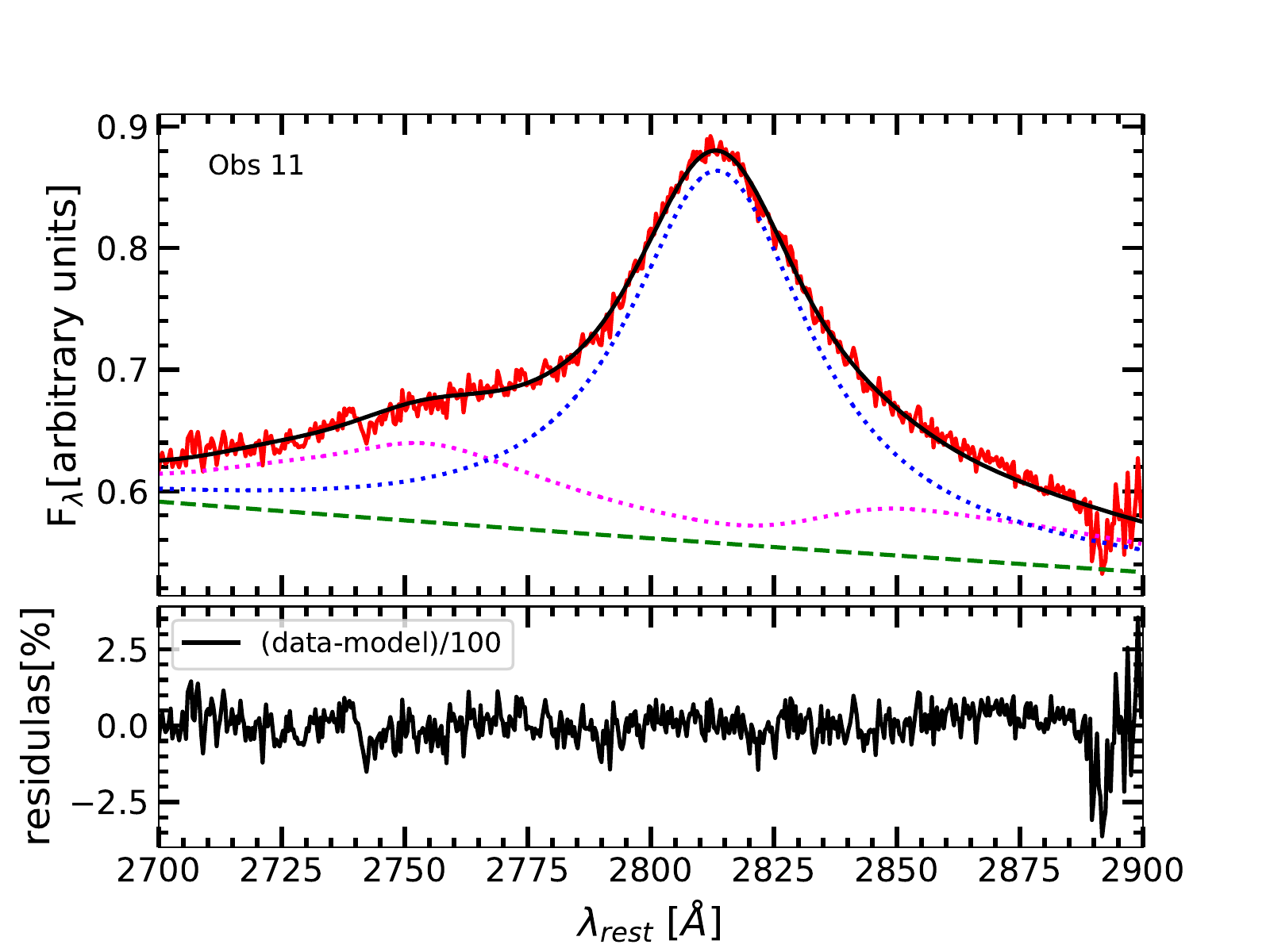}
 \includegraphics[width=0.49\textwidth]{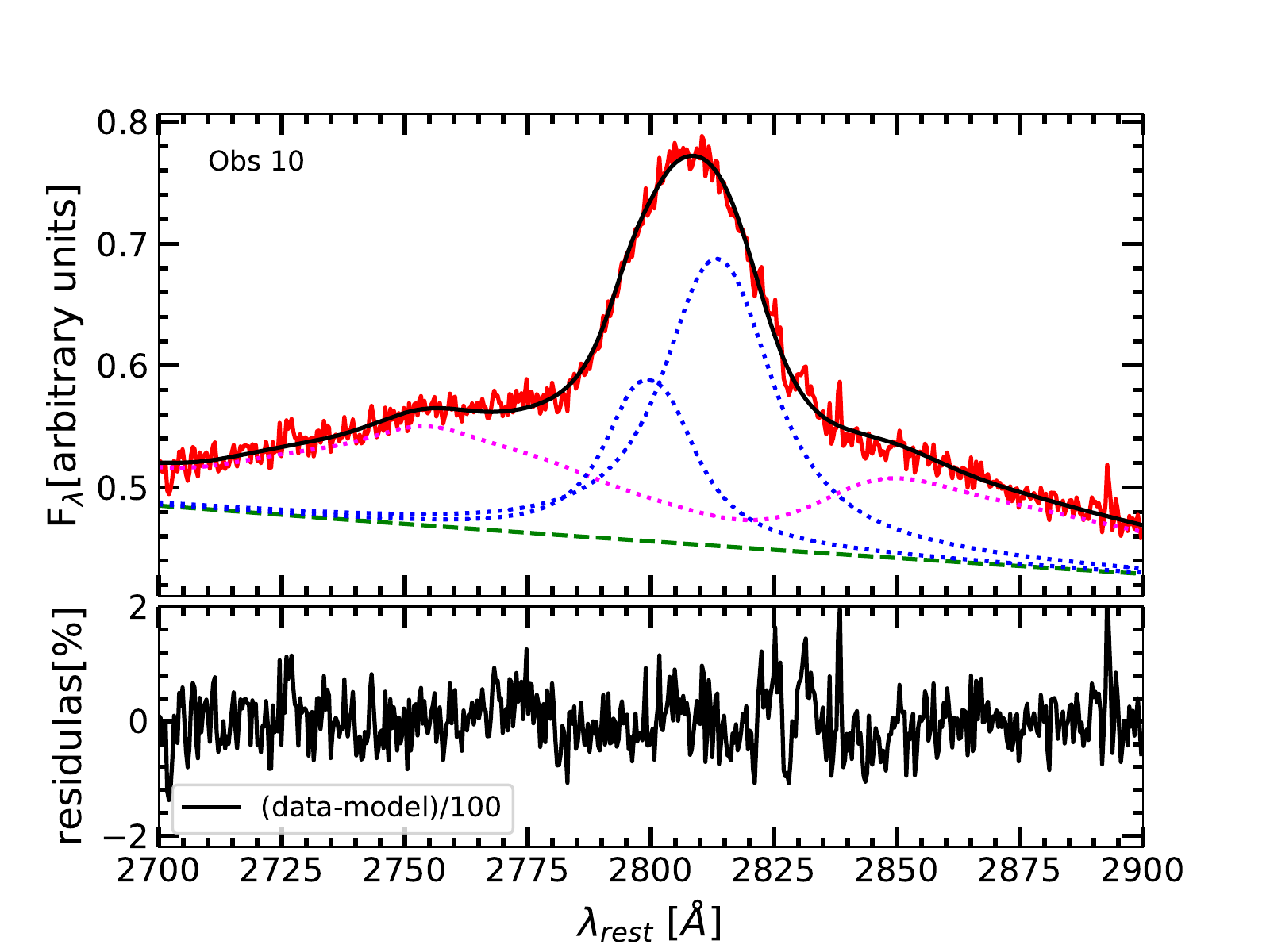}
 \caption{Left panel: Observation 11 from the SALT telescope (red) for HE 0413-4031 fitted with one component (blue) of MgII, FeII pseudo-continuum (magenta), and the power-law (green) representing the continuum from the accretion disk. The lower panel shows the residuals of the fit. Right panel: Observation 10 from the SALT telescope (red) for HE 0435-4312 fitted with two components of MgII (blue) and the rest are the same as the left panel. }
 \label{fig:decomp}
\end{figure*}

The non-normalized spectra were decomposed assuming the following spectral components: a power law with an arbitrary slope and normalization, an Mg II line modeled as two separate kinematic components for HE 0435-4312, and a single kinematic component for HE 0413-4031 since the second kinematic component was not required for this source, as argued by \citet{zajacek2021}. Each kinematic component is modeled as a doublet, with a doublet ratio of 1.9, exhibits a Lorentzian shape, and arbitrary normalization. If two components are present, one of the components is assumed to be at the systematic redshift, and the second one is arbitrarily shifted. If one component is assumed, the arbitrary shift of this component is allowed. We also include Fe II pseudo-continuum emission. This last component was actually assumed to determine the systemic redshift as we could not identify any narrow lines which would establish the redshift more reliably. 

For the shape of the Fe II line, there are multiple templates available that one can use such as \citet{Vestergaard_2001, Tsuzuki_2006} \& \citet{Bruhweiler2008}. The FeII template derived by \citet{Vestergaard2001} is based on Hubble Space Telescope (HST) spectra of the narrow line Seyfert-1 galaxy whereas \citet{Tsuzuki_2006} derived FeII template by cloudy modeling. 
We adopted one of the recent template by \citet{Bruhweiler2008}, \texttt{d12-m20-20-5.dat} model, which assumes the cloud number density of $10^{12}$cm$^{-3}$, the turbulent velocity of 20 km s$^{-1}$, and the hydrogen ionizing photon flux of $10^{−20.5}$cm$^{−2}$s$^{-1}$ for the quasar HE 0413-4031. The FeII template \texttt{d11-m20-20.5-735.dat} was applied to the quasar HE 0435-4312, corresponding to the lower local density of the cloud of $10^{11}$cm$^{-3}$. The templates have a fixed shape but arbitrary normalization. The FeII transition lines identified from the cloudy simulations are shown in Fig~\ref{fig:FeII-trans} (vertical purple lines) which were broadened by the respective Doppler broadening for both the sources and eventually broad FeII emission lines are obtained which is shown in dotted magenta color in Fig ~\ref{fig:FeII-trans}. The fits are not very sensitive to the specific choice of a template, and we discuss this issue in Appendix D.
The template broadening (those are theoretical templates, with no intrinsic broadening) was kept at 2820 km s$^{-1}$ and 2350 km s$^{-1}$ for HE 0413-4031 and HE 0435-4312, respectively, in all the fits. All the spectral parameters were fitted together. The spectra were fitted only in the narrow wavelength range, between 2700 and 2900 \AA~, in the rest frame. We show the exemplary data fit for the two quasars in Figure~\ref{fig:decomp}. The fit quality is generally good with the residuals mostly within 1 \%, only somewhat higher at the longest wavelengths for the quasar HE 0413 due to strong skylines in this part of the spectrum, which even after subtraction exhibit residual errors. 
The kinematic width of Fe II and Mg II from spectral fitting implied that Fe II lines are narrower than the mean total profile of Mg II (4380 km s$^{-1}$, \citealt{Zajacek2020} and 3695 km s$^{-1}$, \citealt{zajacek2021}). We tested the effect of other kinematic widths of the Fe II templates on the fit quality and the results are not very sensitive to the adopted values within a few hundred km s$^{-1}$ (see Appendix D). This is due to the fact that the number of transitions in the 2700 - 2900 \AA ~range is quite high (37 - 61 transitions, depending on the template) so the component is always very broad. We cannot treat the Fe II width as just another parameter for computational reasons: the Fe II template at each wavelength is calculated as a convolution with a Gaussian, which is time-consuming. If the width is kept constant during the fitting process, this convolution is calculated just once, and only the amplitude varies during iterations.

The decomposition of all data sets for the two quasars shows an interesting difference. For HE 0435, the wings of the Mg II line on both sides are well below the Fe II emission contribution. However, for HE 0413, only the left wing of Mg II is strongly dominated by Fe II, while the right wing is comparable to Fe II. This can potentially affect the wavelength-resolved time delays.

Spectral fits allowed us to determine the equivalent widths (EWs) of the Mg II and Fe II UV in each of the spectra. EW(Fe II) was measured only in the 2700 - 2900 \AA~ band.

\subsection{Photometric monitoring}

Spectroscopic monitoring was supplemented with photometric monitoring, with the use of a number of instruments. Some of the data (monitoring in the V band) were collected as a part of the  OGLE-IV survey \citep{udalski2015} which used the 1.3 m Warsaw telescope at the Las Campanas Observatory, Chile. In the later period, the quasars were observed with the 40 cm  Bochum Monitoring Telescope (BMT), again in the V band. A part of the data, over all years, comes from the SALT measurement with the SALTICAM, however, the SALTICAM exposure was not always available. Those exposures (usually two-three exposures for 45 sec) were done in the g band. Therefore, the data required cross-calibration, as already mentioned in \citet{Zajacek2020} and \citet{zajacek2021}. 

In the case of HE 0435-4312, the data obtained by us were supplemented with the old CATALINA data \footnote{https://catalina.lpl.arizona.edu/}, which were binned to decrease the statistical error. The data were required to cover photometrically the start of the spectroscopic campaign. In the case of the other quasars, the photometric coverage was good enough and the CATALINA data were not needed, so we did not include them in our analysis. The photometric data (V and g bands) of ASAS-SN \footnote{https://www.astronomy.ohio-state.edu/asassn/} are also used for both sources. The magnitude in the V and  g bands is intercalibrated. The light curves are shown in Figure~\ref{total_mg_fe} (upper panel).

\subsubsection{Calibration of the spectra}

\begin{figure*}
    \centering
    \includegraphics[scale=0.36]{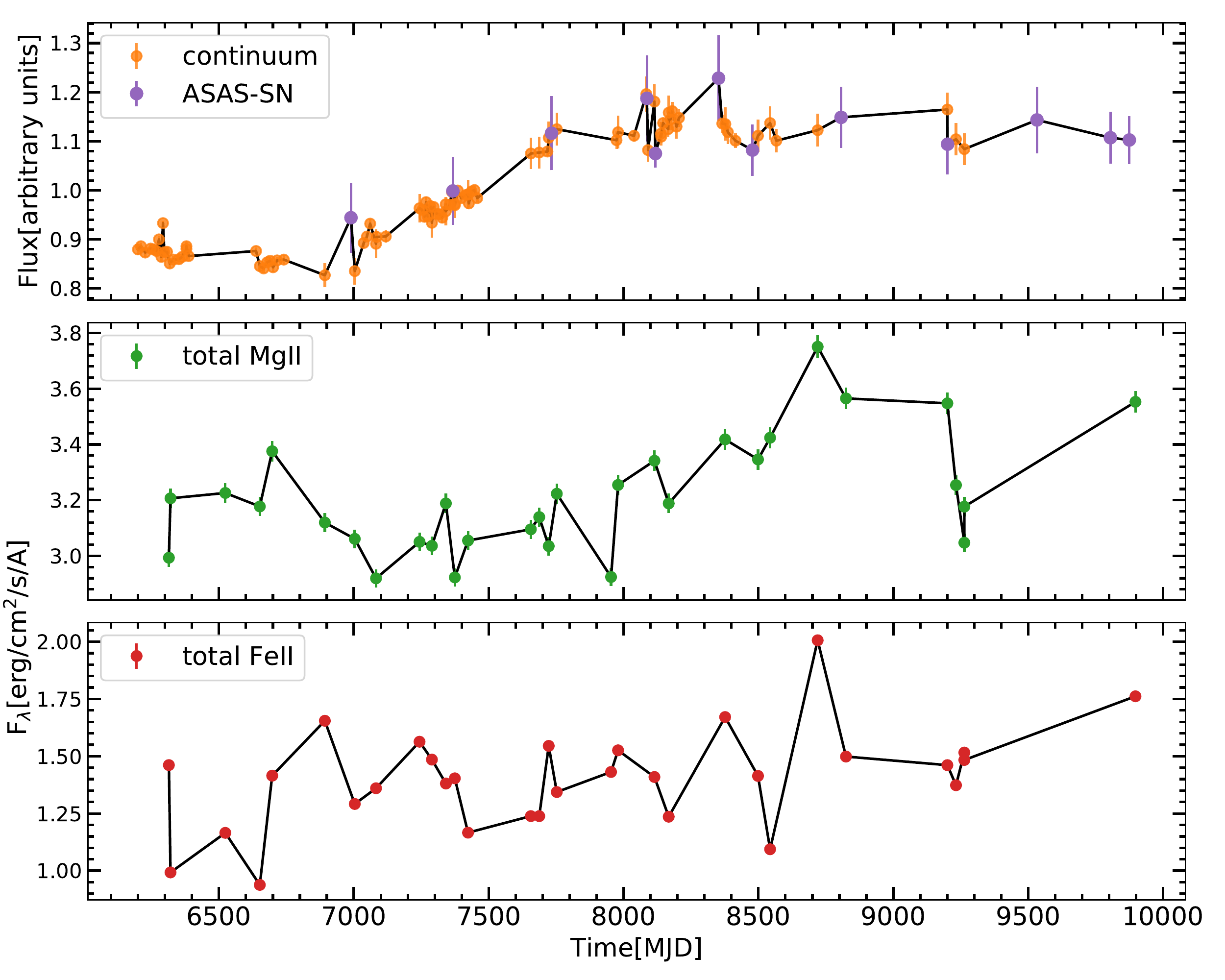}
    \includegraphics[scale=0.36]{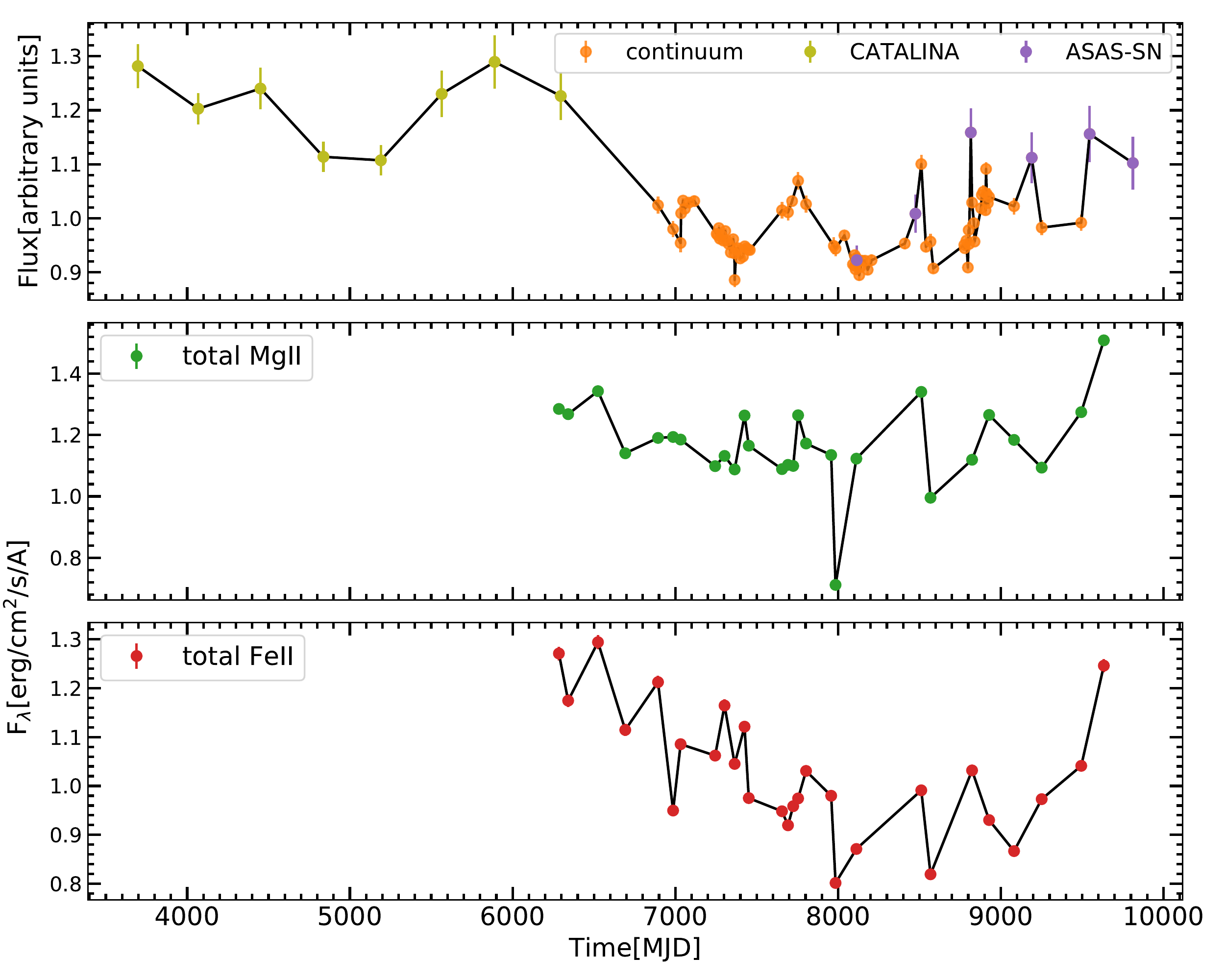}
    \caption{Long-term normalized continuum light curve (top panels) along with total MgII (middle panels) and FeII (bottom panels) line emission light curve. Left panel: HE 0413-4031. Right panel: HE 0435. Total MgII and FeII are in units of 10$^{-16}$ ergs cm$^{-2}$s$^{-1}$ $\AA^{-1}$. The fluxes from archival CATALINA and ASAS-SN are marked in color and the continuum represents the observation from SALT+OGLE+BMT.}
    \label{total_mg_fe}
\end{figure*}

Having the photometry, we calibrated the spectra using the photometric measurements corresponding to the spectroscopic measurement date, or they were interpolated through a weighted spline interpolation of the first order. To connect the photometric flux density with the modeled non-normalized continuum, we took the V magnitudes reported in Table~\ref{tab:basic} to get the continuum flux in the observed and rest-frames. Then, to obtain the continuum flux at 2800\AA, we used the composite quasar spectrum from \citet{vandenberk2001} with a slope of $\alpha_\lambda=-1.56$. The same procedure was followed in \citet{zajacek2021}. This allowed us to obtain the spectra in physical units as well as to determine the time evolution of the Mg II and Fe II intensity. They are presented in Figure~\ref{total_mg_fe}. For HE 0435-4312, we have also used binned CATALINA data since the available photometry from SALT+BMT+OGLE did not cover the spectroscopic points. For HE 0413-4031, we did not have this issue and hence we have not included CATALINA data as mentioned earlier. 

\section{Time-delay measurement methods}
\label{sec_time_delay}

Since our data are of variable quality, generally heterogenous, and irregularly sampled, we made use of several independent methods to determine the time delays, as was previously done in e.g. \citet{Zajacek2020}, \citet{zajacek2021}, \citet{2020A&A...642A..59R}, and \citet{prince_CTS_2022}. Those methods include the Interpolated Cross Correlation Function (ICCF; \citealt{Gaskell1987, Peterson1998, Peterson2004}), Javelin \citep{Zu2011}, $\chi^2$ \citep{Czerny2013}, zDCF \citep{alexander1997}, and von Neumann and Bartels estimators of data regularity/randomness \citep{2017ApJ...844..146C}. \\
{\bf Interpolated Cross-Correlation Function (ICCF):} Line light curves and the continuum variable emission were cross-correlated to measure the centroid and the peak time delays. We search for the time lag between 200 to 1200 days to focus on the main peak in this range. In addition, for the time delays shorter than 200 days and longer than 1200 days, there are additional time-delay peaks that would skew the time-delay peak distributions and thus enlarge the uncertainties to several hundreds of days. We report the two time lags from the ICCF, namely the centroid which is determined for $r > 0.8 $ ($\tau_{cen}$), where $r$ is the correlation coefficient, and the peak of the lag distribution which corresponds to the maximum of the coefficient is denoted as $r_{max}$ ($\tau_{peak}$). The peak and centroid distributions were created by simulating the 1000 realizations of line and continuum light curves using the flux-randomization (FR) and random-subset selection (RSS) methods (see \citealt{Peterson1998}). The errors of the peak and the centroid time lag are measured by considering the entire distribution and hence the asymmetric error bars are large, of the order of 100 days.
Apart from the time delay measurements and the associated errors, we also provide the time delay corresponding to the maximum value of the ICCF considering the original light curves, $r_{\rm max}$, since it informs us directly about the data quality. We also restrict ourselves to $r$ $>$ 0.2 which is generally considered an upper limit for insignificant correlation.
\\ 
\begin{figure}
    \centering
    \includegraphics[width=\columnwidth]{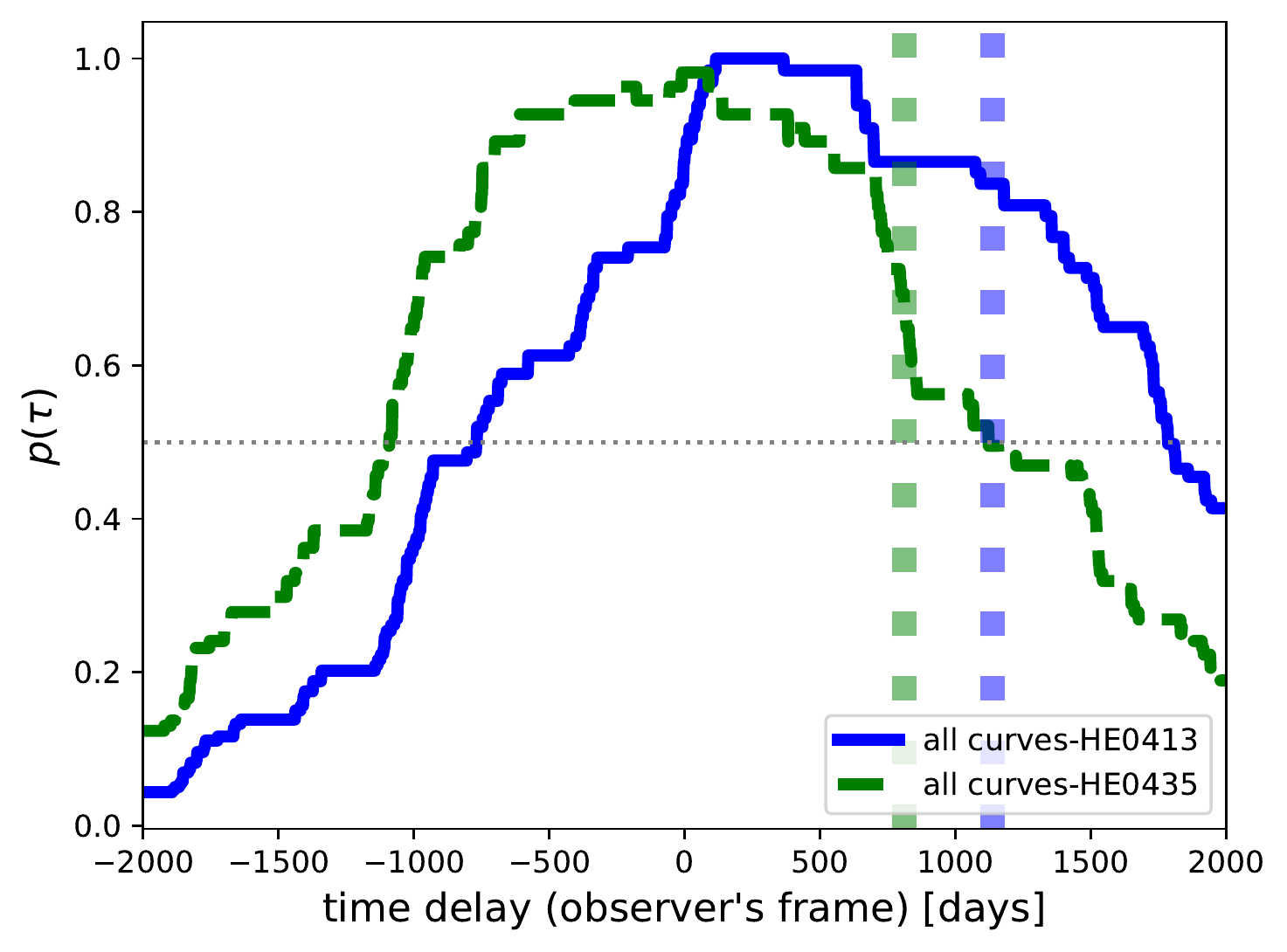}
    \caption{ Light-curve pair distribution function $p(\tau)=[N(\tau)/N(0)]^2$ for the continuum/line-emission light curves of HE0413 (solid blue line) and HE0435 (dashed green line), where $N(\tau)$ is the total number of continuum and line-emission data points for the time delay $\tau$ (time step is 1 day). Vertical dotted lines stand for the expected time delays in the observer's frame inferred from the MgII $R-L$ relation, see Fig.~\ref{fig_MgII_RL} for the best-fit relation based on 94 measurements. The values of $p(\tau)$ at these time delays are above the 0.5 level for both quasars.}
    \label{fig_ptau}
\end{figure}
{\bf Javelin:}  We applied the \texttt{python} version of the Javelin method, whose detailed description can be found in \citet{Zu2011}. It models the continuum variability as a Damped Random Walk (DRW) process \citep{Kelly2009, Zu2013}, and the line emission is considered to be a smoothed, scaled, and time-shifted version of the continuum light curve.  We search the time lag between 0 to 2000 days and also perform 1000 bootstrap realizations to estimate the lag distribution. The peak of the lag distribution and the errors are then calculated. For the error estimation, we consider the entire distribution, and hence the error bars are relatively large. Following \citet{zajacek2021}, we also perform  the alias mitigation \citep{2017ApJ...851...21G} by calculating the light curve pair distribution function, $p(\tau)=[N(\tau)/N(0)]^2$, where $N(\tau)$ is the total number of the continuum-line emission pairs for the time delay $\tau$, and $N(0)$ corresponds to the light curve pair number for the zero time delay. We show the light-curve pair distribution function for both quasars in Fig.~\ref{fig_ptau}, where we also depict the level $p(\tau)=0.5$, which is reached for time delays longer than 1000 days in the observed frame. On the other hand, the expected time delays for the two quasars marked by dashed vertical lines are above the 0.5 level, which implies that in this range the pair statistics are sufficient. The time lag distribution constructed from several bootstrap realizations is then weighted by the light-curve pair distribution function $p(\tau)$, which results in lowering the peaks at long time delays due to a small number of overlapping data points. After applying the alias mitigation, we noticed that for one source it really affects the time lags on longer time scales whereas, for the second source, it does not have much effect. \\
In addition to the above two methods, we also applied four other, model-independent methods to cross-check the robustness and the consistency of the inferred time lags.\\
{\bf $\chi^2$:} The $\chi^2$ method interpolates the continuum and the line-emission light curves and calculates the $\chi^2$ value for a given time lag. For the reverberation mapping measurements, it was applied in \citet{Czerny2013}. We considered the time lag range between 0 and 1500 days. The search range is shorter than for the ICCF method since the $\chi^2$ method exhibits deep $\chi^2$ minima for some of the continuum/line-emission light curve pairs for time delays longer than 1500 days. These would skew the time-delay peak distributions. Based on the comparison with other time-delay methods as well as the light-curve pair distribution $p(\tau)$ (see Fig.~\ref{fig_ptau}), these minima are not statistically significant and arise due to artifacts/aliases in the light curves. Therefore we exclude them from the time-delay analysis. The best time lag was subsequently determined from the time-lag peak distribution constructed from 10000 bootstrap realizations. The uncertainty was determined from the surroundings of the peak value ($\pm 50\%$ of the time-lag peak), for which we calculated the asymmetric $1\sigma$ uncertainties.\\
{\bf zDCF:} The $z$-transformed discrete correlation function (zDCF) is an improvement of the classical discrete correlation function \citep[DCF;][]{Edelson1988}, for which the equal-time binning is replaced by the equal-population binning \citep{alexander1997}. The final time delay is determined from the maximum-likelihood method, and the uncertainties correspond to 1$\sigma$ uncertainties determined from 1000 Monte Carlo flux-randomization realizations.\\ 
{\bf Von Neumann estimator:} Von Neumann estimator belongs to the estimators of data regularity. It estimates the regularity of the continuum and line-emission light curves combined together with a certain time delay $\tau$. The minimum value of the estimator may be considered to represent the best candidate time delay \citep{2017ApJ...844..146C}. The final time delay was determined from the peak of the distribution of the von Neumann estimator minima. To obtain the distribution, we generated 1000 light curve pairs using random subset selection or bootstrap for the wavelength-resolved analysis, while we generated 1500 pairs for the total time delay analysis. The uncertainties of the time lag peak were determined in the same way as for the $\chi^2$ method.\\ 
{\bf Bartels estimator:} It is a ranked version of the von Neumann estimator of the data regularity \citep{2017ApJ...844..146C}. To determine the final time delay peak and its 1$\sigma$ uncertainty, we performed 1000 bootstrap realizations of the continuum and the line-emission light curves for the wavelength-resolved analysis, while we generated 1500 light-curve pairs for the total time-delay analysis. The details about these methods and their application to reverberation mapping of intermediate-redshift quasars can be found in \citet{Zajacek2020} and \citet{zajacek2021}. 
\\

\section{Results}
\label{sec_results}

We perform the time delay analysis for both the total Mg II and Fe II components as well the wavelength-resolved time delays. We do not convert the wavelength to velocity space since our aim, as in \citet{prince_CTS_2022}, is to analyze the delays of the combination of both the Mg II and the Fe II emission, instead of just analyzing the change of the delay along the line profile, as was, for example, performed in \citet{Hu_2020}. 

\subsection{Mg II and Fe II time delay}

The continuum, the Mg II, and the Fe II light curves determined from the normalized SALT data are presented in Figure~\ref{total_mg_fe}. All the light curves exhibit a reasonable quality, and the level of variability ($F_{\rm var}$) determined for the SALT observations (2012 - 2022) of HE 0413-4031 is 11$\%$ for the continuum, 15$\%$ for the FeII emission, and 6$\%$ for the MgII emission. For HE 0435-4312, the variability of MgII, FeII, and the continuum is 12$\%$, 13$\%$, and 6$\%$, respectively. The $F_{\rm var}$ estimated for HE 0435-4312 does not include the CATALINA data. The quasar HE 0413-4031 exhibits a long-term variation corresponding to a higher F$_{\rm var}$ rather than a short-term variation that is more characteristic of the HE 0435-4312 continuum emission, which exhibits a rather low level of variation.  
The continuum flux density variation of the quasar HE 0413-4031 shows a clear rising trend followed by a slightly decreasing trend, while in the case of the second quasar, the SALT data covers the minimum. The flux variation of the line emission follows the continuum with a time shift and these trends help to establish the time delay relatively well.

\begin{table*}[]
    \centering
     \caption{Overview of the time-delay determinations for the total MgII and FeII line emissions. The time delays are expressed in days with respect to the observer's frame of reference unless otherwise stated. The estimated errors are $1\sigma$ standard deviations. For ICCF, we also show the maximum correlation coefficient in the parenthesis.}
    \begin{tabular}{c|c|c}
    \hline \hline \noalign{\smallskip}
    Method  & MgII total [days] & FeII total [days] \\
    \hline \noalign{\smallskip}
     & HE 0435 & \\
      \hline \noalign{\smallskip}
    Javelin (peak) & 489.0$^{+6.1}_{-15.7}$ & 473.0$^{+39.4}_{-3.4}$ \\
    Javelin (mean) & 863.7$^{+123.8}_{-126.2}$ & 722.8$^{+42.3}_{-106.4}$ \\
     \hline 
    \noalign{\smallskip} 
     ICCF ($r_{\rm max}$) & 620.0 (0.45) &  494.0 (0.64)  \\
     ICCF (centroid) &   $580.5^{+64.5}_{-66.3}$ &  $600.3^{+302.9}_{-90.9}$ \\
    ICCF (peak) & $620.0^{+20.4}_{-159.0}$  &  $620.0^{+255.0}_{-178.0}$\\ 
      \hline 
    \noalign{\smallskip} 
    $\chi^2$ (min) &  $516.30$ & $526.26$ \\
    $\chi^2$ (bootstrap) & $636^{+177}_{-247}$    & $636^{+50}_{-181}$   \\
   \hline \noalign{\smallskip}
        zDCF & $629.0^{+78.6}_{-152.4}$  & $629.0^{+83.0}_{-129.6}$   \\
     \noalign{\smallskip}   
    \hline \noalign{\smallskip}
        von Neumann (min) & 560.0    & 501.0   \\ 
    von Neumann (bootstrap) & $615.0^{+58.0}_{-73.8}$ & $685.0^{+4.0}_{-223.0}$  \\
      \noalign{\smallskip}
      \hline \noalign{\smallskip}
     Bartels (min) & 594.0   &  538.0  \\ 
     Bartels (bootstrap) & $ 575.0^{+20}_{-74}$ & $685.0^{+22.0}_{-214.0}$ \\ 
     \noalign{\smallskip}
    \hline
    Mean observed time delay  &  $626.0^{+133.0}_{-166.3}$  &  $631.4^{+162.1}_{-172.0}$ \\
    \noalign{\smallskip}
    Mean rest-frame time delay &  $281.6^{+59.8}_{-74.8}$ & $284.0^{+72.9}_{-77.4}$ \\
      \hline \hline
       \noalign{\smallskip} 
     & HE 0413 & \\
    \hline \noalign{\smallskip}
    Javelin (peak) & 925.4$^{+68.6}_{-143.1}$ & 957.7$^{+125.3}_{-259.9}$  \\
    Javelin (mean) & 774.3$^{+82.9}_{-117.8}$ & 931.3$^{+113.3}_{-248.8}$ \\
   \hline \noalign{\smallskip}
    ICCF ($r_{\rm max}$) & $644.0$ (0.82) & $761.0$ (0.52)     \\
     ICCF (centroid) &  $684.0^{+73.8}_{-73.8}$ &   $718.5^{+108.8}_{-70.2}$
            \\
    ICCF (peak) &  $673.0^{+120.0}_{-40.0}$ & $717.0^{+79.0}_{-81.0}$           \\ 
          \hline 
    \noalign{\smallskip} 
     $\chi^2$ (min) &   $718.36$        &  $751.35$    \\ 
     $\chi^2$ (bootstrap) & $716.0^{+23.4}_{-92.7}$ & $652.0^{+43.3}_{-50.7}$ \\
      \hline
       \noalign{\smallskip} 
     zDCF & $654.4^{+133.1}_{-100.0}$  & $738.7^{+39.1}_{-136.0}$   \\
      \hline \noalign{\smallskip}
    von Neumann (min) & 721.0    & 734.0   \\ 
    von Neumann (bootstrap) & $815.0^{+26.0}_{-141.0}$ & $785.0^{+27.0}_{-274.0}$  \\    
     \noalign{\smallskip}
       \hline \noalign{\smallskip}
     Bartels (min) & 721.0    &  737.0  \\ 
     Bartels (bootstrap) & $735.0^{+76.0}_{-70.0}$ & $ 785.0^{+27.0}_{-170.0}$  \\  
    \noalign{\smallskip} 
      \hline
    Mean observed time delay & $747.1^{+118.6}_{-132.8}$  & $785.7^{+128.2}_{-208.0}$ \\
    \noalign{\smallskip}
    Mean rest-frame time delay & $314.4^{+49.9}_{-55.9}$ & $330.6^{+54.0}_{-87.5}$ \\   
   \hline \hline
   \noalign{\smallskip} 
    \end{tabular}
    \label{tab_totalMgII_FeII}
\end{table*}

We used all the methods mentioned in section 3 to derive the time lags between the continuum and the total Mg II and the Fe II line emissions.
For the quasar HE 0435, the measured Mg II time delay depends on the method, but the results are within the values 489 - 636 days in the observed frame, corresponding to 220 - 287 days in the quasar rest frame. The time delay values for the quasar HE0413 are 654 - 925 days in the observed frame and 275 - 389 days in the rest frame. The results for each method are reported in Table~\ref{tab_totalMgII_FeII}. The non-weighted average rest-frame time delay of the MgII emission is $281.6^{+59.8}_{-74.8}$ days and $314.4^{+49.9}_{-55.9}$ days for HE 0435-4312 and HE 0413-4031, respectively. The uncertainties were inferred by the sum of variances corresponding to the mean-variance determined from all the methods and the standard deviation of the mean value. They are generally consistent with the rest frame values determined by \citet{zajacek2021} for the quasar HE 0435-4312 - ${296}_{-14}^{+13}$ days - and by \citet{Zajacek2020} for the quasar HE 0413-4031 - ${302.6}_{-33.1}^{+28.7}$ days.  

The FeII line emission time lag was not previously reported since the number of spectroscopic measurements was not high enough to obtain a significant result. Here we report the first UV Fe II time delay measurements for the quasars HE 0435-4312 and HE 0413-4031. For HE 0435-4312, the time delays lie between 473 - 685 days in the observed frame and between 212 - 308 days in the rest frame. For the other quasar HE 0413-4031, the time delays are a bit higher, between 652 - 957 in the observed frame, and 274 - 402 days in the rest frame. The total Mg II and Fe II time delays for these two quasars are thus consistent within uncertainties. We note that the Javelin method applied to both sources yielded a longer time delay than those implied by all the other methods. However, the alias mitigation, i.e. weighting of time-delay peak distributions by the light curve pair-number distribution, affected the longer time delay (see Figures of Appendix A) for HE 0435-4312, and therefore we report a shorter time delay of about 489 (Mg II) days and 473 (Fe II) days. The pair-weighting does not have much effect on the other quasar, and hence has longer Javelin time delays of about 925 and 957 days for Mg II and Fe II line emissions, respectively.
As we see in Table~\ref{tab_totalMgII_FeII},
the time lag results vary among the methods, though they are generally consistent within the uncertainties. For the total FeII line emission, we obtain the mean rest-frame value of $284.0^{+72.9}_{-77.4}$ days and $330.6^{+54.0}_{-87.5}$ days for  HE 0435-4312 and HE 0413-4031, respectively. 

The mean MgII and FeII time delays are in generally consistent within their respective uncertainties. For HE 0435-4312, the mean value of FeII is larger only by $\sim 0.85\%$ in comparison with the mean MgII time lag. For this quasar, only ICCF (centroid), von Neumann and Bartels estimators indicate that the UV FeII emission has a slightly longer time delay with respect to the MgII emission. For HE 0413-4031, the difference between the mean values is larger --  the mean FeII time delay is by $\sim 5.15\%$ longer than the MgII time delay. For this quasar, Javelin, ICCF (centroid and peak), zDCF, and the Bartels estimator support this trend.

\subsection{Wavelength-resolved time delays}

\begin{figure*}
    \centering
    \includegraphics[scale=0.45, trim={0 3cm 0 4cm},clip]{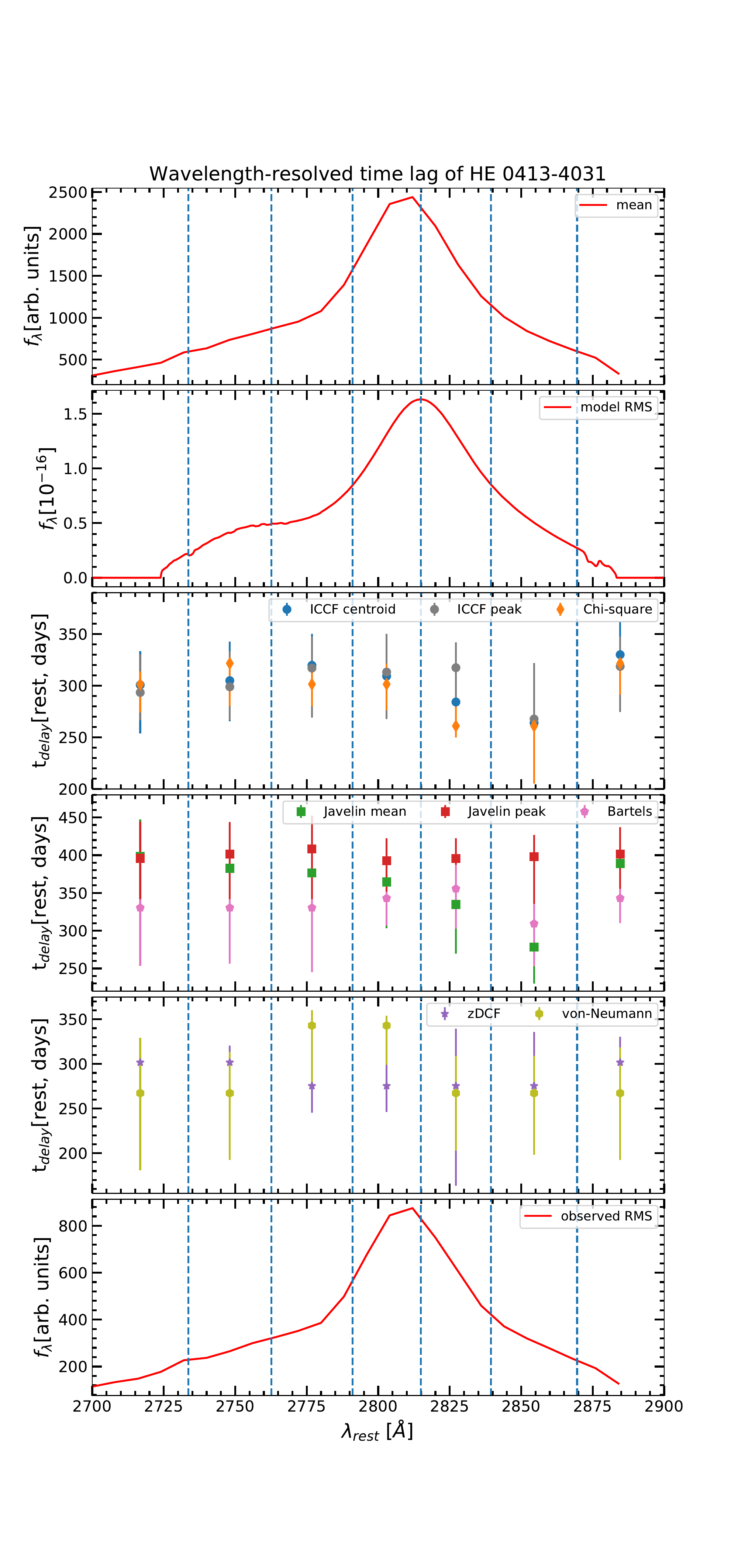}
    \includegraphics[scale=0.45, trim={0 3cm 0 4cm},clip]{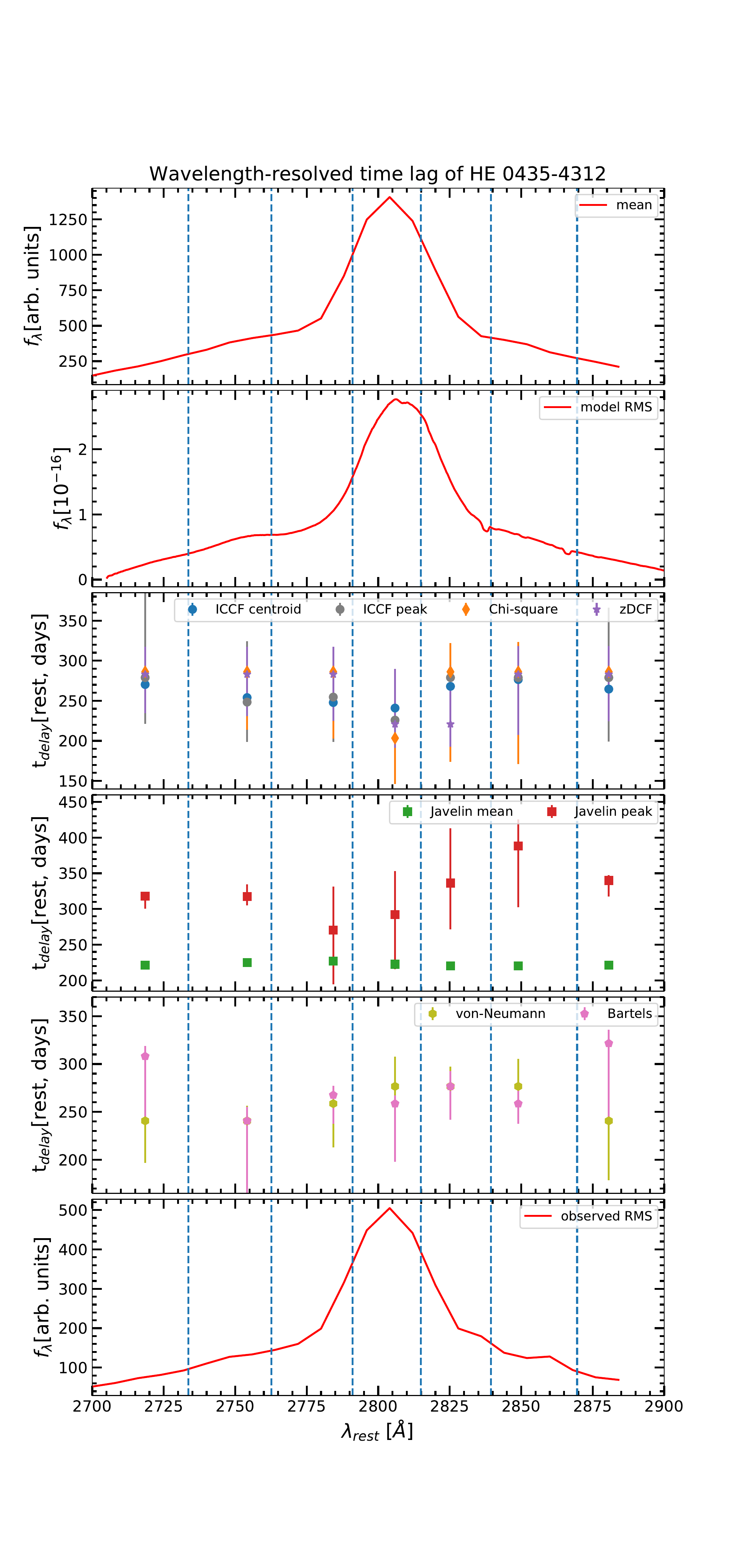}
    \caption{Mean and RMS (model) spectra for the two quasars, with dashed lines indicating the division of the wavelength range. Time delays in the rest frame are plotted in different panels. }
    \label{fig:mean_rms}
\end{figure*}

\begin{figure*}
    \centering
    \includegraphics[scale=0.41, trim={3 2cm 2 4cm},clip]{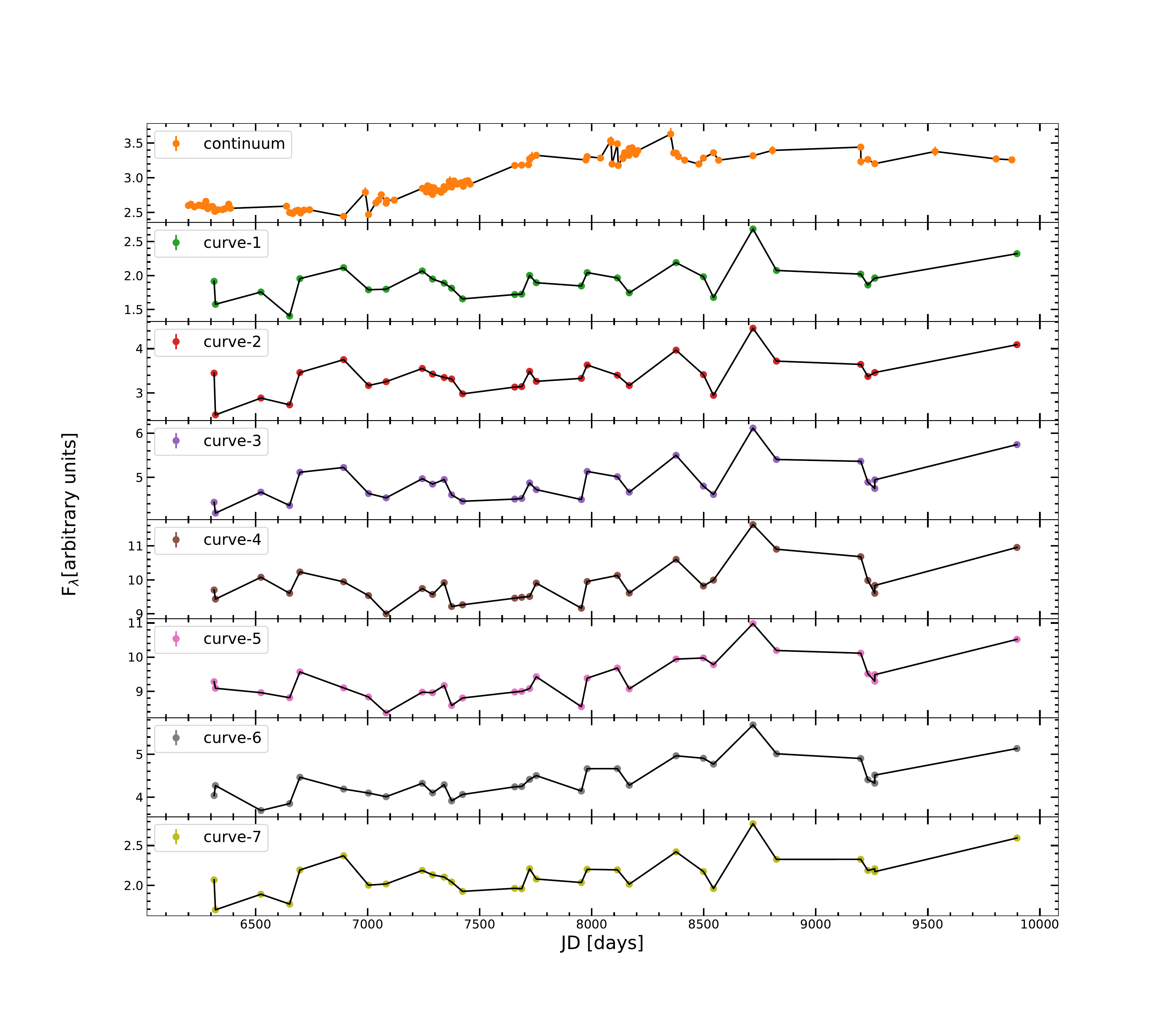}
    \includegraphics[scale=0.41, trim={3 2.5cm 2 4cm},clip]{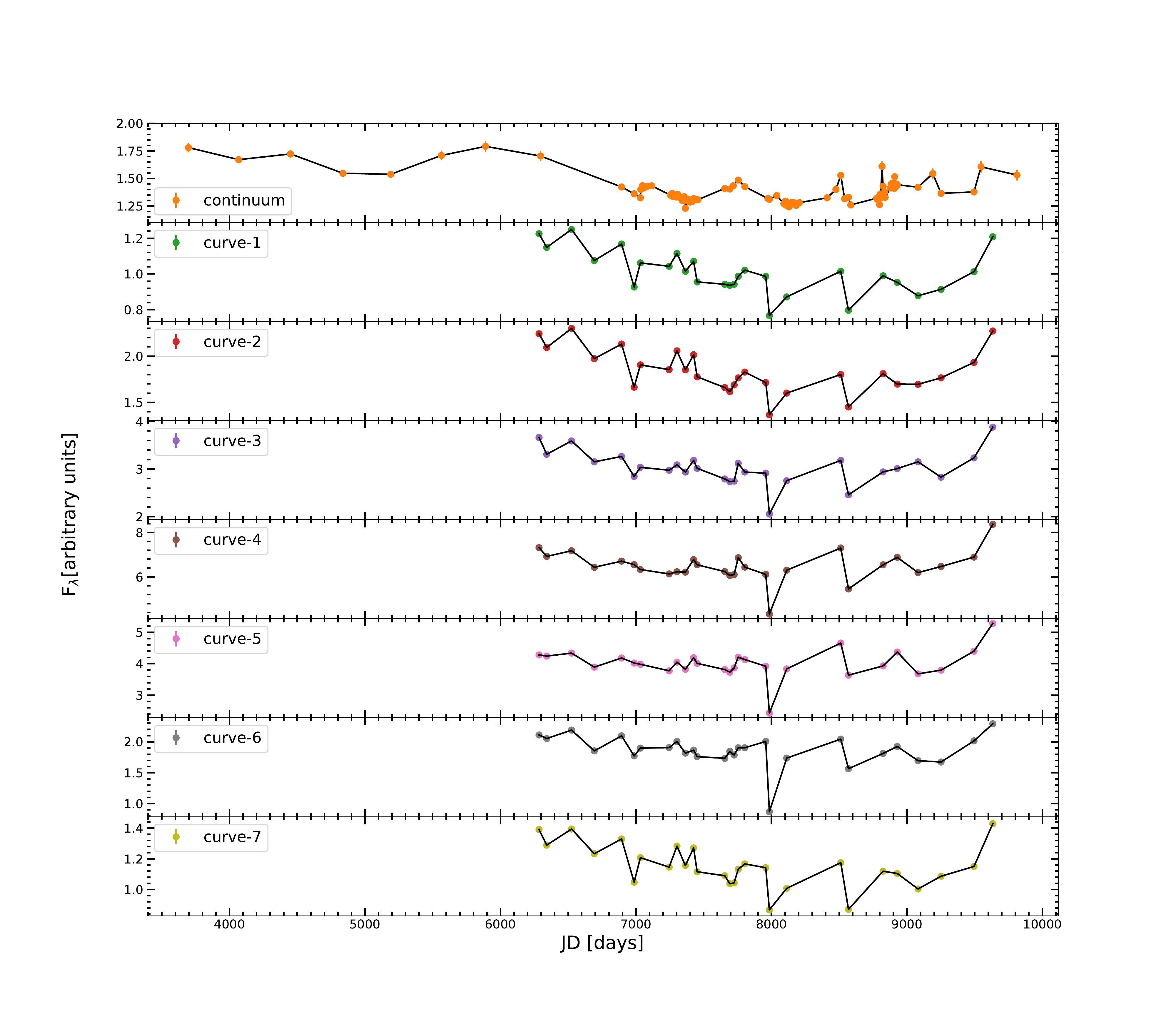}
    \caption{The light curves extracted for the different parts of the RMS spectra. The top panel is for quasar HE 0413-4031 and the bottom one is for HE 0435-4312. }
    \label{fig:all_LC}
\end{figure*}

We follow the approach used by \citet{Hu_2020} in their wavelength-resolved studies, which we also applied previously to analyze the properties of the CTS C30.10 source \citep{prince_CTS_2022}. We subtract the continuum component from each spectrum, which leaves the combined contribution of Mg II and FeII pseudo-continuum. Next, we create individual light curves by dividing the studied spectrum in the 2700 - 2900 \AA ~rest frame range into seven bins, which are not of equal separation in the wavelength but of an equal integral of the RMS spectrum. This choice is much better since it allows a better resolution where the signal-to-noise ratio is higher. We cannot use more bins since the overall data quality does not allow for denser wavelength sampling. The mean quasar spectrum in the fitted range and the RMS spectrum (observed and model) indicating the spectral ranges are presented in Figure~\ref{fig:mean_rms}. The vertical blue dashed lines divide the mean and RMS spectra into different parts that were utilized for wavelength-resolved reverberation mapping. The derived light curves corresponding to different parts of the RMS spectrum are shown in Fig ~\ref{fig:all_LC} for both the quasars.
This division is crucial to understand the detailed kinematics, i.e. the inflow and the outflow on top of the dominant Keplerian motion of the BLR clouds, where the broad-line emission is produced, around the SMBH. 

Since the FeII pseudo continuum lies in the two wings of Mg II, the side curves (1,2,6,7) are expected to be dominated by the Fe II emission, while the central curves (3,4,5) are dominated by the Mg II emission. Subsequently, we cross-correlate all the curves with the continuum emission to measure the corresponding time delays for each wavelength bin. The time delay recovery from all the methods is summarised in Table ~\ref{tab:curves1_7}. Below we provide the notes concerning the time-delay results of individual methods.\\

{\bf Javelin:} We searched for the time delay between 0 to 1500 days for both quasars. The bootstrap methodology was applied to estimate the mean and peak time delays and the total time-delay peak distribution is used for the error estimation. For HE 0435-4312, the time delay distributions for all the curves are shown in Fig ~\ref{fig:javelin_he0435} where multiple peaks can be seen in the whole search range. In some of the cases, a prominent peak is observed at a longer time delay of around $\sim$1000 days and in some cases the peak around $\sim$ 500 days is dominant. We applied an alias mitigation technique based on the number of overlapping light-curve pairs and as a result, it suppressed the longer time delay peaks where the number of overlapping light-curve pairs is rather small. In Fig ~\ref{fig:javelin_he0435}, the black distribution is the original one and the red distribution corresponds to the one weighted by the distribution of overlapping light-cure pairs. The final time delays for all the curves are found to be of the order of $\sim$ 500 days and the exact values with error bars for all the curves are listed in Table~\ref{tab:curves1_7}. However, the alias mitigation has a negligible effect on the time delays of HE 0413-4031 (Fig~\ref{fig:javelin_he0413}), and the peak at around $\sim$1000 days persists to be dominant. However, we do see a wavelength-dependent time delay in the mean distribution. The exact time delays for all the curves are presented in Table~\ref{tab:curves1_7}.\\

\begin{figure*}
    \centering
    \includegraphics[width=0.49\textwidth]{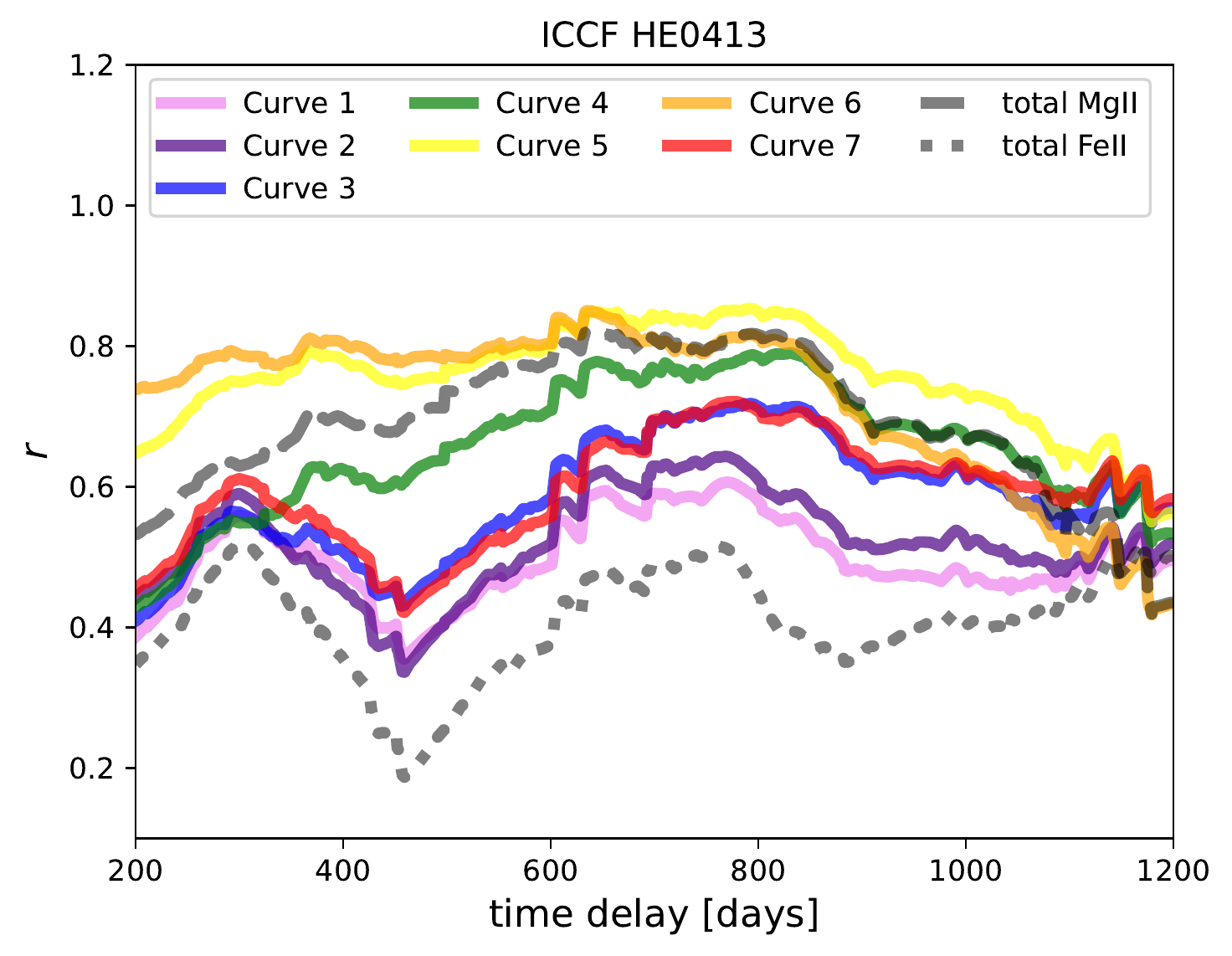}
    \includegraphics[width=0.49\textwidth]{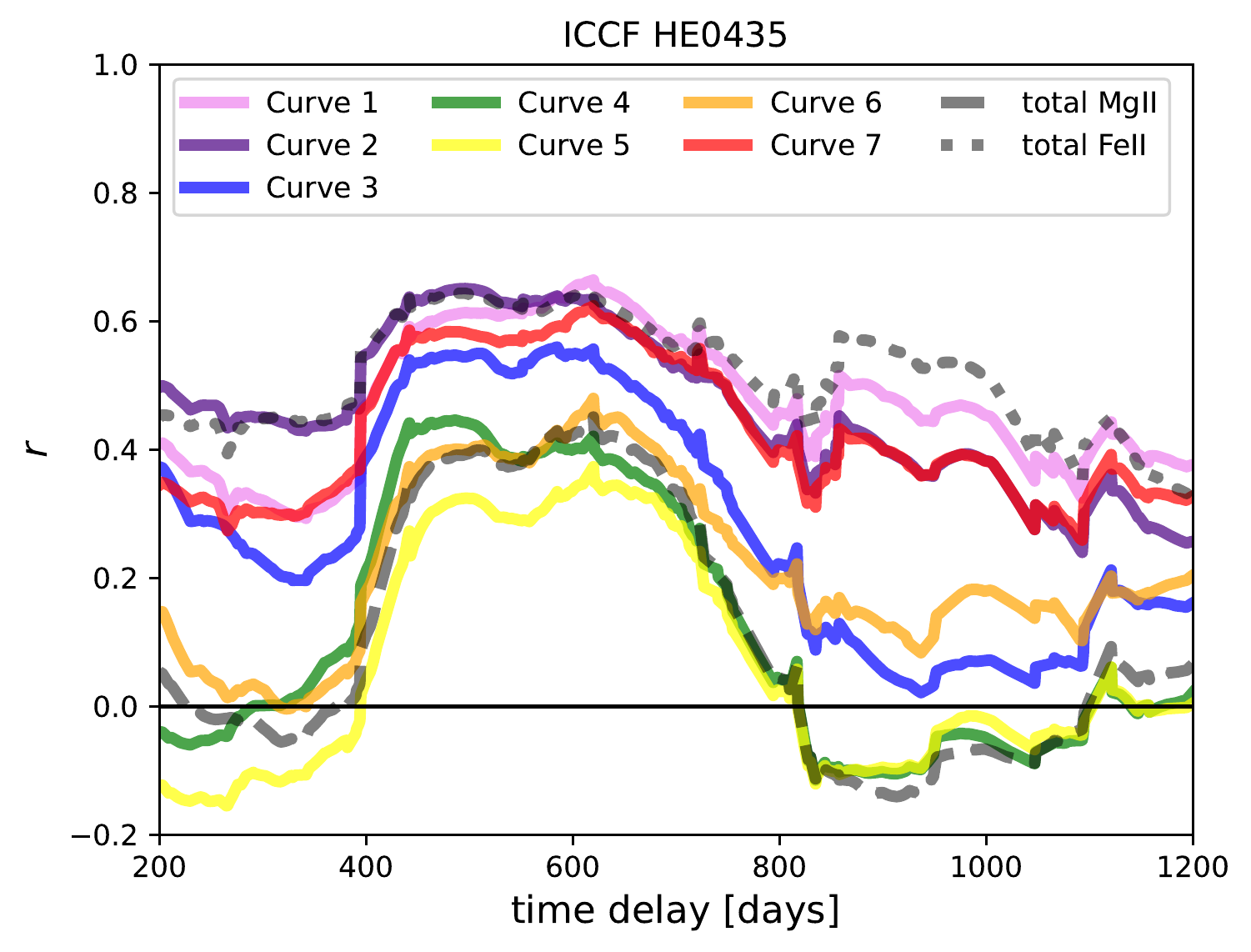}
    \caption{ \textit{Left panel:} ICCFs results for seven light curves of HE 0413-4031 (rainbow-colored solid lines). The total FeII emission is depicted by a gray dotted line, while the total MgII emission is represented by a gray dashed line. \textit{Right panel:} The same for HE 0435-4312.}
    \label{fig_ICCF_all}
\end{figure*}

{\bf ICCF:} For the ICCF method, the time delays were searched for between 200 to 1200 days in the observer's frame. The correlation coefficient ($r$) for all the light curves for both quasars is shown in Fig~\ref{fig_ICCF_all}. 
For HE 0435-4312, the centroid and the peak time delays are found between 535-614 days and 502-620 days in the observed frame, respectively. For the light curves 1, 2, 3, \& 7, the correlation coefficient is above 0.5 
 and for the other curves around 0.4, which is considered to be significant (see Fig~\ref{fig_ICCF_all}, right panel). For HE 0413-4031, the
 centroid and the peak time delays are between 627-784 days and 636-757 days in the observed frame, respectively.
The correlation coefficient for all the cases is above or around 0.6. The exact values for time delays with error bars are presented in Table~\ref{tab:curves1_7}. \\

\begin{figure*}
    \centering
    \includegraphics[width=0.49\textwidth]{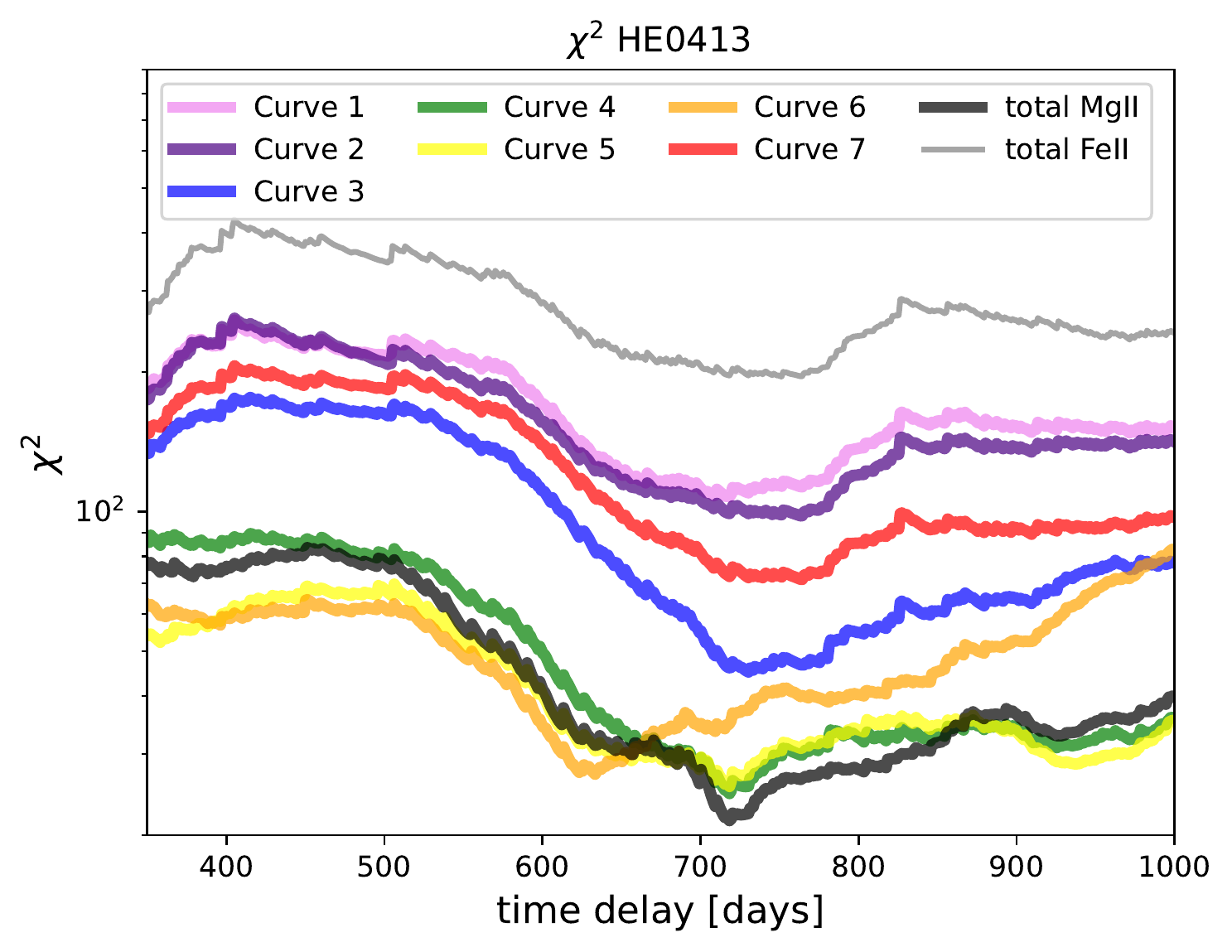}
    \includegraphics[width=0.49\textwidth]{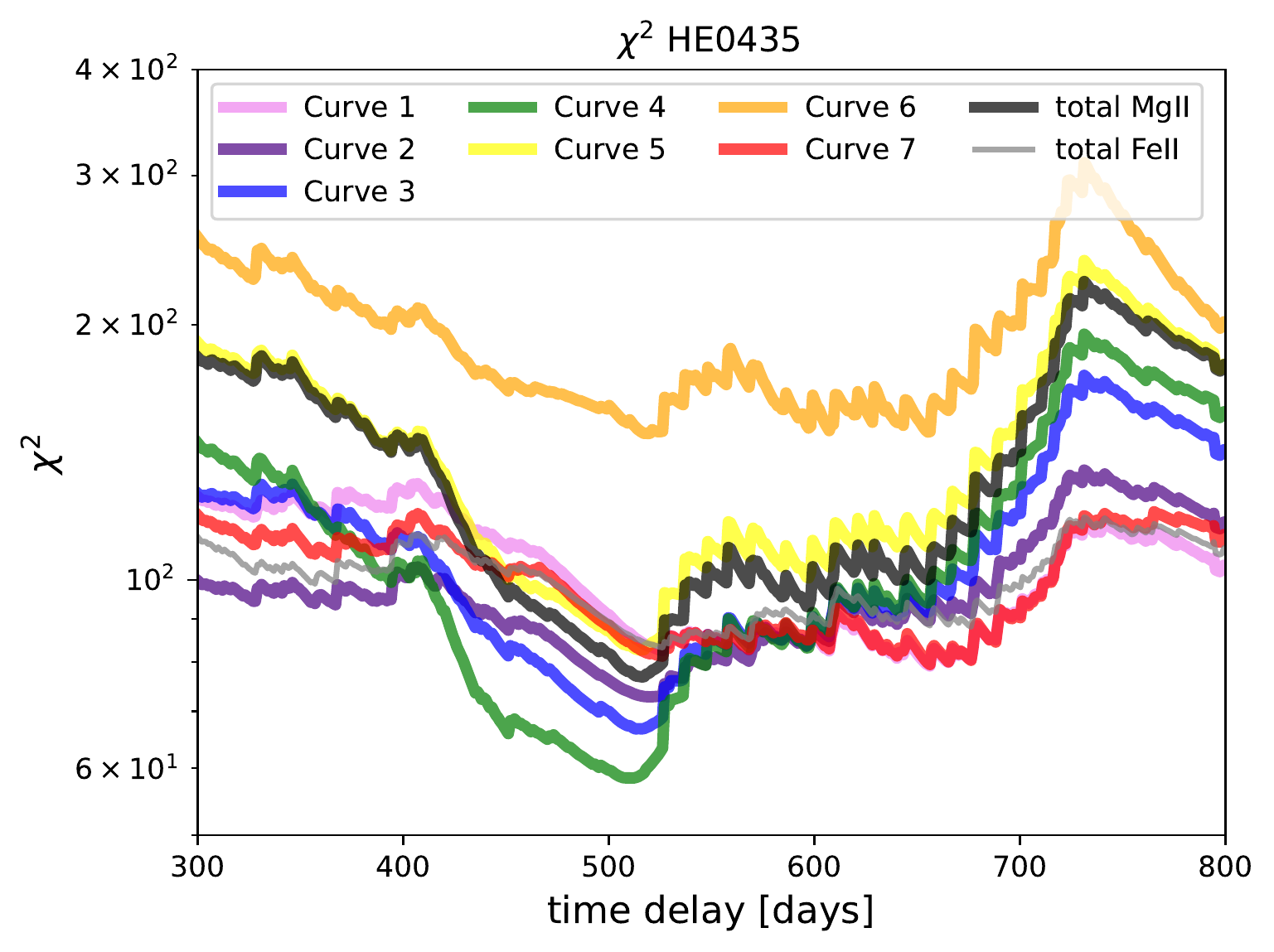}
    \caption{$\chi^2$ results for HE 0413-4031 and HE 0435-4312. \textit{Left panel:} $\chi^2$ for seven light curves of HE 0413-4031 (rainbow-colored solid lines). The total MgII emission is represented by a black solid line, while the total FeII emission is depicted by a gray solid line. \textit{Right panel:} The same for HE 0435-4312.}
    \label{fig_chi2_all}
\end{figure*}

{\bf $\chi^2$ method:} 
The distribution of the $\chi^2$ as a function of the time delay is shown in Figure~\ref{fig_chi2_all} for both the quasars. For HE 0435-4312, the $\chi^2$ distribution exhibits two dips corresponding to $\sim$500 days and $\sim$700 days. In some curves, the dip around $\sim$500 days is prominent, while for the rest of the curves, the dip close to $\sim$700 dominates. However, the $\chi^2$ distribution in HE 0413-4031 shows a single prominent dip in all the cases though the location of the dips varies from one light curve to another. 
The time delay range for HE 0435-4312 was found to be between $\sim$452-636 days in the observers' frame, and for HE 0413-4031, it is between $\sim$620-764 days (see Table~\ref{tab:curves1_7}). 
The observed time delays inferred using the $\chi^2$ method are consistent with ICCF results within the uncertainties.  \\

\begin{figure*}
    \centering
    \includegraphics[width=0.49\textwidth]{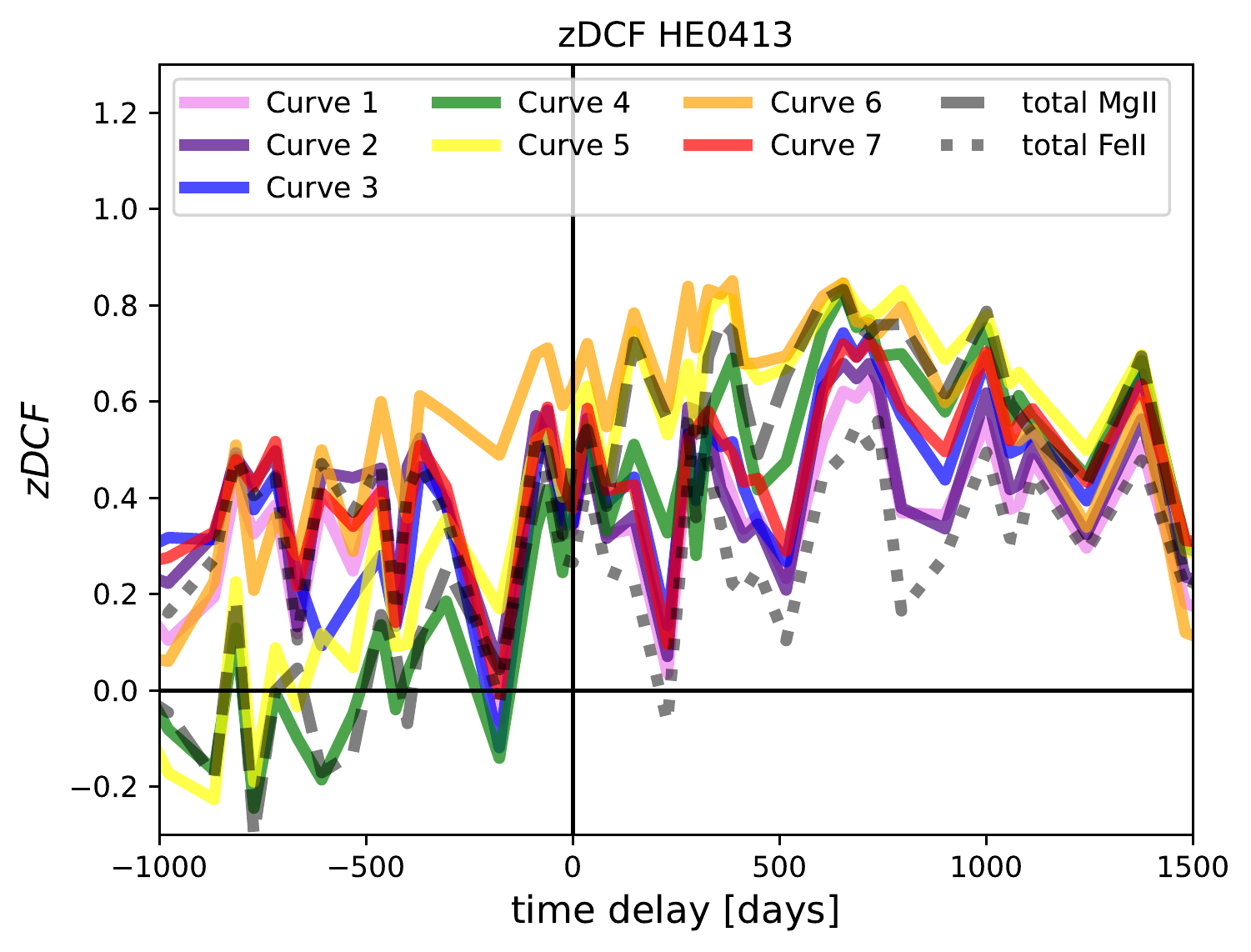}
    \includegraphics[width=0.49\textwidth]{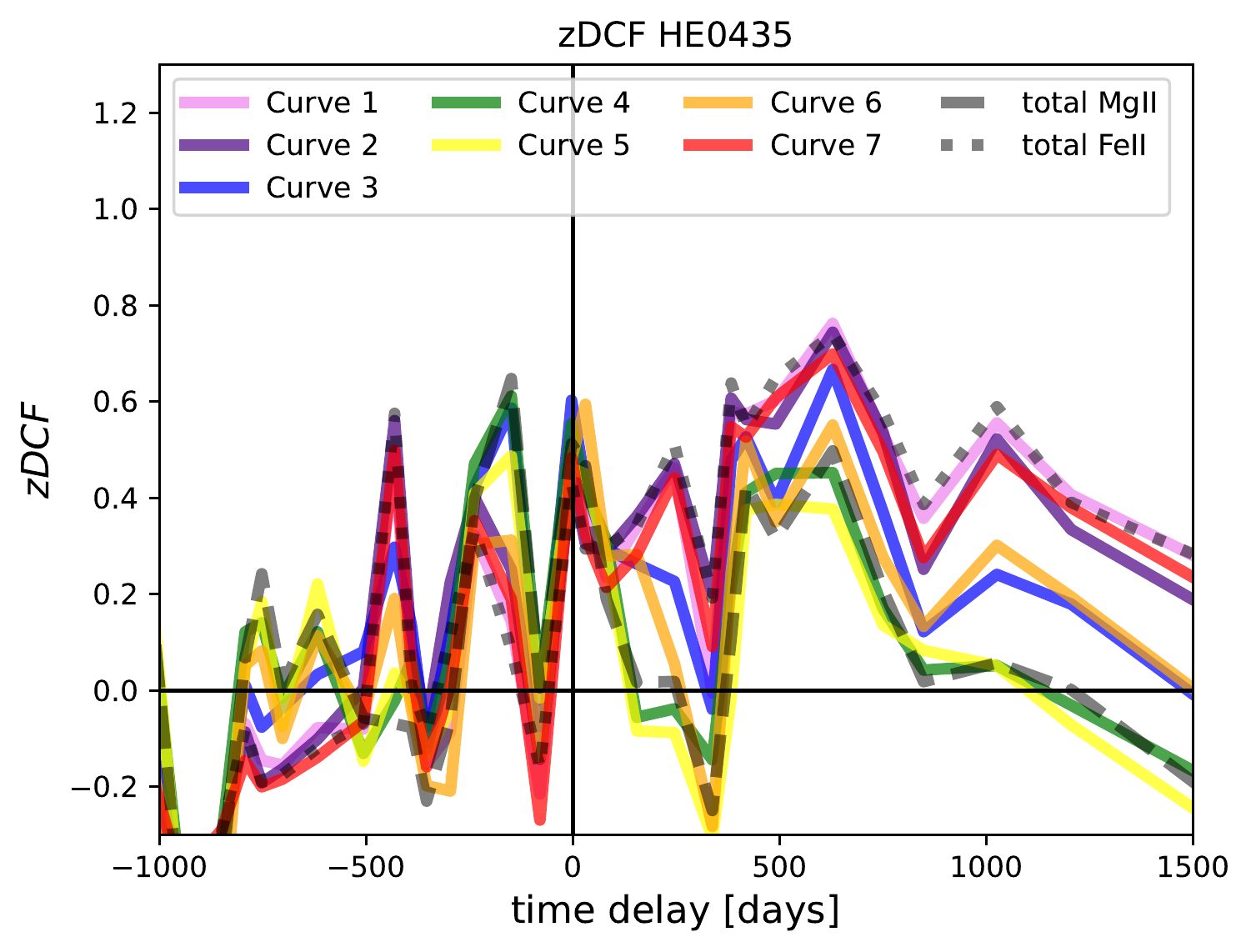}
    \caption{zDCF results for HE 0413-4031 and HE 0435-4312. \textit{Left panel:} zDCF for seven light curves of HE 0413-4031 (rainbow-colored solid lines). The total MgII emission is represented by a black dashed line, while the total FeII emission is depicted by a dotted black line. \textit{Right panel:} The same for HE 0435-4312.}
    \label{fig_zdcf_all}
\end{figure*}

{\bf zDCF:} The time delays of all the light curves were searched for between -1000 to 1500 days for both sources. Multiple peaks were detected on both positive and negative sides, however, the positive peaks have systematically larger correlation coefficients (see Fig.~\ref{fig_zdcf_all}). For HE 0435-4312, the largest correlations are detected at $\sim$629 days for curves 1, 2, 3, and 6, 7. For light curves 4 and 5, the largest correlation is detected at shorter time delays of the order of 490 days. For HE 0413-4031, multiple peaks were detected on the positive delay side and the most prominent one lies between $\sim$600 and 700 days. The time delays for curves 1, 2, and 7 have the largest correlation coefficient of 716 days, while curves 3, 4, 5, and 6 have a bit lower time delay of 654 days in the observer's frame. The best time delay for each light curve was chosen by inferring the time lag with the maximum likelihood and the corresponding uncertainties are 1$\sigma$ errors. \\

\begin{figure*}
    \centering
    \includegraphics[width=0.49\textwidth]{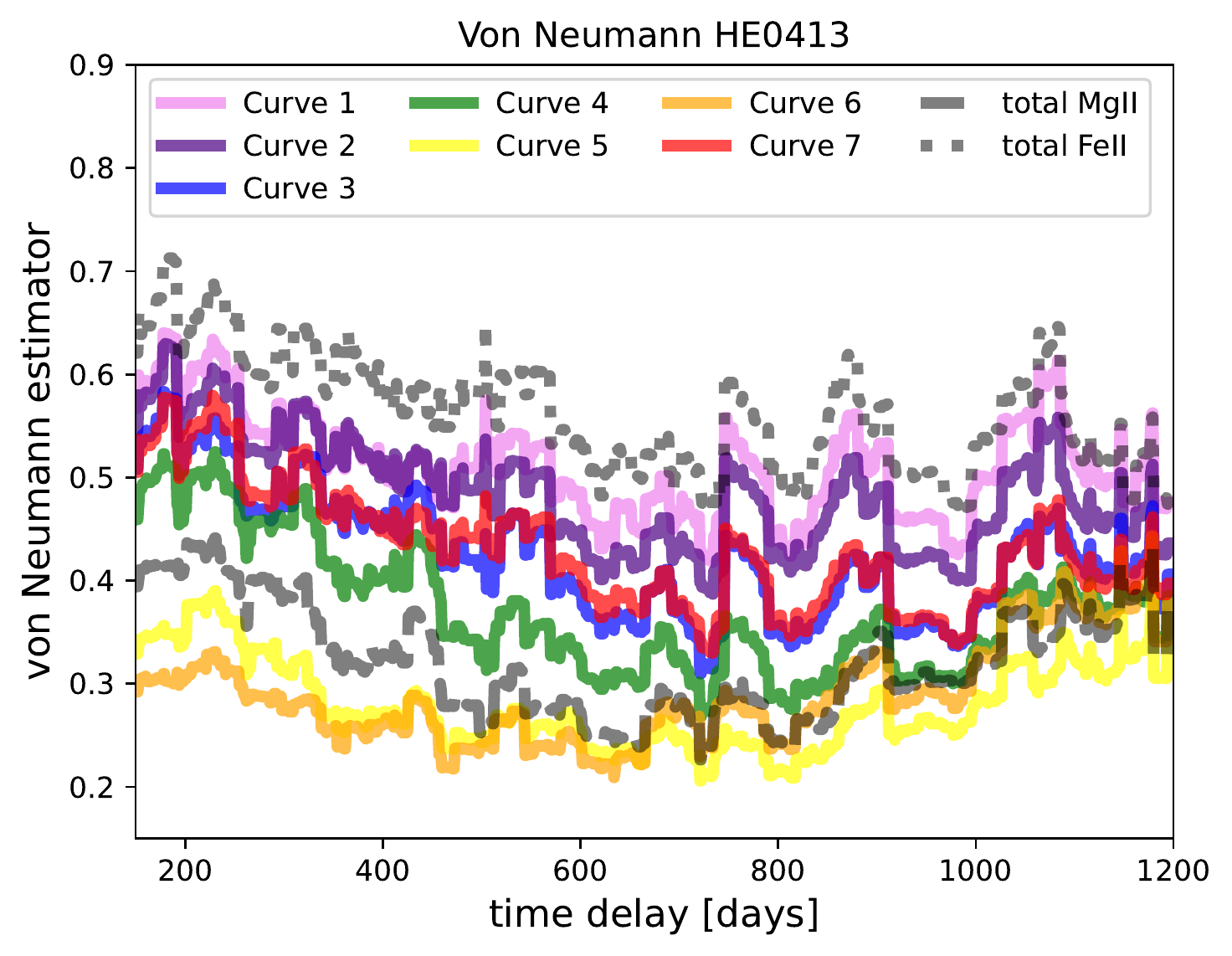}
    \includegraphics[width=0.49\textwidth]{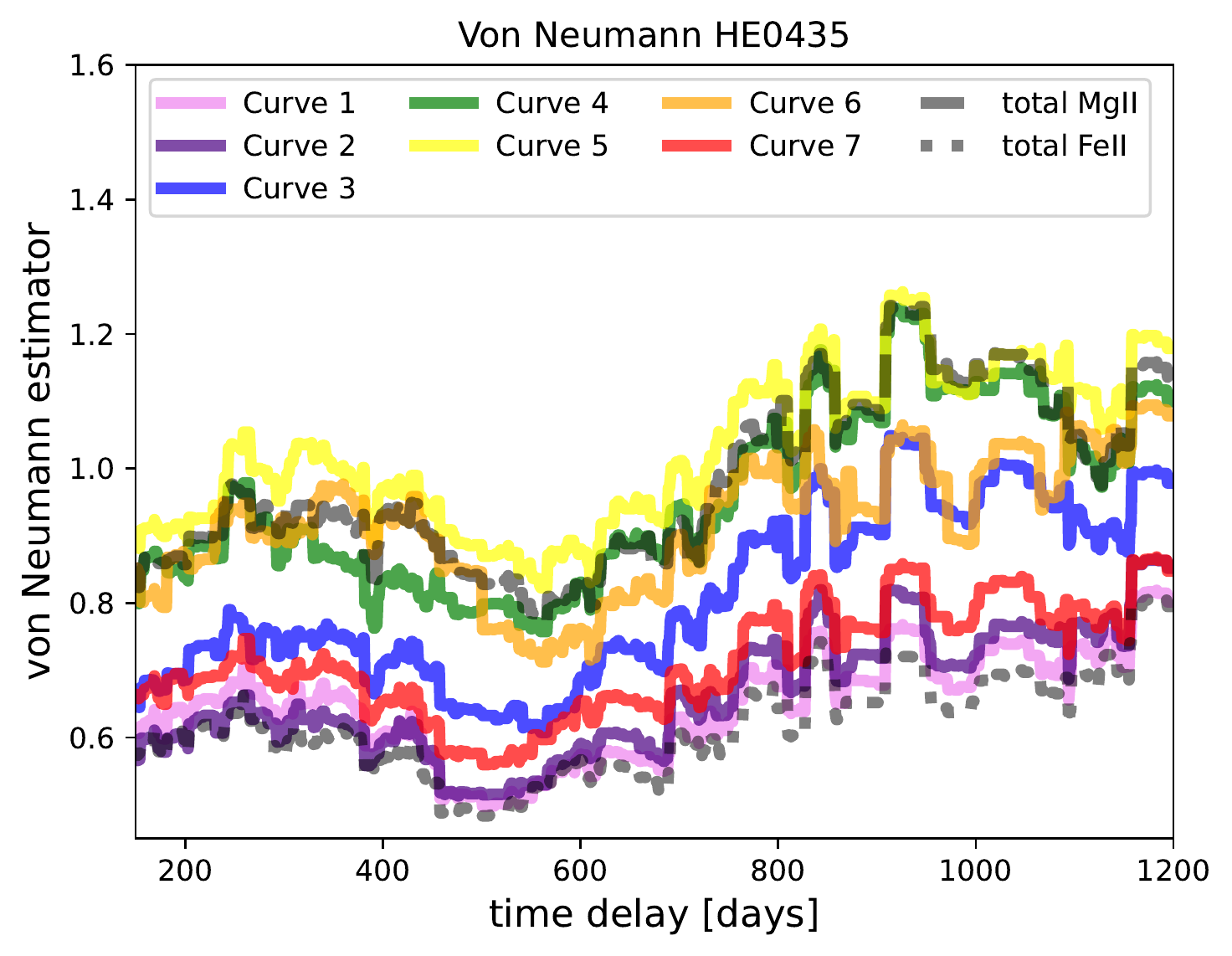}
    \caption{Von Neumann estimator results for HE 0413-4031 and HE 0435-4312. \textit{Left panel:} The temporal evolution of the von Neumann estimator for seven light curves of HE 0413-4031 (rainbow-colored solid lines). The total MgII emission is represented by a black dashed line, while the total FeII emission is depicted by a dotted black line. \textit{Right panel:} The same for HE 0435-4312.}
    \label{fig_neumann_all}
\end{figure*}

\begin{figure*}
    \centering
    \includegraphics[width=0.49\textwidth]{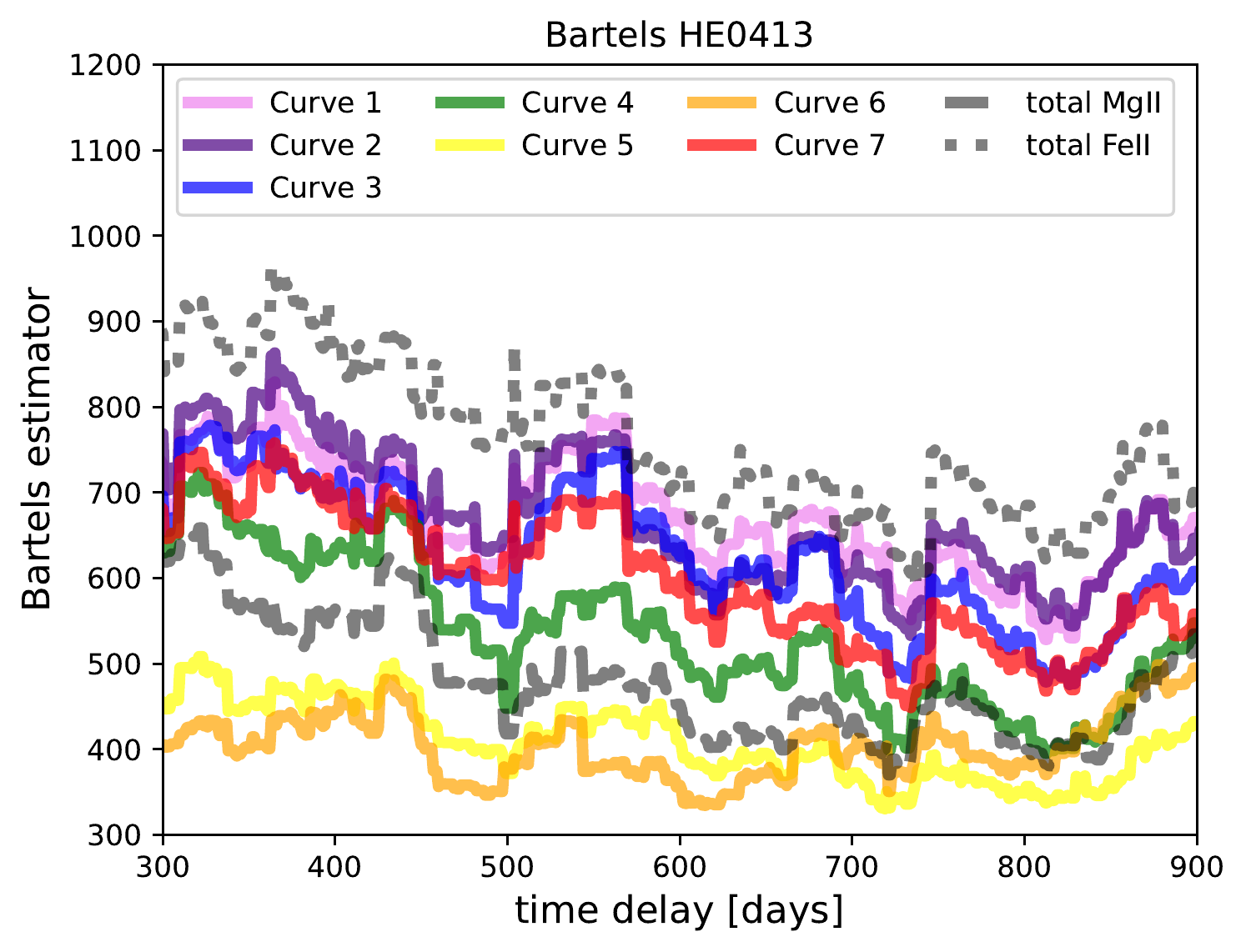}
    \includegraphics[width=0.49\textwidth]{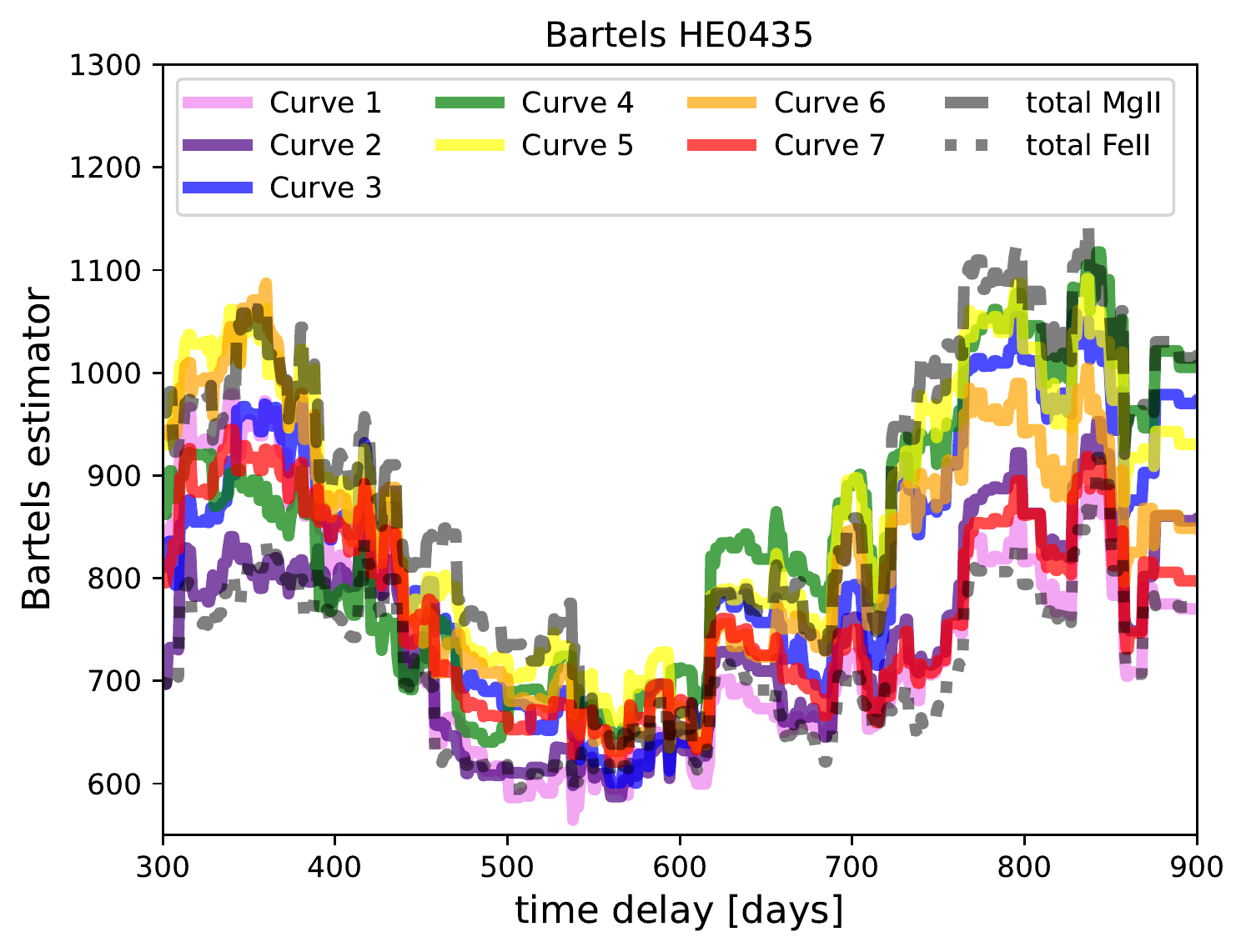}
    \caption{Bartels estimator results for HE 0413-4031 and HE 0435-4312. \textit{Left panel:} The temporal evolution of the Bartels estimator for seven light curve bins of HE 0413-4031 (rainbow-colored solid lines). The total MgII emission is represented by a black dashed line, while the total FeII emission is depicted by a dotted black line. \textit{Right panel:} The same for HE 0435-4312.}
    \label{fig_bartles_all}
\end{figure*}

{\bf Von Neumann \& Bartels estimators:} 
We searched for time delays between 0 to 1200 days for both the quasars and the results are shown in Figs.~\ref{fig_neumann_all} and \ref{fig_bartles_all}. 
The best time delays for HE 0435-4312 are found to be between 535-615 days for the von-Neumann estimator and between 535-715 days for the Bartels estimator. For HE 0413-4031, the observed time delays are between 635-815 days and 735-845 days for von Neumann and Bartels estimators, respectively. The peak time delays with the uncertainties that correspond to each light curve are listed in Table~\ref{tab:curves1_7}.

\begin{table*}[]
    \centering
    \caption{Time-delay measurements for seven wavelength bins containing a combination of MgII and FeII emission, after subtracting the power-law component. Curves 1 and 7 mostly contain FeII, and curves 4 and 5 are strongly dominated by MgII. For ICCF, we also show the maximum correlation coefficient in the parenthesis.}
\resizebox{\textwidth}{!}{      
    \begin{tabular}{c|ccccccc}
\noalign{\smallskip}
\hline \hline \noalign{\smallskip}
& & & HE 0435-4312 & & & & \\
\hline\noalign{\smallskip}
Methods & Curve 1 & Curve 2 & Curve 3 & Curve 4& Curve 5 & Curve 6 & Curve 7 \\
\noalign{\smallskip}
\hline \noalign{\smallskip}
Javelin (peak) & 492.0$^{+10.5}_{-1.3}$&500.0$^{+12.0}_{-4.1}$&505.0$^{+22.3}_{-8.5}$&495.0$^{+20.9}_{-14.3}$&490.0$^{+5.6}_{-10.0}$&490.0$^{+5.3}_{-13.3}$   &492.0$^{+12.7}_{-3.0}$ \\ \noalign{\smallskip}
 Javelin (mean) & 706.7$^{+4.8}_{-38.6}$&705.7$^{+37.5}_{-27.8}$&600.9$^{+136.3}_{-168.2}$&649.7$^{+135.0}_{-170.2}$&747.6$^{+171.2}_{-144.2}$&863.1$^{+82.7}_{-190.4}$&755.9$^{+16.0}_{-49.6}$ \\
\noalign{\smallskip} \hline \noalign{\smallskip}
ICCF ($r_{\rm max}$) & $620.0$ (0.66) & $496.0$ (0.65) & $585.0$ (0.56) & $486.0$ (0.45) &  $620.0$ (0.37) & $620.0$ (0.48) & $620.0$ (0.62) \\  \noalign{\smallskip}
ICCF (centroid) & $600.8^{+231.8}_{-58.4}$& $564.6^{+130.6}_{-59.0}$ & $550.8^{+52.0}_{-47.3}$& $535.5^{+90.3}_{-75.3}$&$595.7^{+65.5}_{-112.6}$&$614.5^{+ 57.9}_{-78.3}$ & $588.1^{+221.8}_{-74.1}$ \\  \noalign{\smallskip}
ICCF (peak) & $620.0^{+238.0}_{-128.8}$& $552.0^{+169.0}_{-  110.0}$& $566.0^{+54.0}_{-124.0}$& $502.0^{+118.0}_{-60.0}$&$620.0^{+58.0}_{-169.0}$& $620.0^{+ 59.4}_{-143.0}$ &  $620.0^{+194.1}_{-177.8}$\\ 
\noalign{\smallskip}
\hline \noalign{\smallskip}
$\chi^2$ (min) & $656.29$ & $519.26$ & $513.26$ &  $511.26$ & $515.26$ & $519.28$  &  $656.33$   \\
$\chi^2$ (bootstrap) & $636^{+50}_{-106}$ &  $636^{+42}_{-161}$ & $636^{+70}_{-185}$ & $452^{+85}_{-127}$ & $636^{+80}_{-250}$ & $636^{+83}_{-256}$ & $636^{+50}_{-120}$  \\
 \noalign{\smallskip}
\hline \noalign{\smallskip}
     zDCF &  $629.0^{+77.0}_{-108.0}$ &  $629.0^{+77.0}_{-115.8}$ & $629.0^{+77.0}_{-129.0}$ & $490.9^{+153.7}_{-66.0}$ & $490.9^{+162.6}_{-62.4}$ & $629.0^{+79.1}_{-168.0}$ &  $629.0^{+78.8}_{-130.3}$ \\  
\hline \noalign{\smallskip}
von Neumann (min)  & 501.0  & 476.0 & 560.0  & 560.0   & 560.0 &  560.0    & 501.0 \\
von Neumann (bootstrap) & $535.0^{+36.6}_{-97.2}$   & $ 535.0^{+35.0}_{-153.0}$ & $575.0^{+10.0}_{-102.0}$ & $ 615.0^{+69.0}_{-144.0}$  & $615.0^{+46.3}_{-77.0}
$  & $615.0^{+64.0}_{-56.0}$  & $535.0^{+42.3}_{-137.7}$ \\
 \noalign{\smallskip}
\hline \noalign{\smallskip}
Bartels (min)  & 538.0  & 560.0 & 574.0  & 487.0 & 610.0  & 560.0   & 560.0 \\
Bartels (bootstrap) & $685.0^{+24.0}_{-149.0}$   & $535.0^{+33.0}_{-201.6}$   & $595.0^{+21.1}_{-67.0}$   &  $575.0^{+19.0}_{-135.0}$  & $615.0^{+36.0}_{-77.0}$ & $575.0^{+35.0}_{-47.2}$ & $715.0^{+31.5}_{-177.0}$ \\
 \noalign{\smallskip} 
\hline \hline \noalign{\smallskip}
& & & HE 0413-4031 & & & & \\
\noalign{\smallskip}
\hline \noalign{\smallskip}
Javelin (peak) & 940.5$^{+115.2}_{-190.9}$&953.9$^{+101.4}_{-168.6}$&970.4$^{+104.9}_{-168.5}$&932.7$^{+70.8}_{-146.1}$&939.7$^{+64.4}_{-204.0}$&945.6$^{+69.2}_{-156.5}$&953.7$^{+85.5}_{-176.5}$  \\\noalign{\smallskip}
Javelin (mean) & 946.6$^{+116.2}_{-196.6}$&909.2$^{+95.9}_{-135.8}$&895.1$^{+91.5}_{-157.6}$&866.4$^{+71.0}_{-145.9}$&795.4$^{+81.0}_{-154.5}$&661$^{+99.7}_{-114.1}$&924.3$^{+75.8}_{-174.0}$ \\
\noalign{\smallskip}
\hline \noalign{\smallskip}
ICCF ($r_{\rm max}$) & $770.0$ (0.61) & $768.0$ (0.64) &$792.0$ (0.72) & $831.0$ (0.79) & $792.0$ (0.85) &$635.0$ (0.85) &$774.0$ (0.72)  \\
\noalign{\smallskip}
ICCF (centroid) &$714.4^{+77.3}_{-111.7}$&$724.3^{+89.7}_{- 93.6}$&$759.5^{+72.0}_{-91.8}$& $734.7^{+72.4}_{-76.6}$&$675.1^{+74.7}_{-52.6}$&$627.1^{+52.9}_{-55.7}$ &$784.0^{+76.1}_{-98.2}$\\ 
 \noalign{\smallskip}
ICCF (peak) & $697.0^{+88.0}_{-64.0}$&$710.0^{+82.0}_{-76.4}$&$753.0^{+  73.0}_{-114.0}$&7$44.0^{+88.0}_{-108.0}$&$754.0^{+58.0}_{-119.0}$&
 $636.0^{+128.6}_{-8.3}$& $757.0^{+69.0}_{-105.0}$\\
\noalign{\smallskip}
\hline \noalign{\smallskip}
$\chi^2$ (min) & $718.36$ & $763.38$ & $730.37$ & $718.36$ & $ 718.36$ & $ 633.32$& $764.38$ \\
\noalign{\smallskip}
$\chi^2$ (bootstrap) & $716.0^{+29.4}_{-64.7}$ & $764.0^{+10.4}_{-98.7}$ & $716.0^{+29.4}_{-50.7}$ & $716.0^{+47.4}_{-59.7}$ & $620.0^{+60.3}_{-26.7}$ & $ 620.0^{+17.3}_{-132.8}$ & $764.0^{+17.4}_{-71.7}$\\ \noalign{\smallskip}
\hline
 \noalign{\smallskip}
zDCF & $716.9^{+50.7}_{-121.2}$ & $716.9^{+44.5}_{-109.8}$ & $654.4^{+89.8}_{-71.2}$ & $654.4^{+101.0}_{-69.1}$ &  $654.4^{+152.1}_{-265.6}$ & $654.4^{+144.2}_{-81.6}$ & $716.9^{+68.6}_{-97.3}$ \\ \noalign{\smallskip}
\hline  \noalign{\smallskip}
von Neumann (min)  & 734.0 & 734.0 & 721.0  & 721.0  & 721.0  & 634.0 &  734.0   \\
von Neumann (bootstrap) & $635.0^{+147.0}_{-205.4}$ & $ 635.0^{+110.2}_{-178.0}$   & $815.0^{+41.0}_{-162.0}$  & $815.0^{+26.0}_{-104.6}$  & $635.0^{+99.0}_{-153.0}$  & $635.0^{+99.0}_{-164.0}$ & $635.0^{+121.6}_{-178.0}$ \\
 \noalign{\smallskip}
\hline \noalign{\smallskip}
Bartels (min)  & 812.0  & 734.0 & 826.0 & 812.0 & 719.0   & 615.0   & 734.0\\
Bartels (bootstrap) & $785.0^{+27.0}_{-183.2}$  & $785.0^{+27.0}_{-175.7}$  & $785.0^{+27.0}_{-203.0}$  & $ 815.0^{+21.0}_{-86.0}$  & $845.0^{+75.3}_{-126.0}$ & $735.0^{+62.8}_{-134.0}$  & $815.0^{+30.0}_{-78.0}$
 \\
 \noalign{\smallskip}  
\hline \noalign{\smallskip}
\end{tabular}
}      
\label{tab:curves1_7}
\end{table*}

\section{Discussion}
\label{sec_discussion}

Our monitoring of the two quasars HE0413-4031 and HE0435-4312 over 11 years allowed us to report the measurements of the Mg II time delay as well as the UV Fe II time delay. We also obtained wavelength-resolved time delays for the combined Fe II and Mg II overlapping emissions.

The newly reported Mg II time delays are consistent with those previously obtained for HE0413-4031 and HE0435-4312 by \citet{Zajacek2020} and \citet{zajacek2021}, respectively. The two measurements are also consistent with other Mg II time delays, predominantly obtained for lower-luminosity sources. In fact, the two quasars currently belong to the highest-luminosity sources of the MgII sample.

\begin{figure*}
    \centering
    \includegraphics[width=0.33\textwidth]{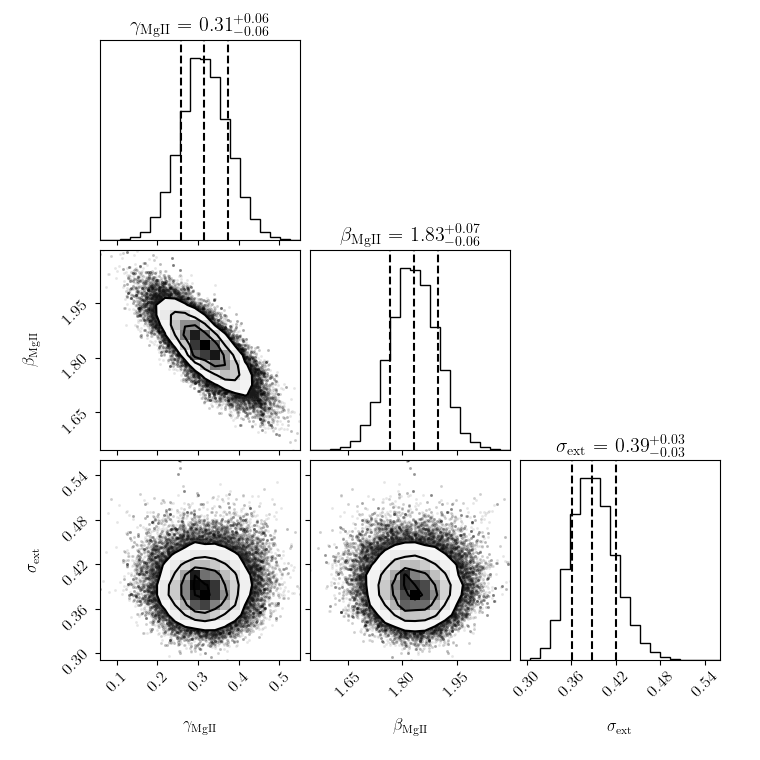}
    \includegraphics[width=0.47\textwidth]{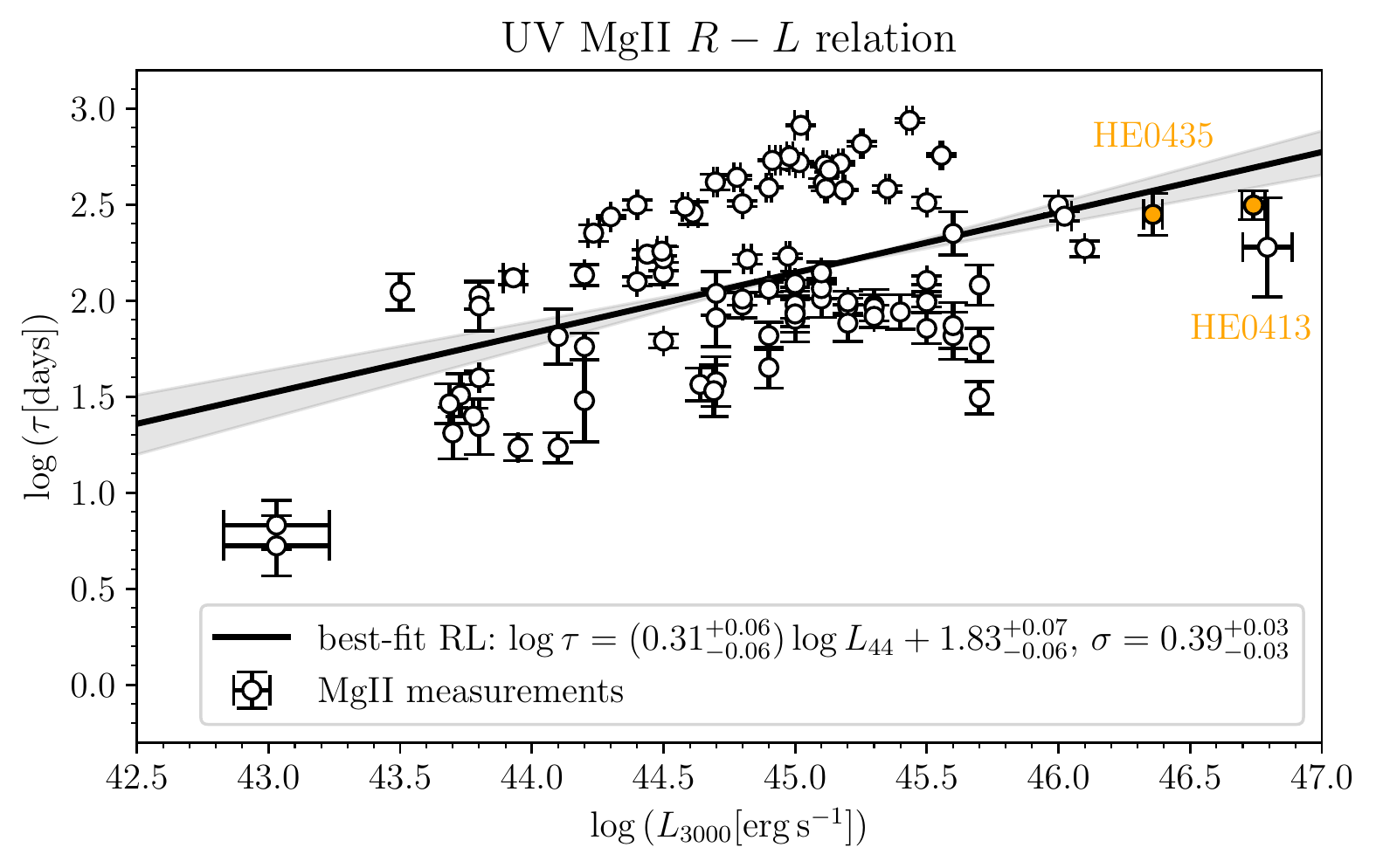}
    \caption{The MgII radius-luminosity relation parameters (see the corner plot to the left) and the comparison of the maximum-likelihood R-L relation with 94 measurements. The scatter of the relation is $\sigma=0.39^{+0.03}_{-0.03}$. The two quasars studied in this paper are marked as orange circles and belong to the highest luminosity MgII sources.}
    \label{fig_MgII_RL}
\end{figure*}

The first MgII radius-luminosity relation was presented in \citet{Czerny2013}, which was followed by the updated MgII R-L relations for an increasing number of sources \citep{Zajacek2020,zajacek2021,2020ApJ...903...86M, Homayouni2020, 2022arXiv220805491Y}. With the gradual increase in the number of sources, the slope decreased from $\gamma\sim 0.5$ to $\gamma\sim 0.3$, hence being in tension with the simple photoionization arguments. At the same time, the scatter significantly increased and is larger than for the H$\beta$ sample (for discussion see \citealt{Homayouni2020}). Here we determine the R-L relation parameters for the currently largest sample of 94 MgII sources (considering our measurements of total Mg II for quasars HE 0413-4031 and HE 0435-4312), whose flux densities and rest-frame time delays are taken from \citet{2020ApJ...903...86M}, \citet{zajacek2021}, \citet{2022arXiv220805491Y}, and references therein. The correlation between the rest-frame time delay and the monochromatic luminosity at 3000\AA\ is significant with the Pearson correlation coefficient of $r=0.50$ ($p=2.47 \times 10^{-7}$) and the Spearman rank-order correlation coefficient of $s=0.36$ ($p=3.07 \times 10^{-4}$). The MCMC inference of the R-L parameters is presented in Fig.~\ref{fig_MgII_RL}, including the likelihood distributions for the slope, the intercept, and the scatter in the left panel and the best-fit MgII R-L relation alongside 94 measurements in the right panel of Fig.~\ref{fig_MgII_RL}. The best-fit MgII relation is,
\begin{equation}
    \log{\tau}=(0.31^{+0.06}_{-0.06})\log{\left(\frac{L_{3000}}{10^{44}\,{\rm erg\,s^{-1}}}\right)}+(1.83^{+0.07}_{-0.06})\,,
    \label{eq_MgII_RL}
\end{equation}
with the intrinsic scatter of $\sigma=0.39^{+0.03}_{-0.03}$ dex. The slope of the MgII R-L relation $\gamma=0.31 \pm 0.06$ is significantly shallower than the one for the H$\beta$ \citep{Bentz2013} and MgII R-L relation derived in \citep{Homayouni2020}. From the 2D likelihood contours in Fig.~\ref{fig_MgII_RL} (left panel), we can see a degeneracy between the slope and the intercept, i.e. within 3$\sigma$ confidence interval, a steeper slope of $\gamma\sim 0.40$ in combination with the smaller intercept $\beta\sim 1.70$ is possible as well as a very shallow slope of $\gamma\sim 0.20$ in combination with the larger intercept of $\beta\sim 1.95$.    

Thus, the Mg II broad-line emission that is well constrained for intermediate-redshift sources forms an R-L relation analogous to the R-L relation for H$\beta$ line well established for low-redshift sources. Such a relation can be used both for the determination of the black hole mass as well as for cosmological applications. The Mg II measurements (78 QSOs spanning the redshift
range 0.0033 - 1.89 in the sample analyzed by \citealt{Cao2022}) does not yet give strong constraints but it implies that the MgII R-L relation is standardizable and the weak cosmological constraints derived from it are consistent with better established cosmological probes \citep[see also][]{khadka2021,2022MNRAS.515.3729K}. 

\subsection{BLR kinematics in HE 0413-4031 and HE 0435-4312}
The measured delays of the Fe II and the wavelength-resolved time delay studies give additional insight into the structure of the BLR. The measurement of the integrated Fe II time delay is rather similar to the measured Mg II time delay. The mean
Fe II delay (averaged over the methods) for HE 0435 is nearly the same as for Mg II (just $\sim 0.85\%$ larger), while for HE 0413 the Fe II delay is longer by $\sim 5.15\%$. The small time-delay difference implies the spatial proximity of MgII and FeII line-emitting regions. Although it is expected that the generally narrower FeII line should be positioned further from the SMBH than MgII line-emitting material \citep{gaskell2022}, the FeII line widths are not constrained well for both the quasars based on the comparison of different FeII templates and line widths that yield a comparable $\chi^2$ statistic in terms of the spectral fitting, see Appendix~\ref{appendix_FeII_templates}. 


The decomposition of the spectrum into Mg II and Fe II in 2700 - 2900 \AA~wavelength range is not unique as it depends on the adopted templates. We used theoretical templates of \citet{Bruhweiler2008}, and the quality of the fits seemed satisfactory.

Wavelength-resolved measurements of the combined Fe II and Mg II emission do not suffer from this issue but the division of the light curves into 7 wavelength bins decreased the data quality. Still, we seem to see a distinct wavy pattern across the wavelength range and it is different for the two sources (see Fig~\ref{fig:mean_rms}).

\begin{figure}
    \centering
    \includegraphics[width=0.5\textwidth]{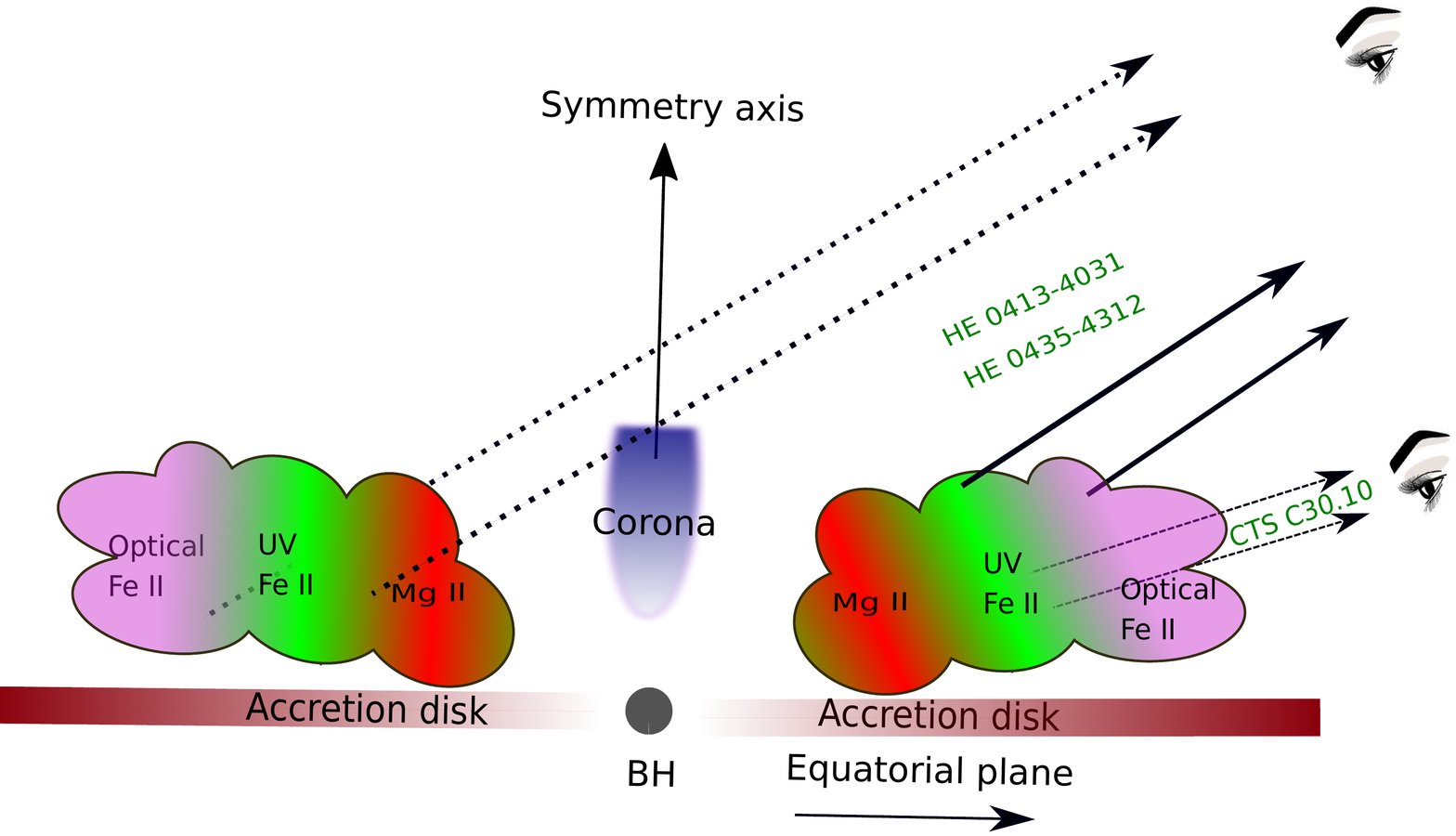}
    \caption{Schematic representation of MgII and FeII emission regions derived from the total time delays in both quasars and comparison with CTS C30.10 \citep{prince_CTS_2022}. We also noticed that in CTS C30.10 the MgII time delay is shorter than the FeII time delay which suggests the higher inclination of the source which leads to the invisibility of the MgII emission coming from the other side of the black hole. 
    }
    \label{fig:cartoon}
\end{figure}

For HE 0435-4312, we see a roughly symmetric behavior, particularly in ICCF,  
 $\chi^2$, and zDCF results, with the shortest time delay in curve 4 (see Figure~\ref{fig:mean_rms}), most dominated by Mg II and a longer time delay of the wings with a strong contribution from Fe II implied by the spectral decomposition (see Figure~\ref{fig:decomp}). In the case of these three methods, the rest frame time delay at the core of Mg II is $\sim 225$ days while Fe II-dominated wings show $\sim 275$ days time delay. This is clearly different from the results for the CTS C30.10 source \citep{prince_CTS_2022} where the UV Fe II delay was shorter and we postulated that the part of the Fe II emitting region more distant from the observer is shielded from the view. We thus conclude that for HE 0435-4312, the viewing angle towards the nucleus, measured from the symmetry axis, is smaller, and such shielding does not occur. The generalized new picture is shown in Figure~\ref{fig:cartoon}. Just for the completeness of the AGN picture we also show the corona as a base of the failed jet but it does not have any scientific relevance here. We would also like to mention the results from the Javelin (mean and peak), von Neumann, and Bartels estimators. The Javelin peak distribution shows a trend mildly correlated with the ICCF but mean time delays are almost constant. This can happen because the original Javelin delay distribution has multiple peaks of different strength and the mean is taken over the entire delay distribution leading to a constant time delay across the light curves. The delays of von Neumann and Bartels are mostly consistent with each other except for the wings but they show a wavy pattern more or less consistent with ICCF.

In the case of the second quasar, HE 0413-4031, the trend seems to be more complicated, rather like a wavy pattern, with the shortest time delay in some methods (ICCF, $\chi^2$, and Javelin mean) detected for curve 6 (see Figure~\ref{fig:mean_rms}). Also, there is an asymmetry between the wings. It likely reflects the asymmetry between the wings in the spectral decomposition, since for this quasar, only the blue wing is Fe II-dominated but the red wing has a comparable contribution from both the MgII and the FeII emission. Thus, we propose that the difference in the Fe II strength is responsible for the effect and the viewing angle is again small, without any detectable shielding effects. The delays derived from the Javelin mean are consistent with the ICCF and the $\chi^2$ methods. The results from the Bartels estimator are rather consistent with the ICCF showing a shorter time delay for curve 6. The delays derived from the zDCF and von Neumann methods are consistent with each other in the wings but differ for curves 3 \& 4.

\begin{figure*}
    \centering
    \includegraphics[width=0.45\textwidth]{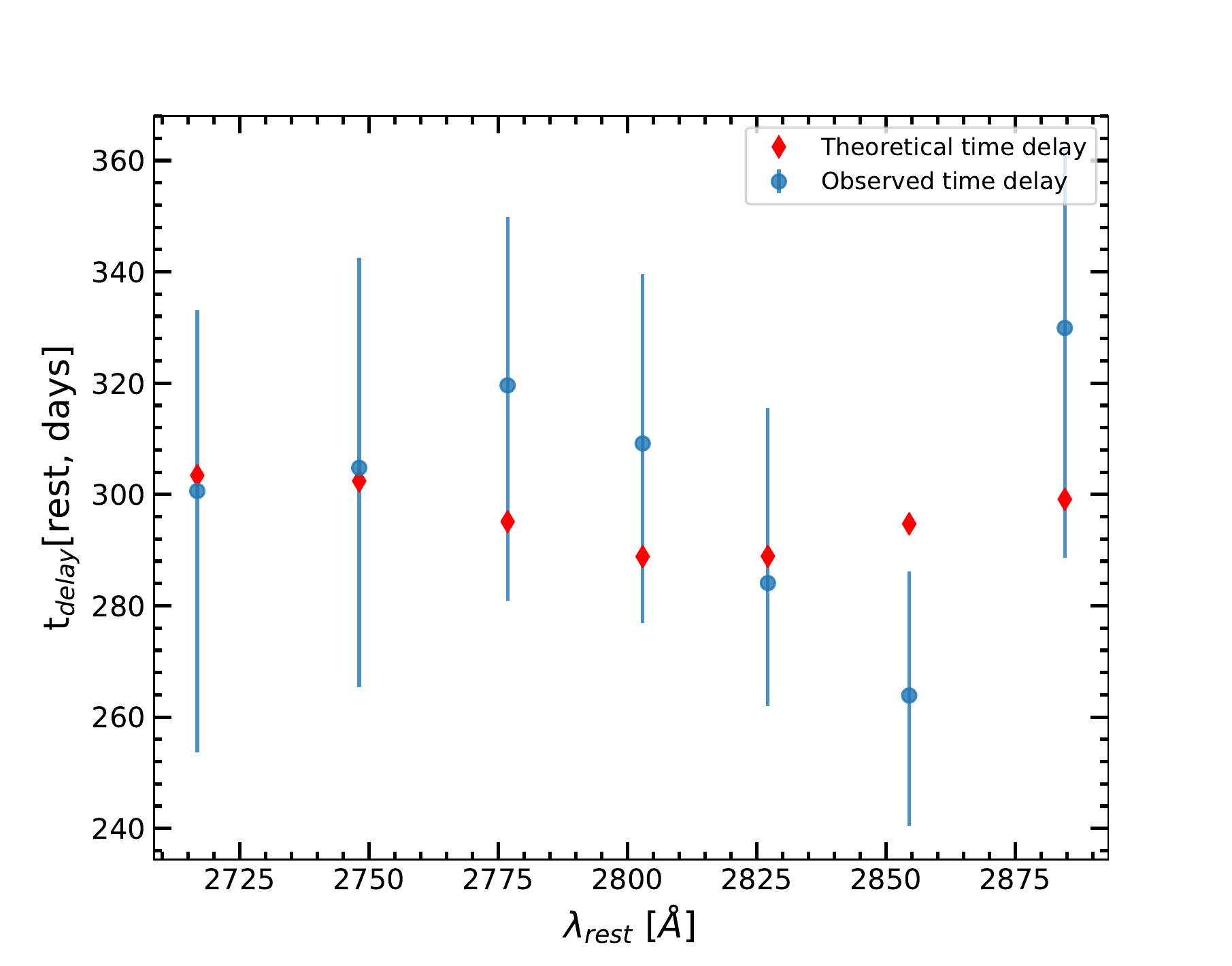}
    \includegraphics[width=0.45\textwidth]{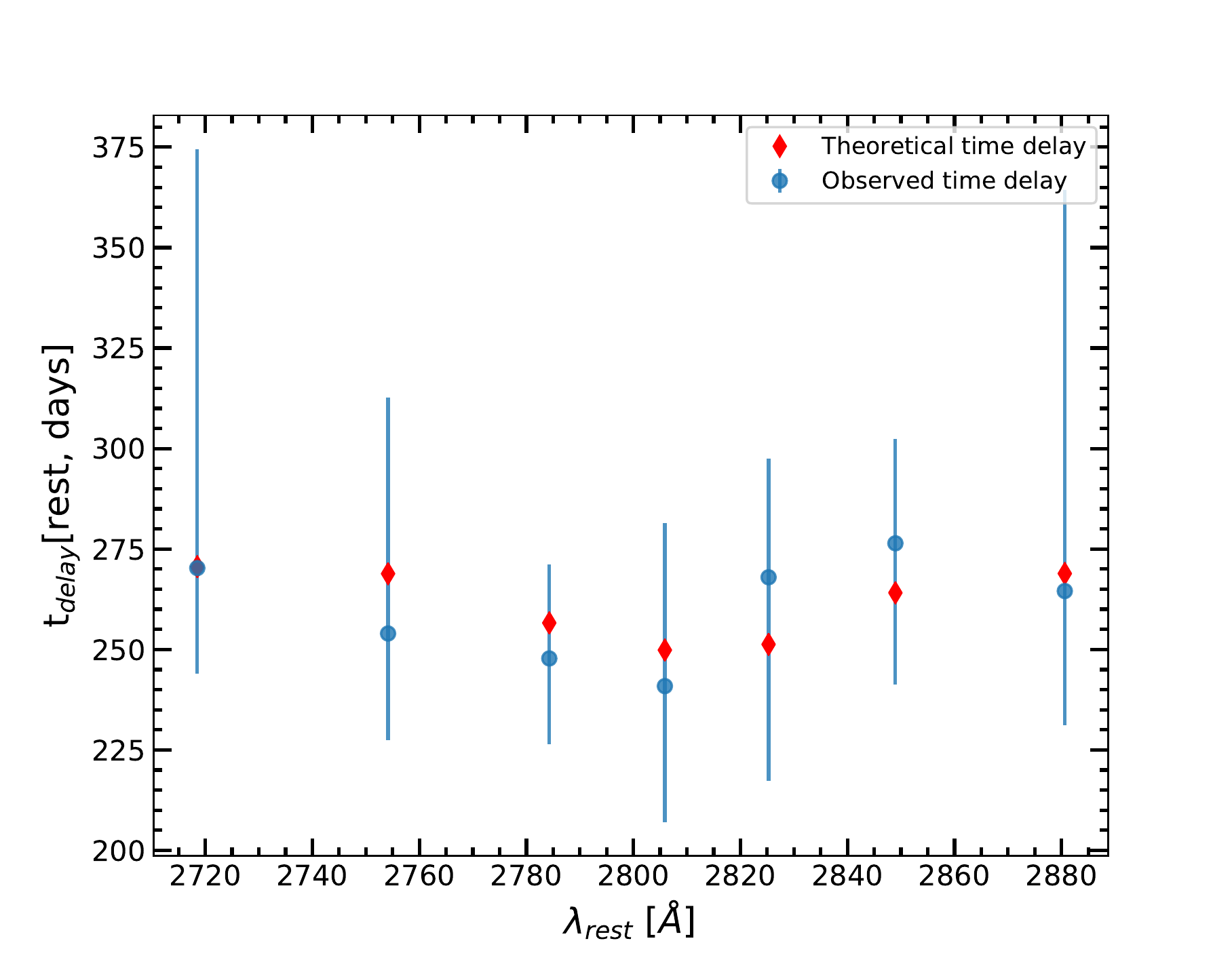}
    \caption{Comparison of observational (ICCF centroid) and theoretical time delay for two quasars (left: HE 0413-4031, right: HE 0435-4312).   }
    \label{fig:theory-delay}
\end{figure*}

Further, we test whether the pattern in wavelength-resolved data is fully explained by the complex contribution of Fe II and Mg II to the spectrum. Following \citet{2022MNRAS.509.2637N} we thus assumed that the delay at a given wavelength should be a mean average of the intrinsic Fe II and Mg II time delays weighted by the relative flux contribution,
\begin{equation}
\tau_i = \tau_{\rm FeII}\left(\frac{\rm FeII_i}{\rm FeII_i + MgII_i}\right) + \tau_{\rm MgII}\left( \frac{\rm MgII_i}{\rm FeII_i + MgII_i}\right)\,,
\end{equation}
where ${\rm FeII_i}$ and ${\rm MgII_i}$ are the corresponding integrated flux contribution to the bin $i$. We fitted such a model for the delays obtained from the ICCF centroid method (see Figure~\ref{fig:theory-delay}). For the quasar HE 0413, the wavy pattern is not fully reproduced. It is likely that there are some kinematic effects that are not included in this simple picture. The mean relative shift of the Fe II and Mg II determined from the spectral fitting ($1595 \pm 74$ km s$^{-1}$ ) also implies such an effect. The agreement seems much better for the quasar HE 0435.

By a factor of $\sim 2$ smaller role of Fe II in the spectrum of HE 0413-4031 is surprising since this quasar has a considerably higher Eddington ratio and a single-component line profile characteristic for type A quasars \citep{Sulentic2000}. In the optical plane (and in the corresponding UV plane), the Fe II equivalent contribution usually rises with the Eddington ratio \citep[e.g.][]{panda2018,panda2019,2019ApJ...875..133P}. This might be consistent with the findings of no correlation between equivalent widths in the optical and UV bands \citep{kovacevic2015}.

\subsection{Radius-Luminosity relation for the UV Fe II and optical Fe II}

Radius-luminosity (R-L) relation offers an important insight into the structure of the BLR. Our new measurements for the delay of the Mg II line nicely supplement the results obtained by us and by several other groups \citep{Czerny2019,Zajacek2020,zajacek2021,2022arXiv220805491Y}, see also Appendix~\ref{fig_MgII_RL}. 

However, the R-L relation for Fe II UV emission is new, and it was introduced for the first time in \citet{prince_CTS_2022}.

Here we update the Fe II UV R-L relation by including two more sources studied in this work. The time delays of the total Fe II emissions are taken from Table \ref{tab_totalMgII_FeII}. Combining these two measurements with the Fe II time delay for CTS C30.10 \citep{prince_CTS_2022} as well as with the old IUE measurement of the Fe II UV time delay for NGC 5548 \citep{Maoz1993}, we can now attempt more reliably to establish radius - luminosity relation for the UV Fe II emission. We list these measurements conveniently in Table~\ref{tab_uv_FeII_sources}. 

The relation has a typical slope close to 0.5 (see Figure~\ref{fig_FeII_RL}, right panel) which is expected either from the scaling with the ionization parameter \citep[e.g.][]{}, or from the dust-based BLR model of \citet{czhr2011}. However, not just a slope but also a vertical shift is important as it informs us about the relative localization of various emission components of the BLR. 

We thus compare the UV Fe II R-L relation with the optical Fe II R-L relation in order to decompose the emission regions and the photoionization processes. The Fe II emission in the optical band has been studied for many years \citep[see][for a recent review]{gaskell2022}. The optical Fe II time delays of a few nearby sources have been estimated by different groups (\citealt{Bian2010, Barth2013, chelouche2014, HuChen2015, Hu_2020, HuChen2020}). In total, we collected 20 measurements from the literature and summarized them in Table~\ref{tab_optical_FeII_sources}. 

The fitted R-L relation for the optical band is shown in Figure~\ref{fig_FeII_RL} (left panel). The slope again is consistent with 0.5, and the dispersion is typical for the delay measurements in single emission lines.

For a better comparison with the UV R-L relation, we convert the 5100\AA~to 3000\AA~ UV monochromatic luminosity and plotted the two relations together for better visualization of the trend (right panel). We clearly see the vertical offset, with UV Fe II being located closer to the SMBH by a factor of $\sim$ 1.8. It is not unexpected since UV lines correspond to larger energies of the atomic transitions. In addition, statistical studies of $\sim$ 300 AGN implied that optical UV Fe II and optical Fe II emissions do not exhibit a simple scaling although they are kinematically connected  \citep{kovacevic2015}. According to their study, the mean FWHM width of optical Fe II is smaller than UV (2360 km s$^{-1}$ vs. 2530 km s$^{-1}$) which would correspond to a separation factor of 1.15, i.e. less than the difference found by us between the two R-L relations but going in the same direction. { In addition, \citet{kovacevic2015} measured a considerable mean outflow velocity in the UV Fe II of 1150 km s$^{-1}$ with respect to [OIII] implying an inflow of materials. \citet{Hu_2008} have studied the optical Fe II emission in a large sample of quasars selected from SDSS and they show that the red asymmetry in the optical Fe II emission profile could represent an inflow of materials.


In Figure~\ref{fig_FeII_RL} (right panel), we also plot the MgII R-L relation, which is clearly flatter than both the optical and the UV FeII R-L relations and has a larger intercept. This results in larger MgII time delays than both optical and UV FeII time delays for lower luminosity sources. At the same time, all three relations appear to converge towards high-luminosity sources, where  our three quasars are also located. This behavior is also observationally confirmed for the low-luminosity source NGC5548, for which the UV FeII time delay is $10\pm 1$ days \citet{Maoz1993}, while the MgII time delay is constrained to be in the range $34-72$ days \citep{1991ApJ...366...64C}. The three luminous quasars -- CTS C30.10, HE0413-4031, and HE0435-4312 -- have their corresponding MgII and FeII time delays generally very close to each other. For CTS 30.10, several methods indicated that the UV FeII time delay is $180.3^{+26.6}_{-30.0}$ days, hence shorter than the MgII time delay of $275.5^{+12.4}_{-19.5}$ \citep{prince_CTS_2022}, again going in the direction predicted by the best-fit MgII and FeII R-L relations.

\begin{figure*}
    \centering
    \includegraphics[width=0.495\textwidth]{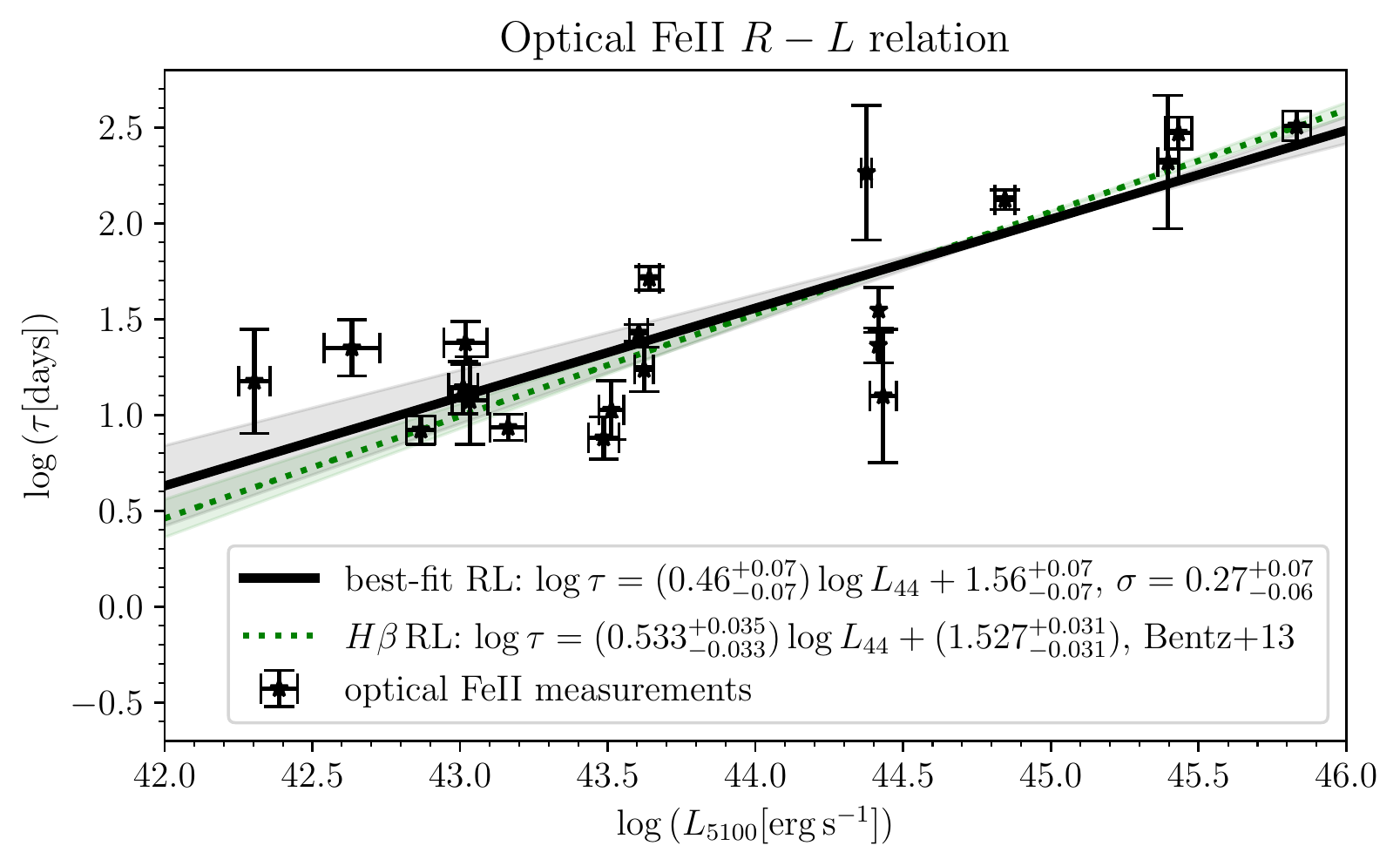}
    \includegraphics[width=0.495\textwidth]{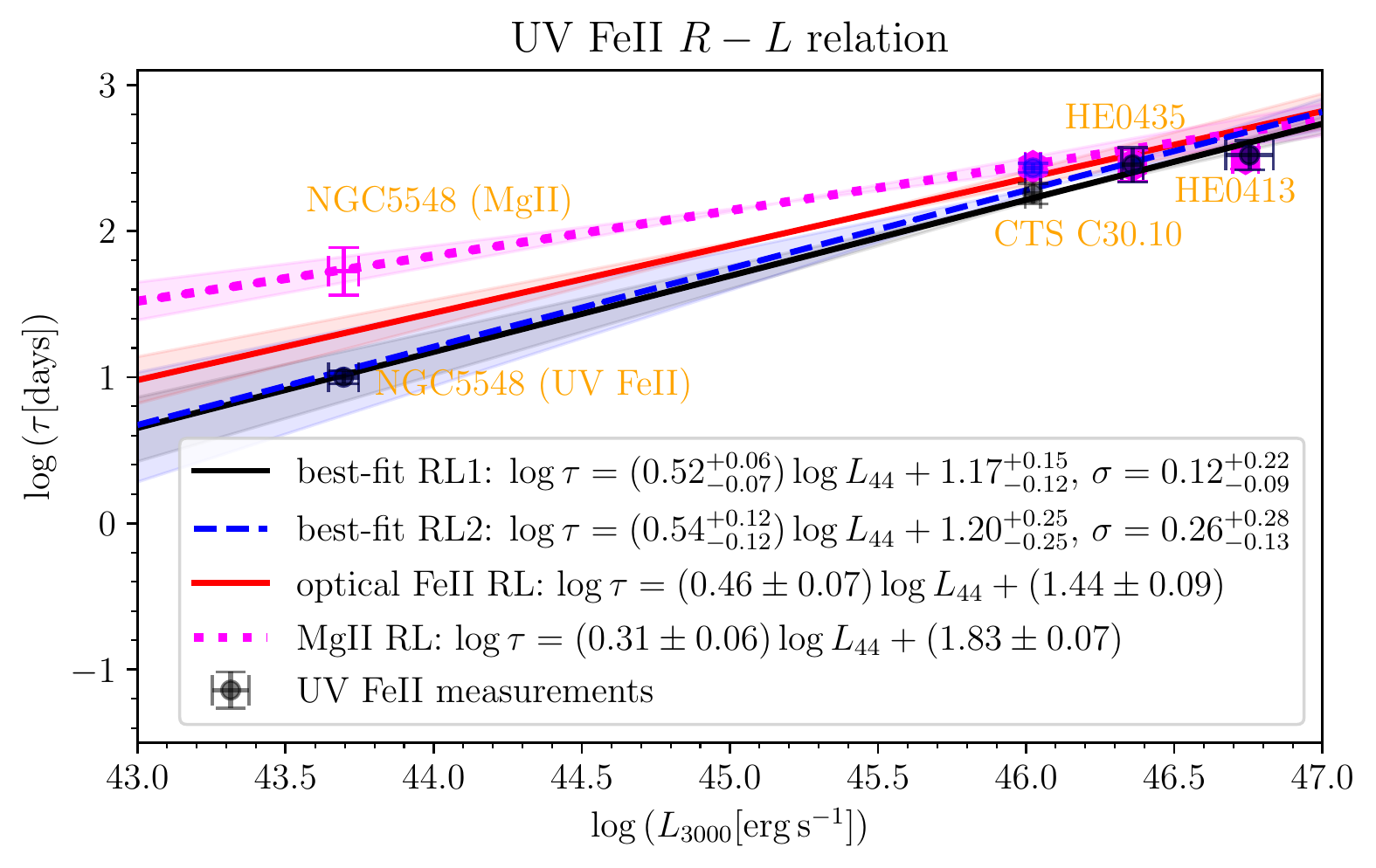}
    \caption{Radius-luminosity relation for the broad FeII complex. \textit{Left panel:} The optical FeII radius-luminosity based on 20 measurements (points with error bars; see Table~\ref{tab_optical_FeII_sources}) and the corresponding best-fit relation (black solid line). The best-fit coefficients are in the legend. The intrinsic scatter is $\sigma=0.27$ dex. For comparison, the H$\beta$ RL relation \citep{Bentz2013} is plotted using a dotted green line. Both the optical FeII and the H$\beta$ RL relations are in agreement with the uncertainties. \textit{Right panel:} The UV FeII radius luminosity relation is based on 4 measurements summarized in Table~\ref{tab_uv_FeII_sources}. The black solid line and the blue dashed line represent the best-fit solutions considering two different FeII time delays for CTS C30.10. The UV FeII radius-luminosity relation is compared to the optical FeII radius-luminosity relation (red solid line), which is rescaled from 5100\AA\, to 3000\,\AA\, using the bolometric corrections of \citet{2019MNRAS.488.5185N}. A vertical offset between the two R-L relations is apparent, which corresponds to the potential mean size difference between the optical FeII-emitting region and the UV FeII-emitting region of $R_{\rm FeII-opt}\sim (1.7-1.9) R_{\rm FeII-UV}$, assuming the same slope of $\gamma=0.5$ within the uncertainties. In addition, we also plot the MgII R-L relation based on 94 measurements (dotted magenta line). This relation is flatter than the UV FeII R-L relation and has a larger intercept. It is, however, consistent with the MgII time-delay measurements (magenta points) for four UV FeII sources.} 
    \label{fig_FeII_RL}
\end{figure*}

\begin{figure*}
    \centering
    \includegraphics[width=0.33\textwidth]{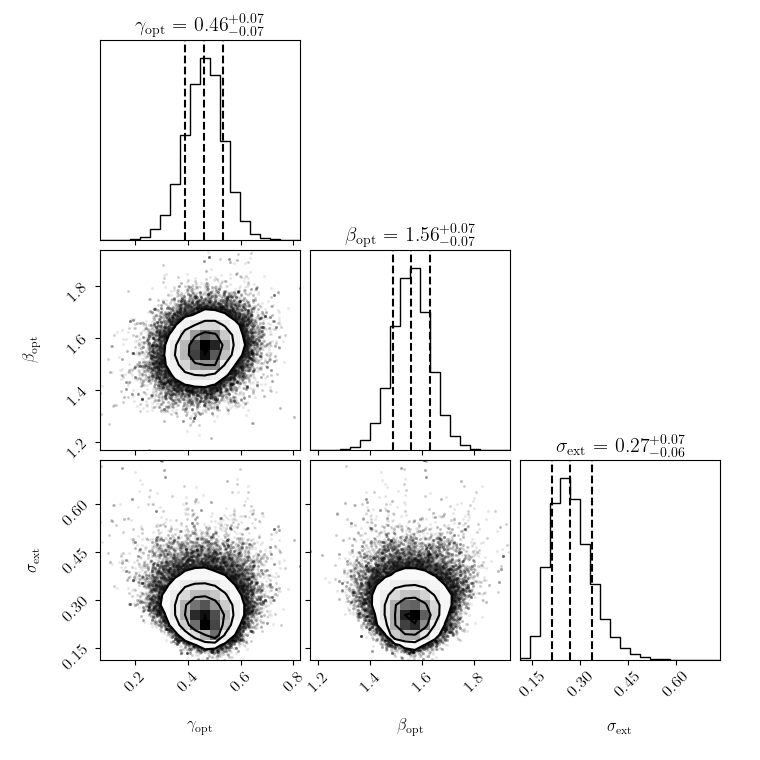}
    \includegraphics[width=0.33\textwidth]{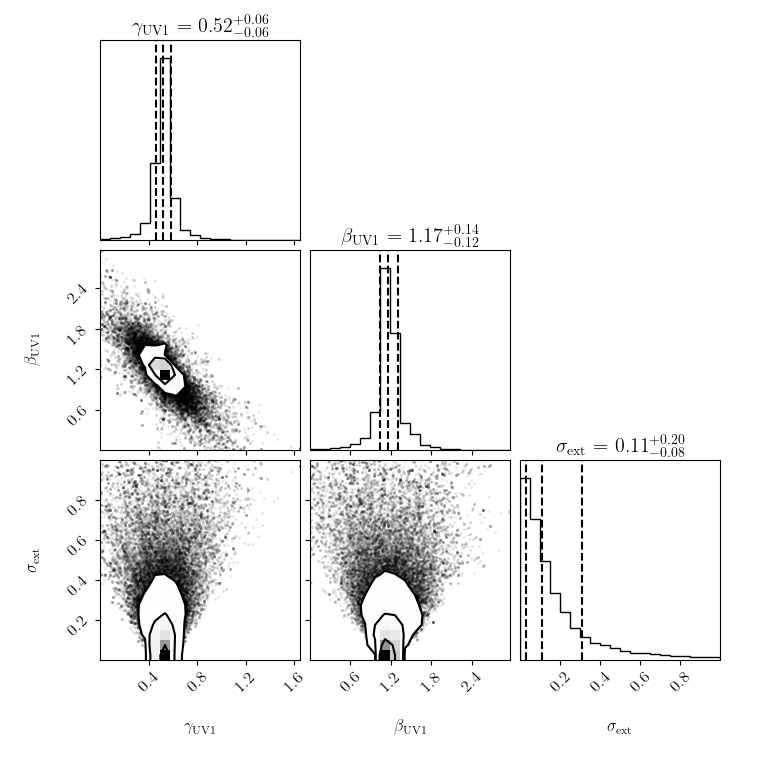}
    \includegraphics[width=0.33\textwidth]{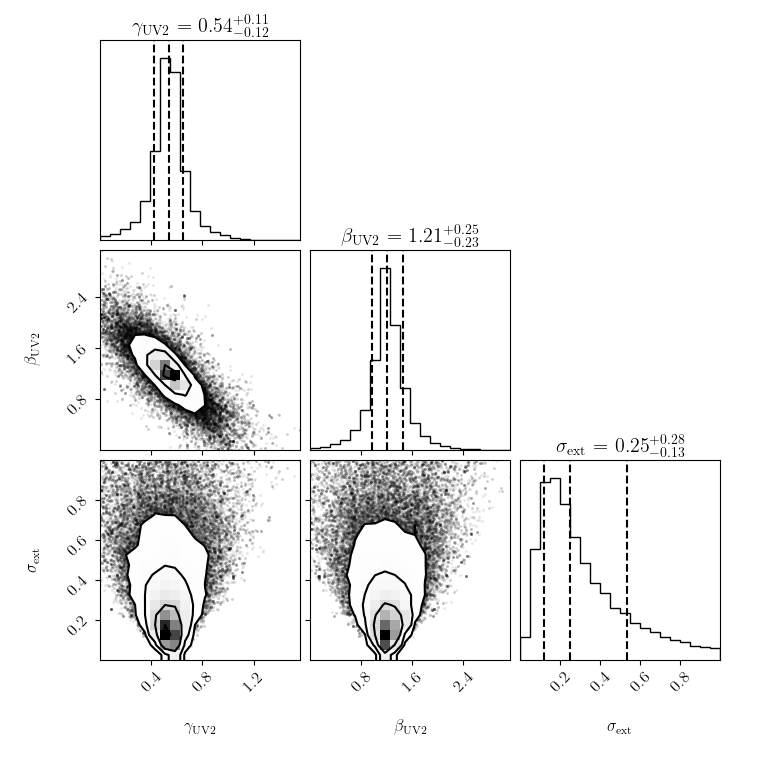}
    \caption{One-dimensional likelihood distributions (along the diagonal) and the two-dimensional likelihood contours for the FeII radius-luminosity relation. \textit{Left panel:} Parameter constraints and the intrinsic scatter $\sigma_{\rm ext}$ for the optical FeII radius-luminosity relation based on 20 measurements. \textit{Middle panel:} Parameter constraints and the intrinsic scatter $\sigma_{\rm ext}$ for the UV FeII radius-luminosity relation based on 4 measurements (for CTS C30.10 we considered the lower value of the FeII time delay, $180.3^{+26.6}_{-30.0}$ days in the rest frame). \textit{Right panel:} Parameter constraints and the intrinsic scatter $\sigma_{\rm ext}$ for the UV FeII radius-luminosity relation based on 4 measurements (for CTS C30.10 we considered the higher value of the FeII time delay, $270.0^{+13.8}_{-25.3}$ days in the rest frame).}
    \label{fig_FeII_RL_corner}
\end{figure*}

\section{Conclusions}
Understanding the BLR kinematics and its geometry is a long-standing problem in AGN physics. The wavelength-resolved reverberation technique has made some progress in improving our understanding of the size and structure of the BLR. We present the long-term spectroscopic monitoring of two intermediate-reshift, luminous quasars - HE 0413-4031 and HE 0435-4312, with the SALT and photometric monitoring with 1-m class telescopes. The reverberation mapping of broad Mg II and pseudo-continuum UV Fe II emissions helps us to disentangle their locations.

The Mg II time delays inferred in both the quasars using several methods are consistent within the uncertainties. However, the mean Fe II time delays are a bit higher in both quasars compared to Mg II time delays, suggesting spatially shifted regions of their emission origin. Since this was the first time the UV Fe II time delays were derived for these two quasars, we combined these results with CTS C30.10 to constrain the UV Fe II R-L relation. A subsequent comparison of the optical Fe II and the UV Fe II R-L relations reveals two spatially separated locations with $R_{\rm FeII-opt}\sim (1.7-1.9) R_{\rm FeII-UV}$.

\label{sec_conclusions}

\begin{acknowledgements}
We thank the anonymous referee for their insightful suggestions and comments. The project is based on observations made with the
SALT under programs 2012-2-POL-003, 2013-1-POL-RSA-002, 2013-2-POL-RSA-001, 2014-1-POL-RSA-001, 2014-2-SCI-004, 2015-1-SCI-006, 2015-2-SCI-017, 2016-1-SCI-011, 2016-2-SCI-024, 2017-1-SCI-009, 2017-2-SCI-033, 2018-1-MLT-004 (PI: B. Czerny). Polish participation in SALT is funded by grant No. MEiN nr 2021/WK/01.
This project has received funding from the European Research Council (ERC) under the European Union’s Horizon 2020 research and innovation program (grant agreement No. [951549]).
The project was partially supported by the Polish Funding Agency National Science Centre, project 2017/26/A/ST9/00756 (MAESTRO 9). BC and MZ acknowledge the OPUS-LAP/GAČR-LA bilateral project (2021/43/I/ST9/01352/OPUS 22. M. L. M.-A. acknowledges financial support from Millenium Nucleus NCN$19\_058$ (TITANs). SK acknowledges the financial support of the Polish National Science Center through the grant no. 2018/31/B/ST9/00334 (OPUS 16)GF22-04053L). C.S.F. and M.H. acknowledge support from DFG programs HA3555/14-2 and CH71/33-1. G.P. acknowledge  the grant of the Polish Ministry of Science and Higher
Education (decision number DIR/WK/ 2018/09).
\end{acknowledgements}

%
%

\bibliographystyle{aa}
\bibliography{reference-list.bib}

\appendix

\section{Alias Mitigation}
{ We applied the alias mitigation technique to examine the reliability of longer-time delays frequently seen in Javelin results. The detail of the technique is described in section 3. The Javelin time delay distribution for both the sources before and after the alias mitigation are shown in Figure \ref{fig:javelin_he0435} \& \ref{fig:javelin_he0413}. }
\begin{figure*}
\centering
\includegraphics[scale=0.35]{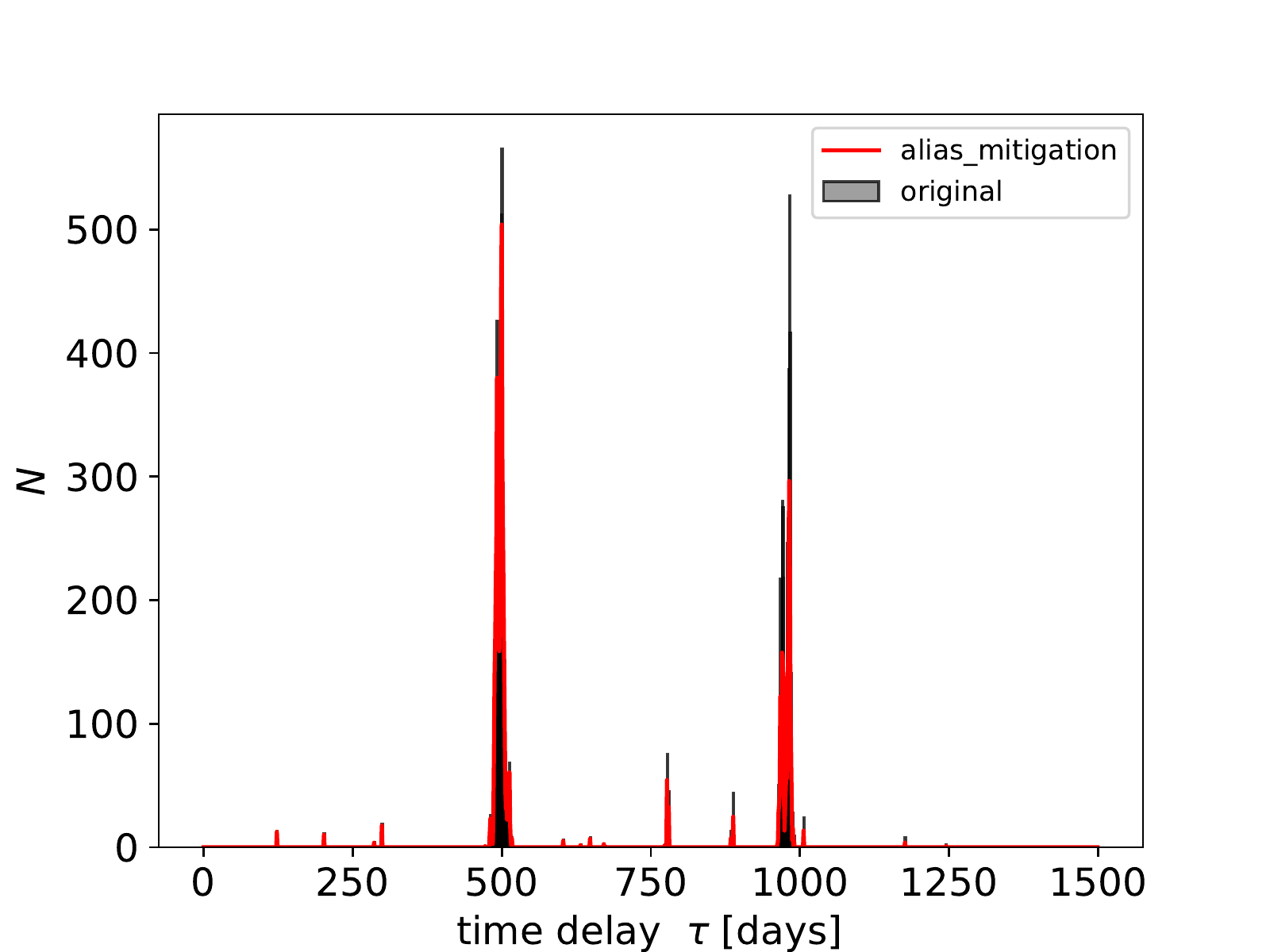}
\includegraphics[scale=0.35]{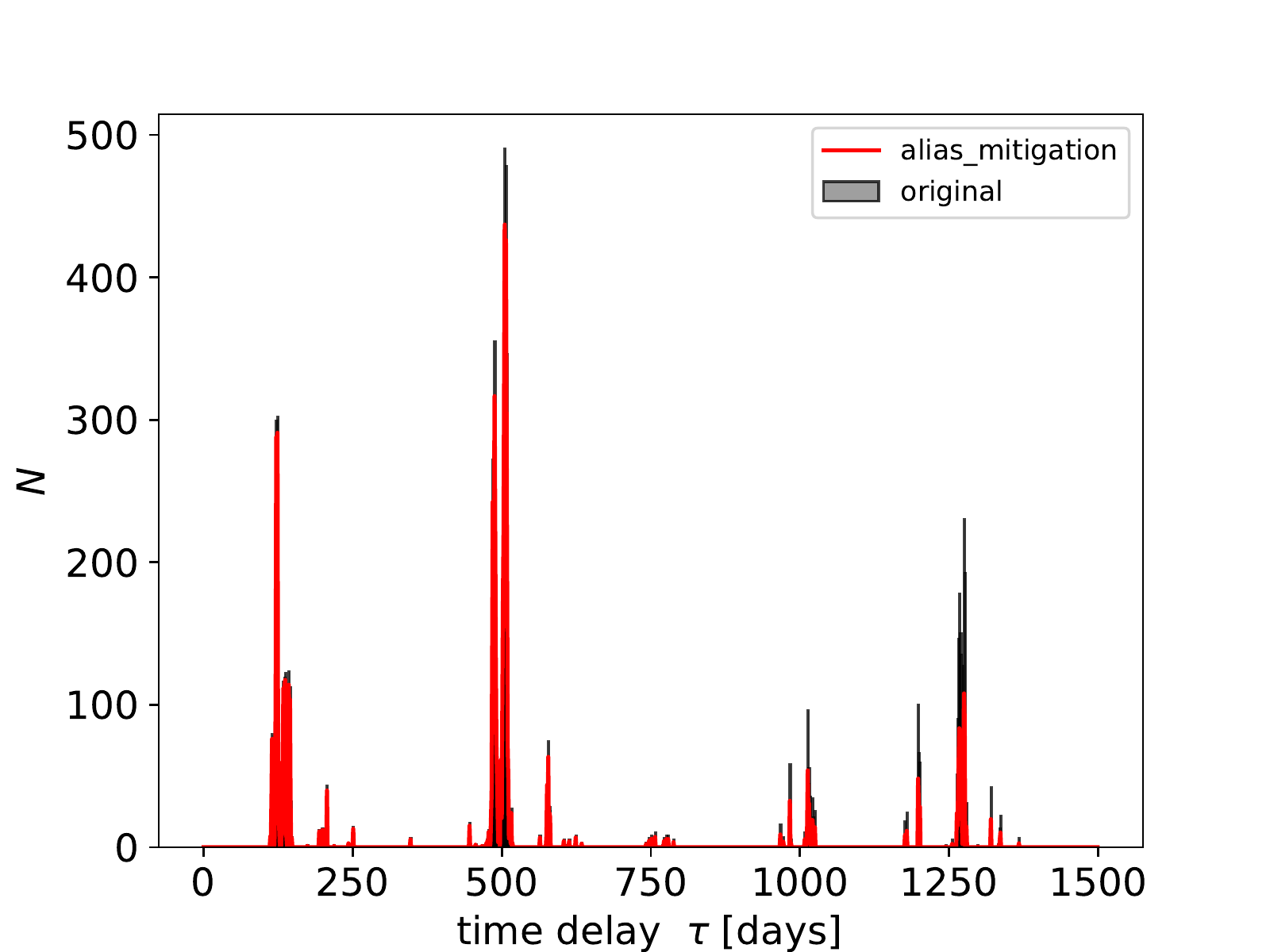}
\includegraphics[scale=0.35]{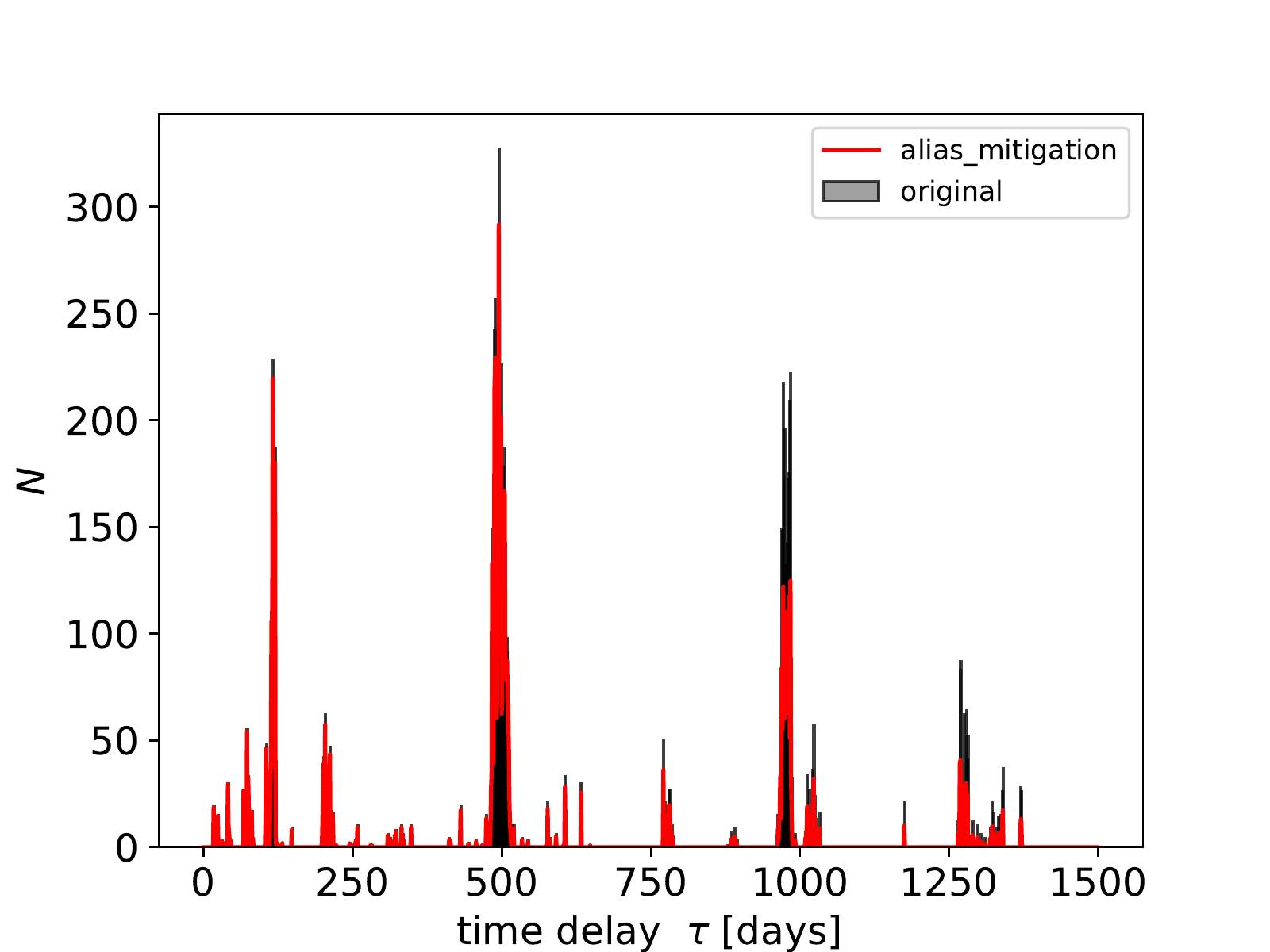}
\includegraphics[scale=0.35]{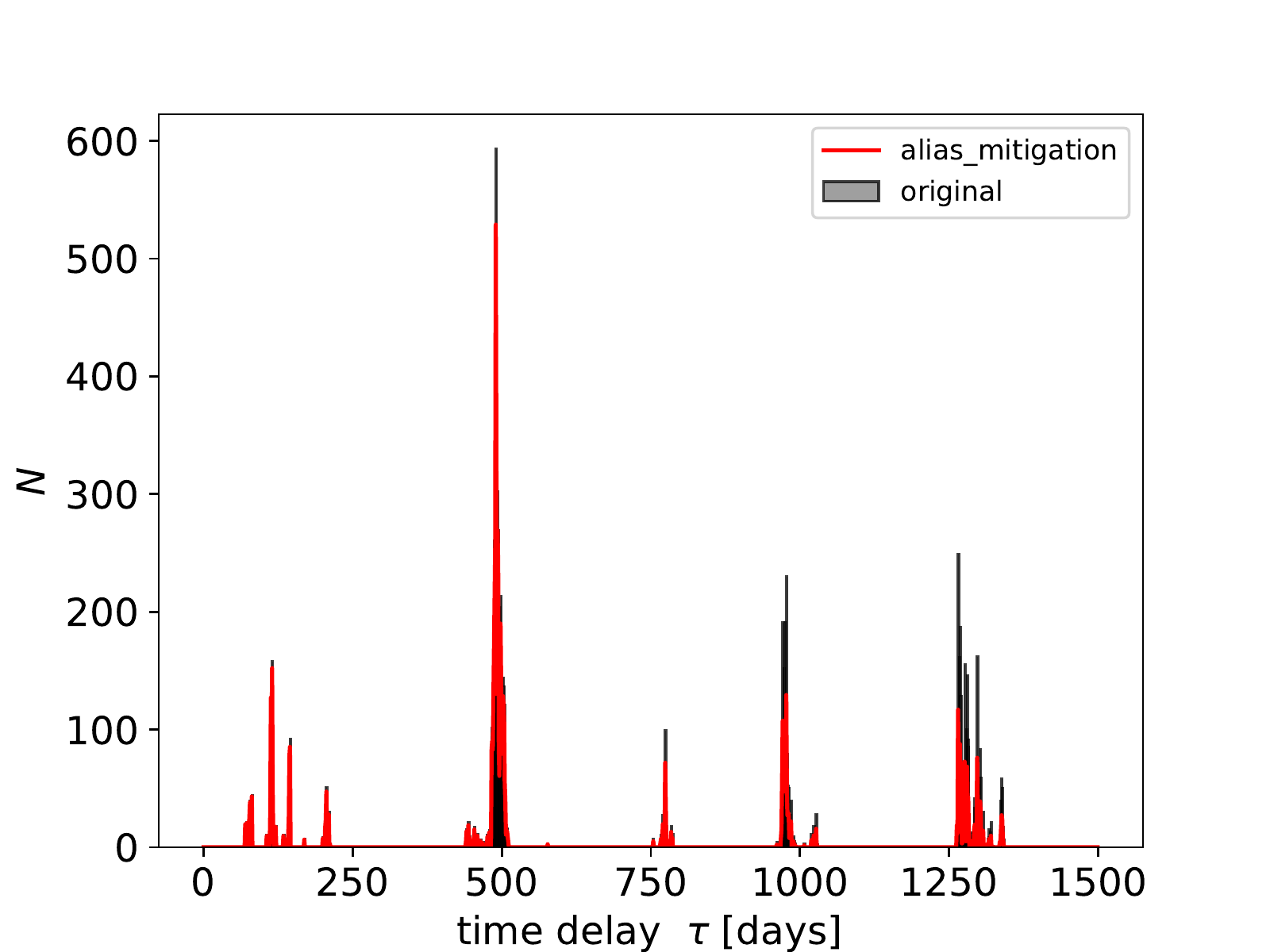}
\includegraphics[scale=0.35]{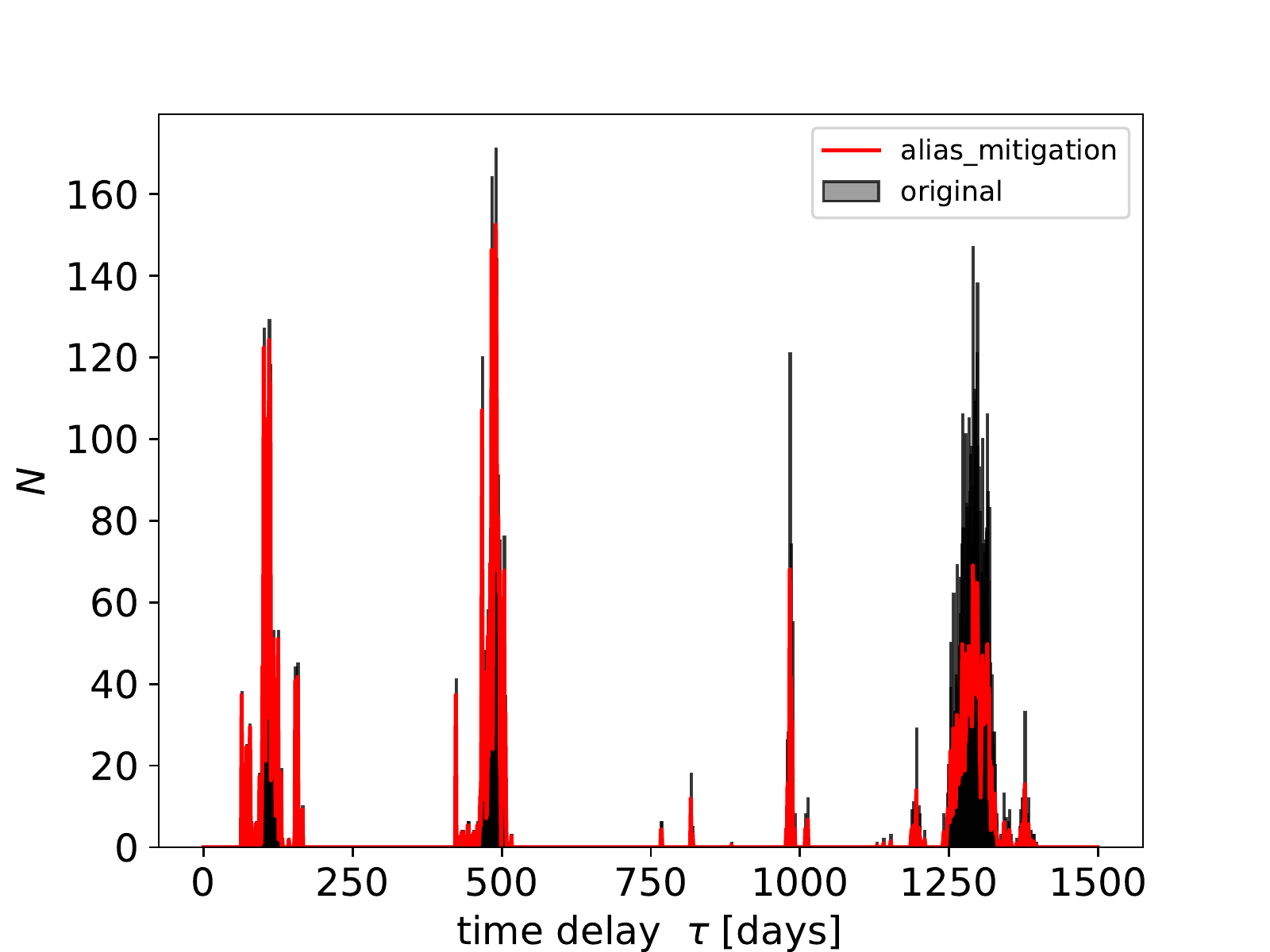}
\includegraphics[scale=0.35]{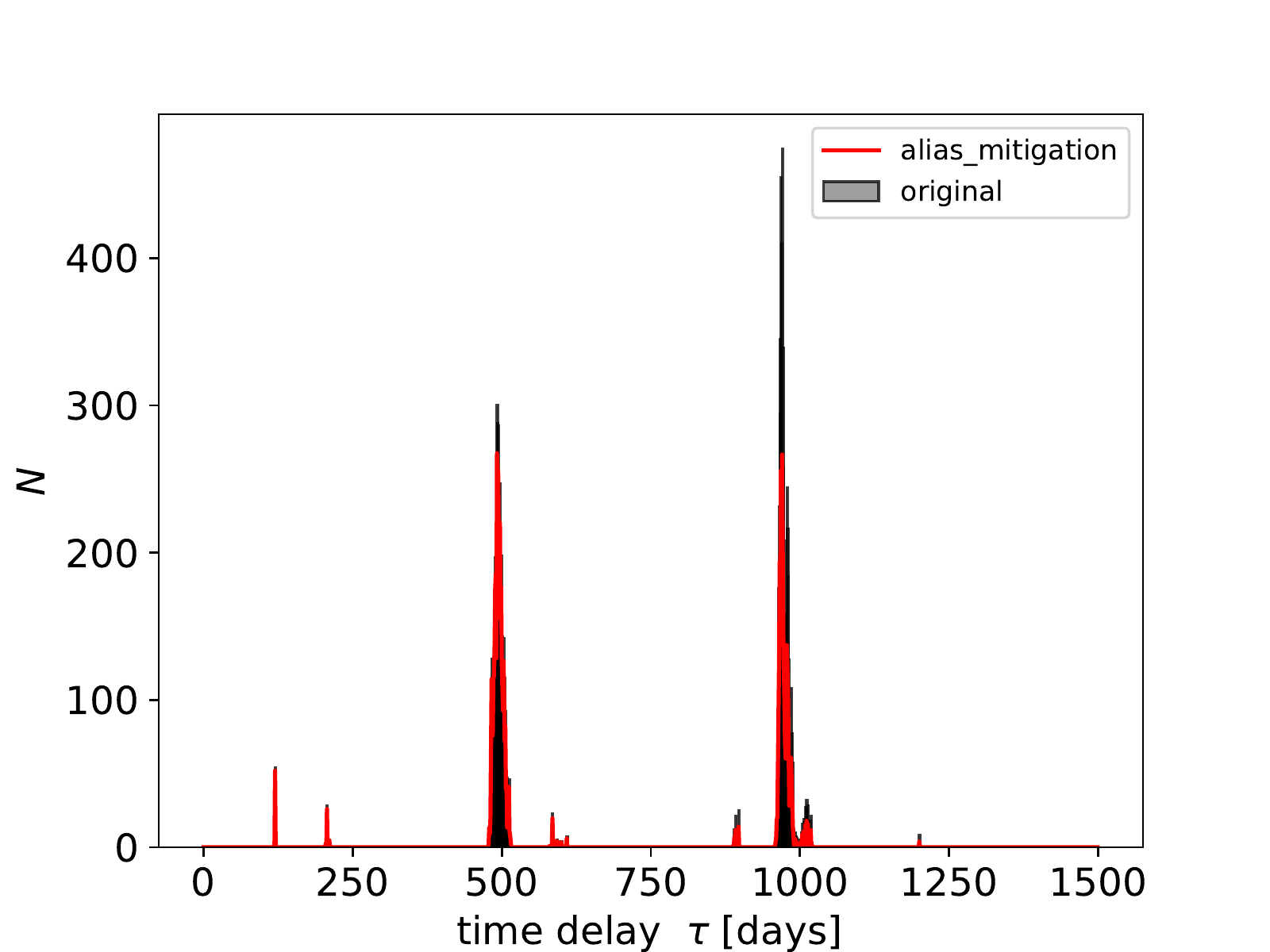}
\includegraphics[scale=0.35]{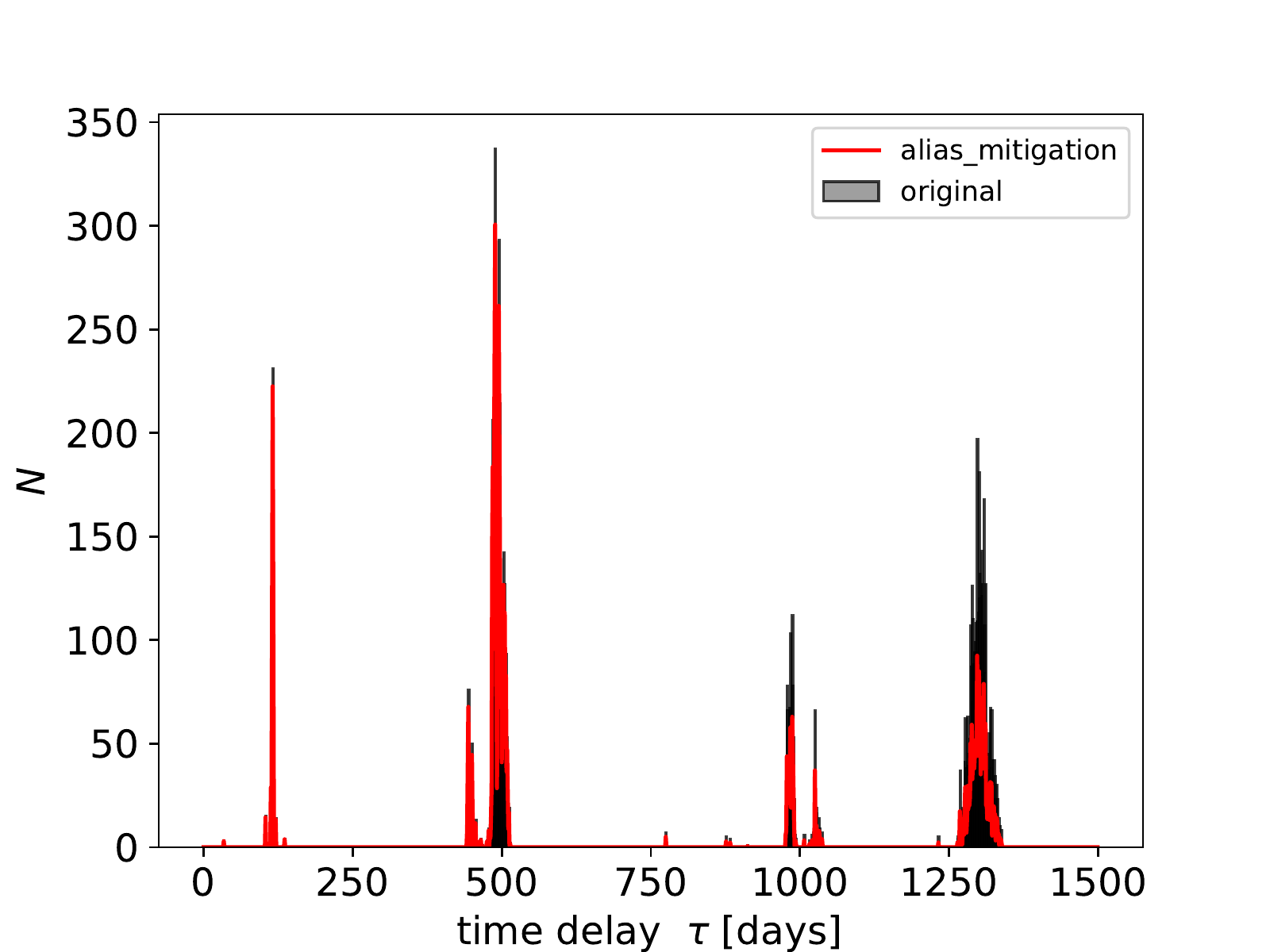}
\includegraphics[scale=0.35]{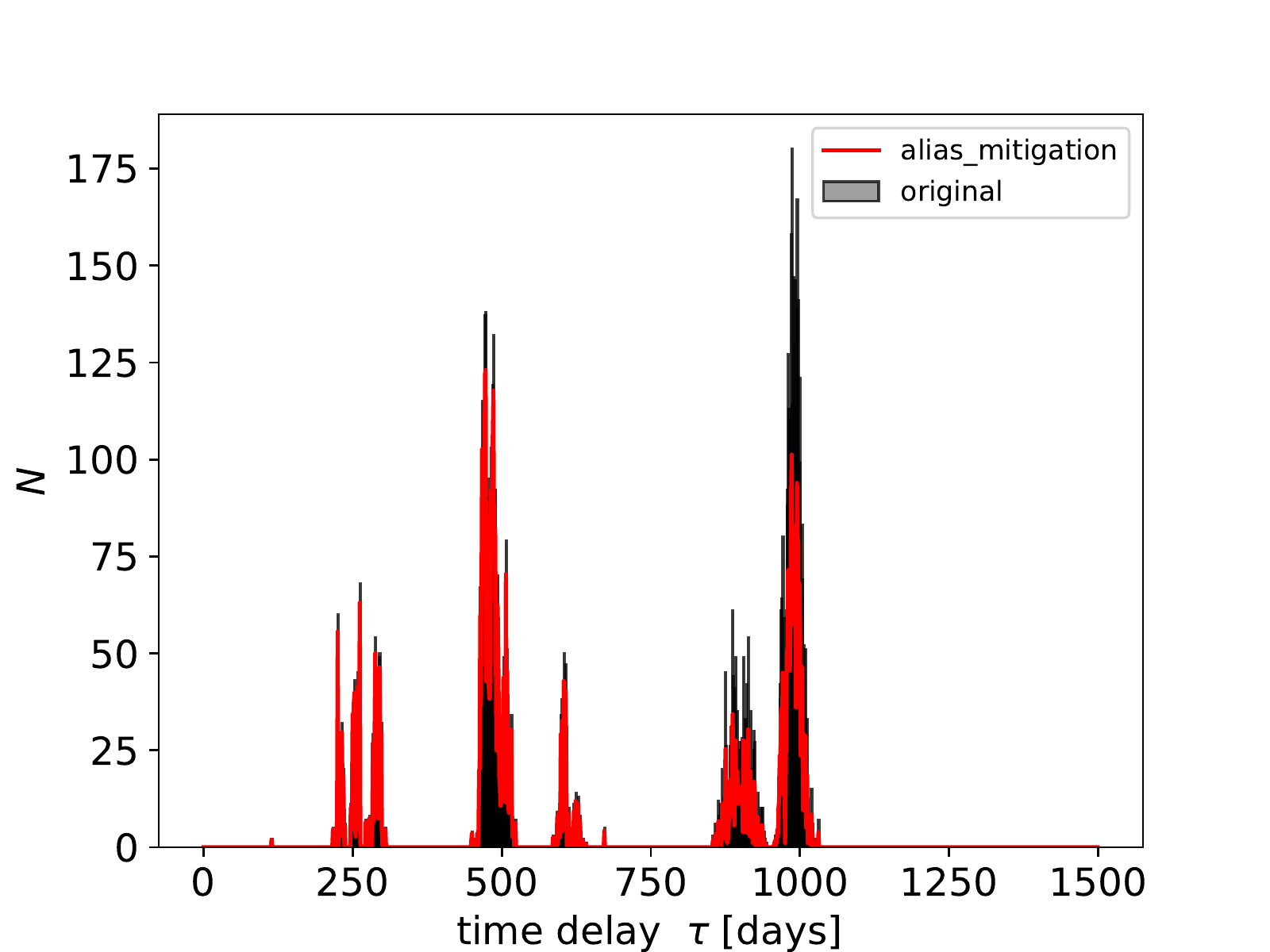}
\caption{Javelin bootstrap results from HE 0435 with 1000 realizations for all the seven curves (from left to right) along with total MgII and FeII (last 2 plots of the lower panel). The peak and  results from this are listed in Table \ref{tab:curves1_7}. We also use the alias mitigation using down-weighting by the overlapping pairs. The black histogram represents the original delay distribution and the red one is after alias mitigation.}
\label{fig:javelin_he0435}
\end{figure*}

\begin{figure*}
\centering
\includegraphics[scale=0.35]{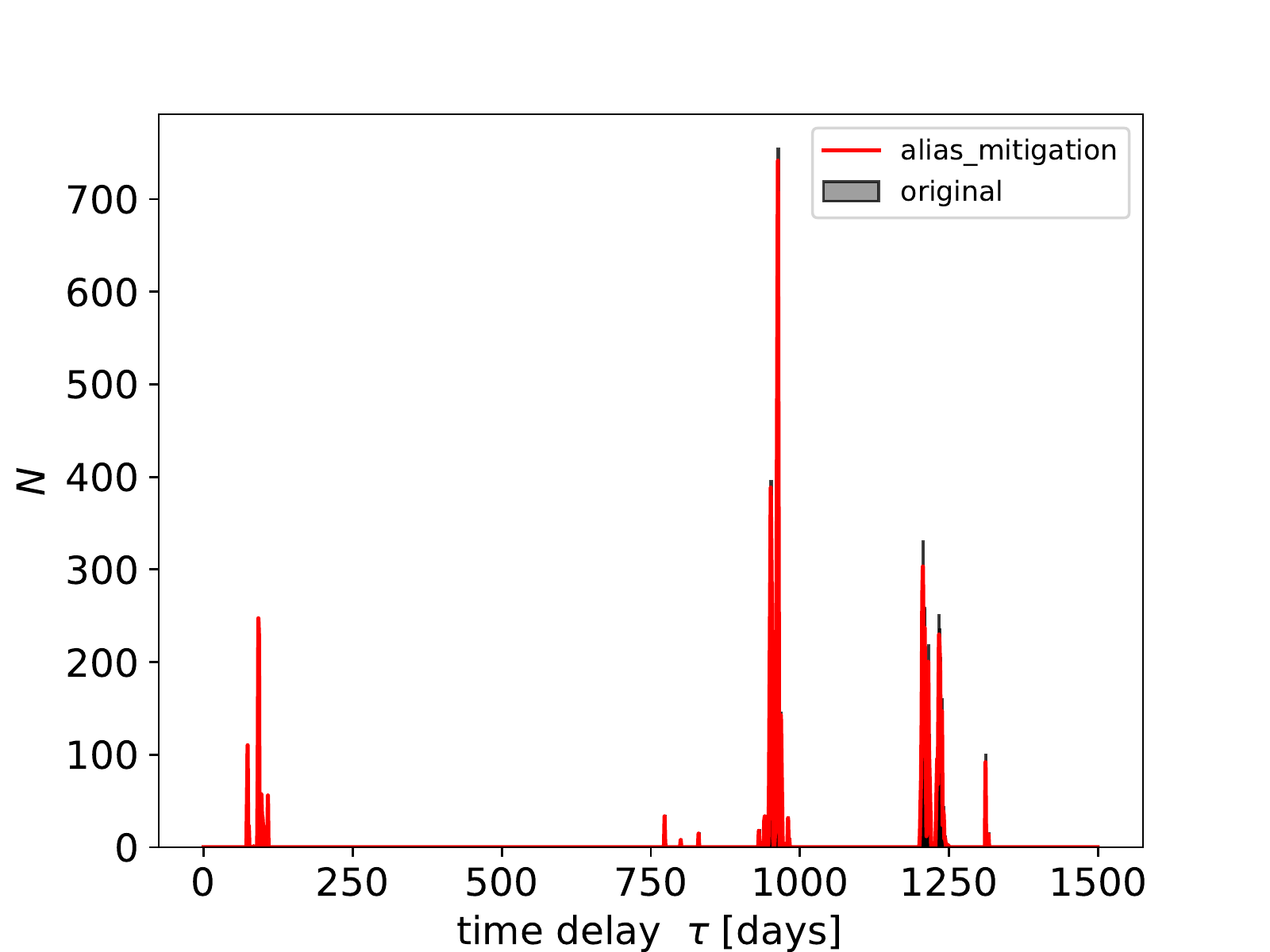}
\includegraphics[scale=0.35]{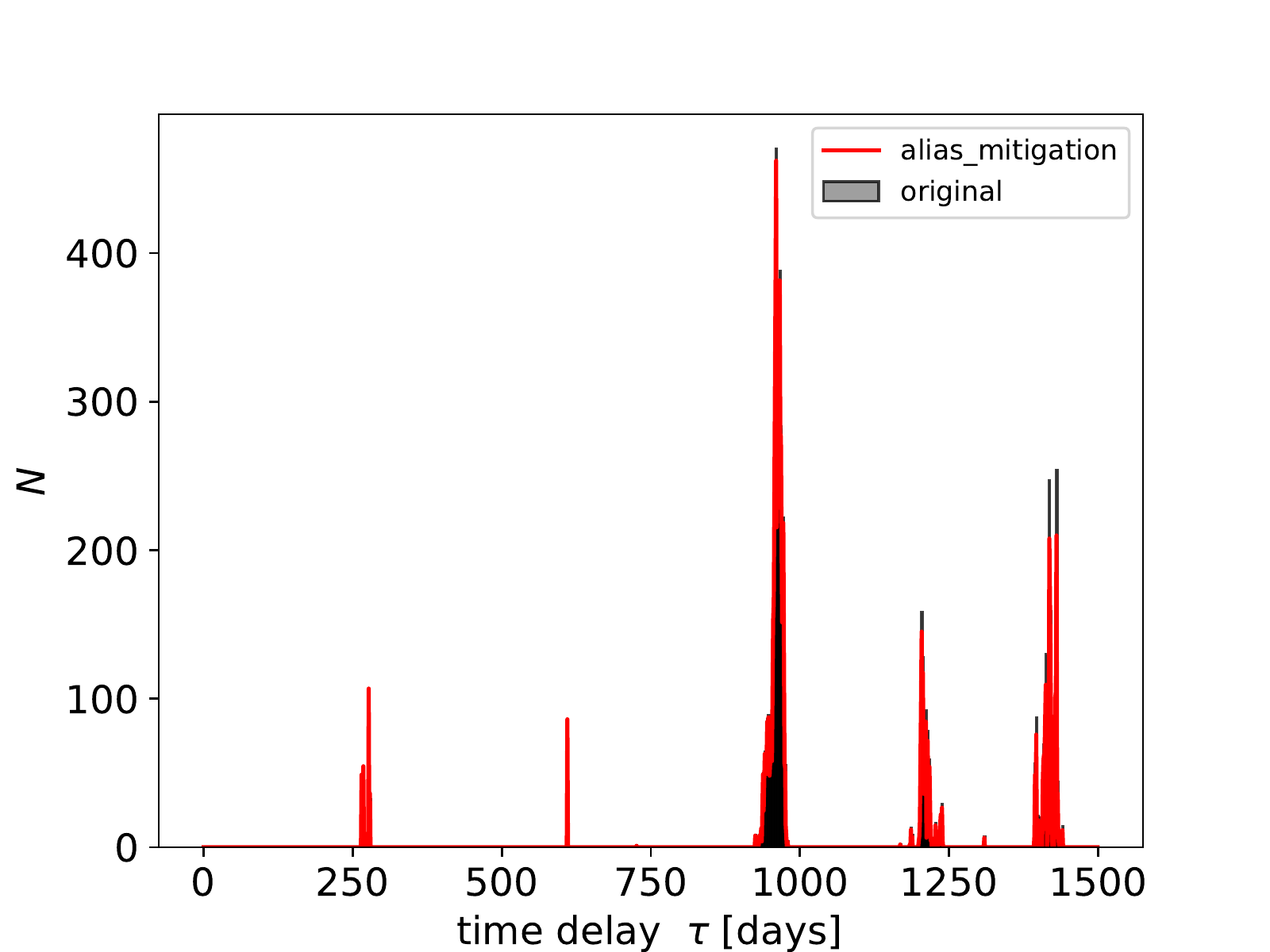}
\includegraphics[scale=0.35]{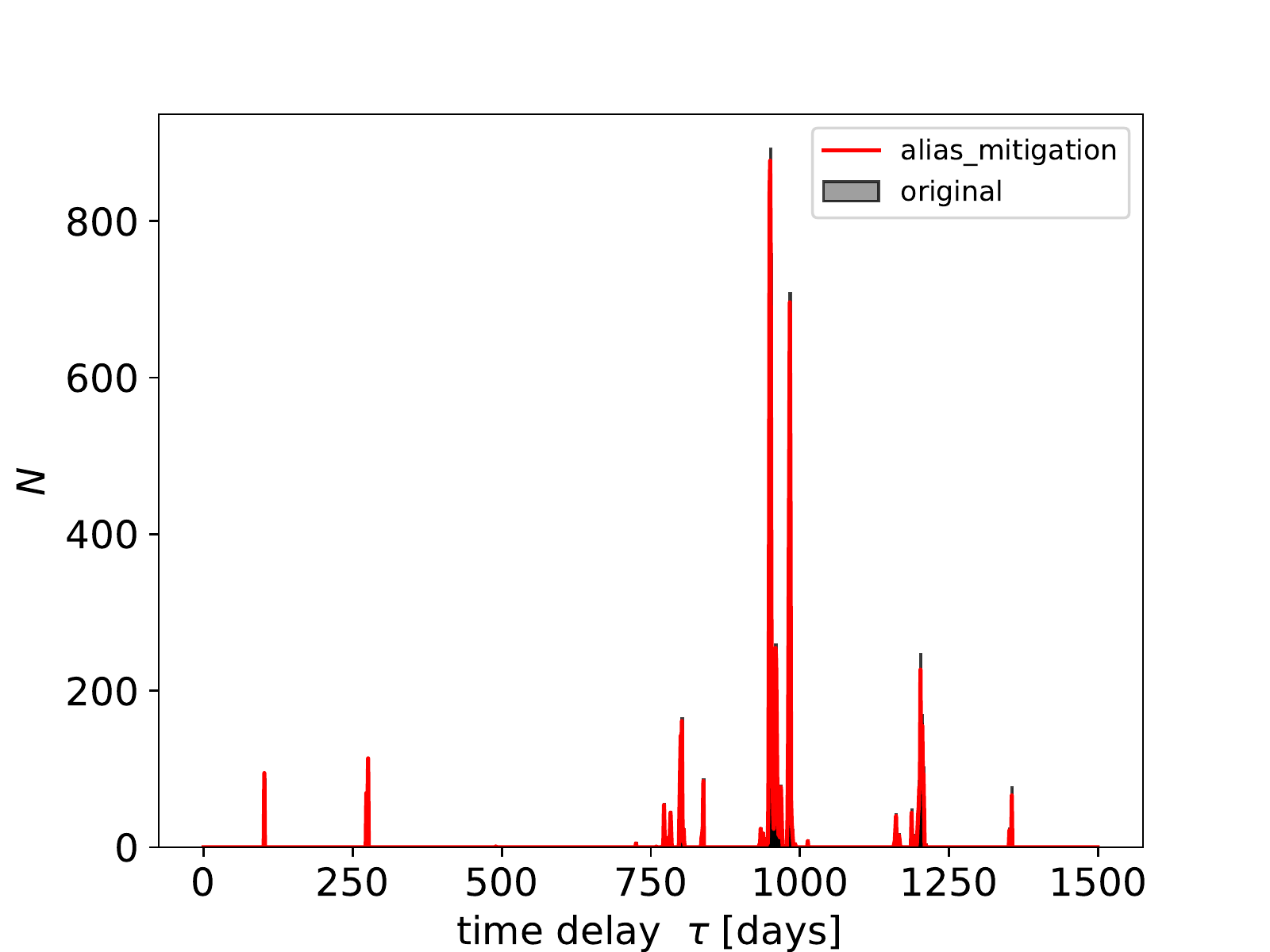}
\includegraphics[scale=0.35]{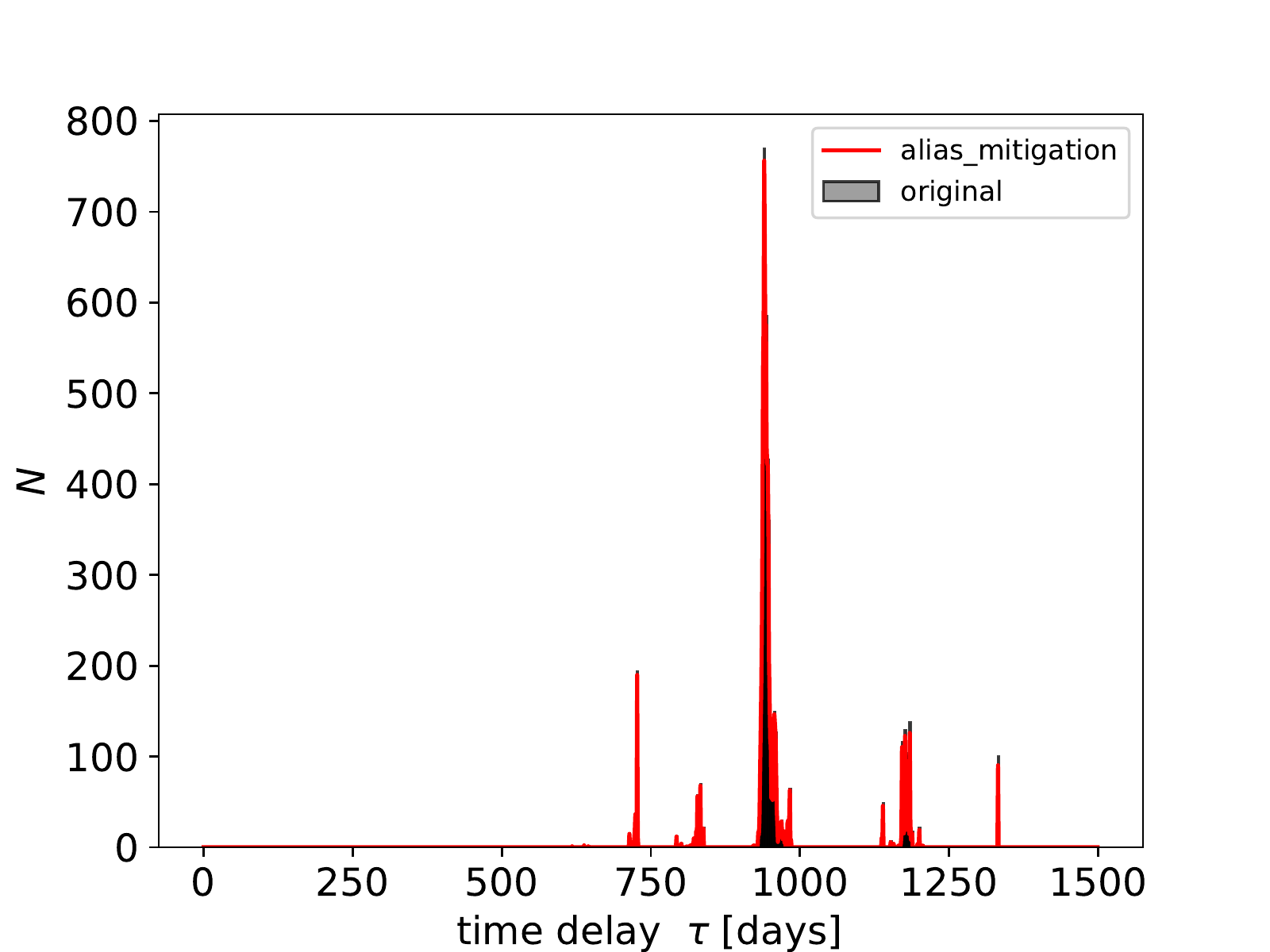}
\includegraphics[scale=0.35]{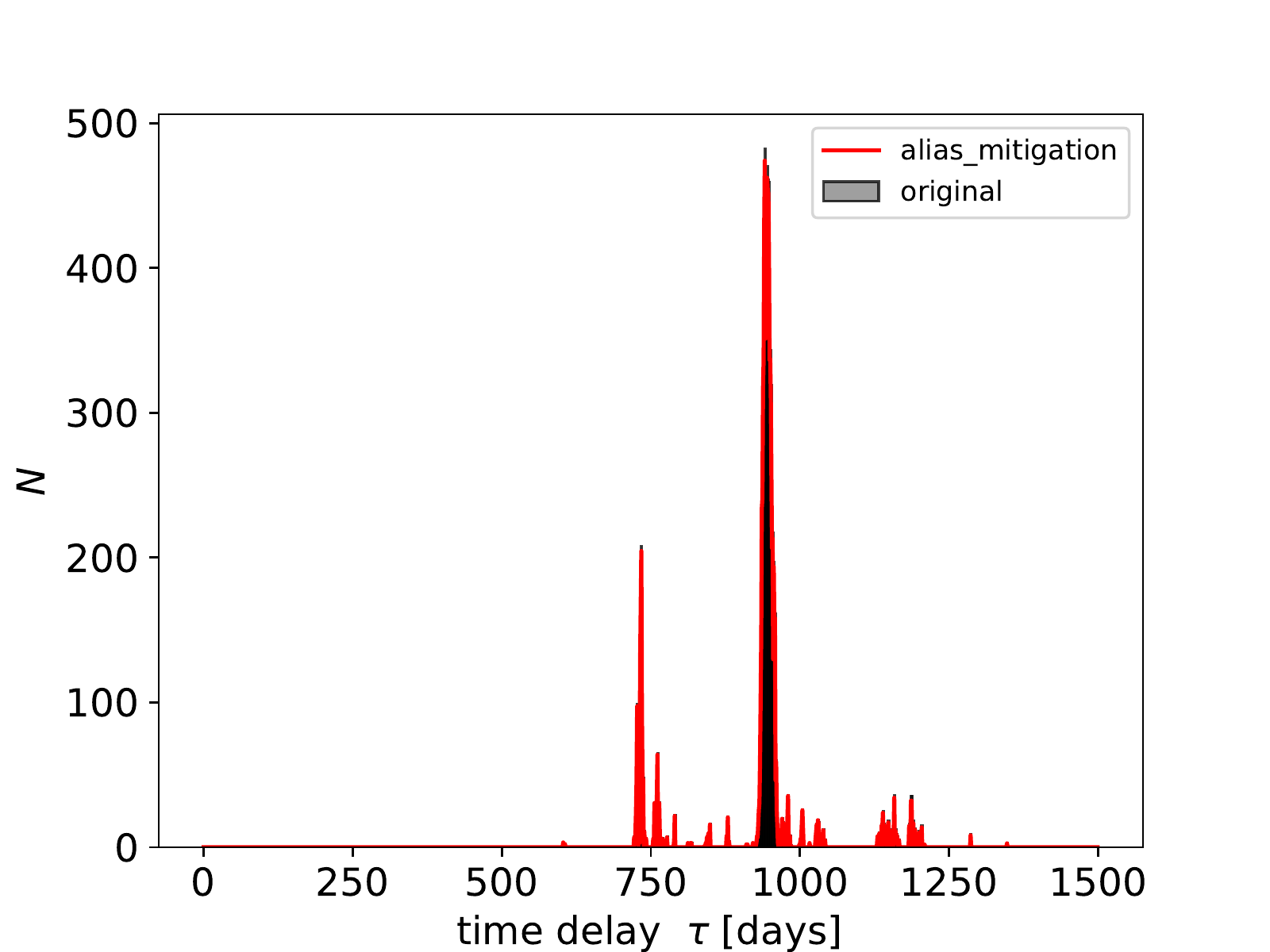}
\includegraphics[scale=0.35]{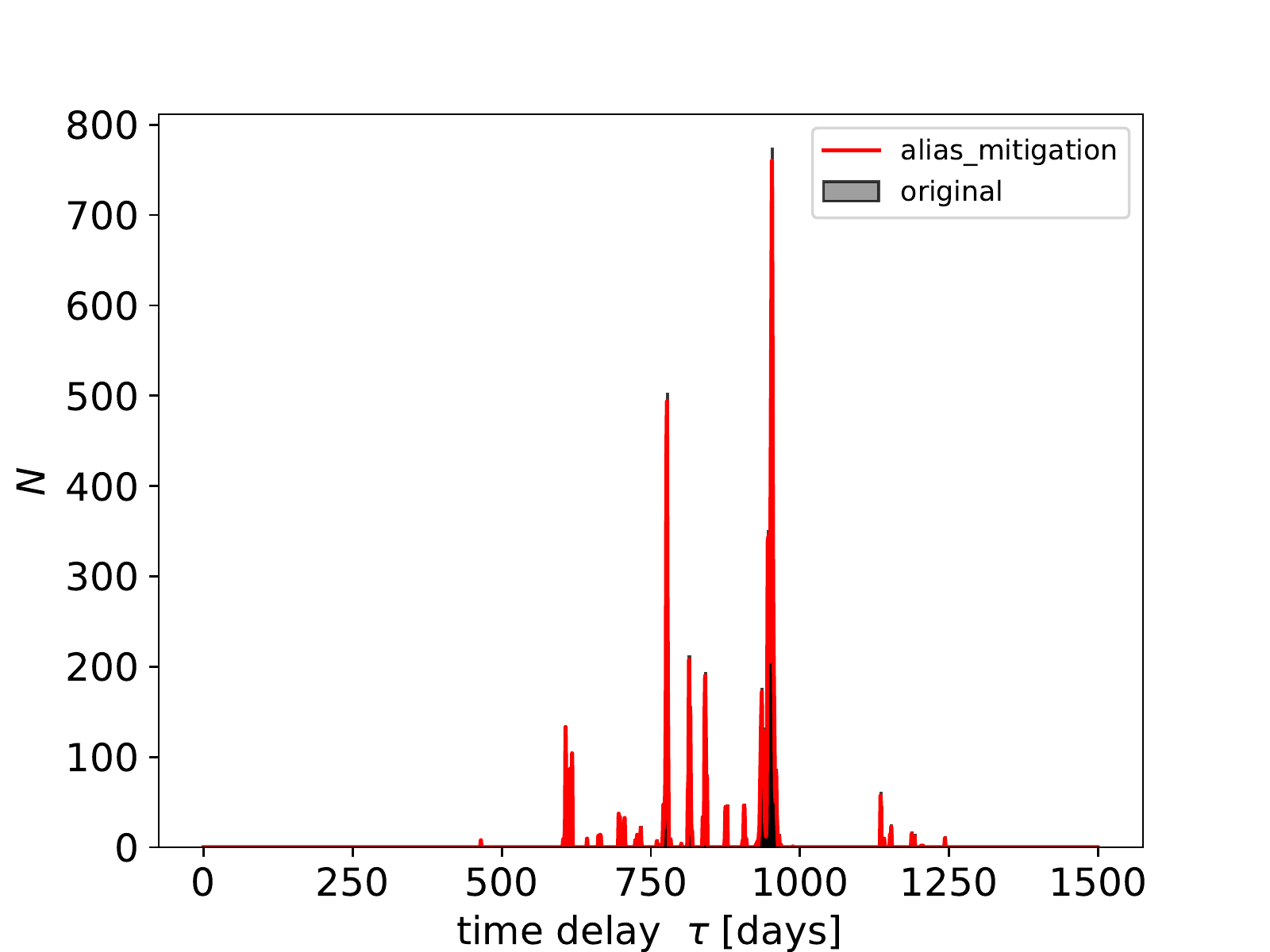}
\includegraphics[scale=0.35]{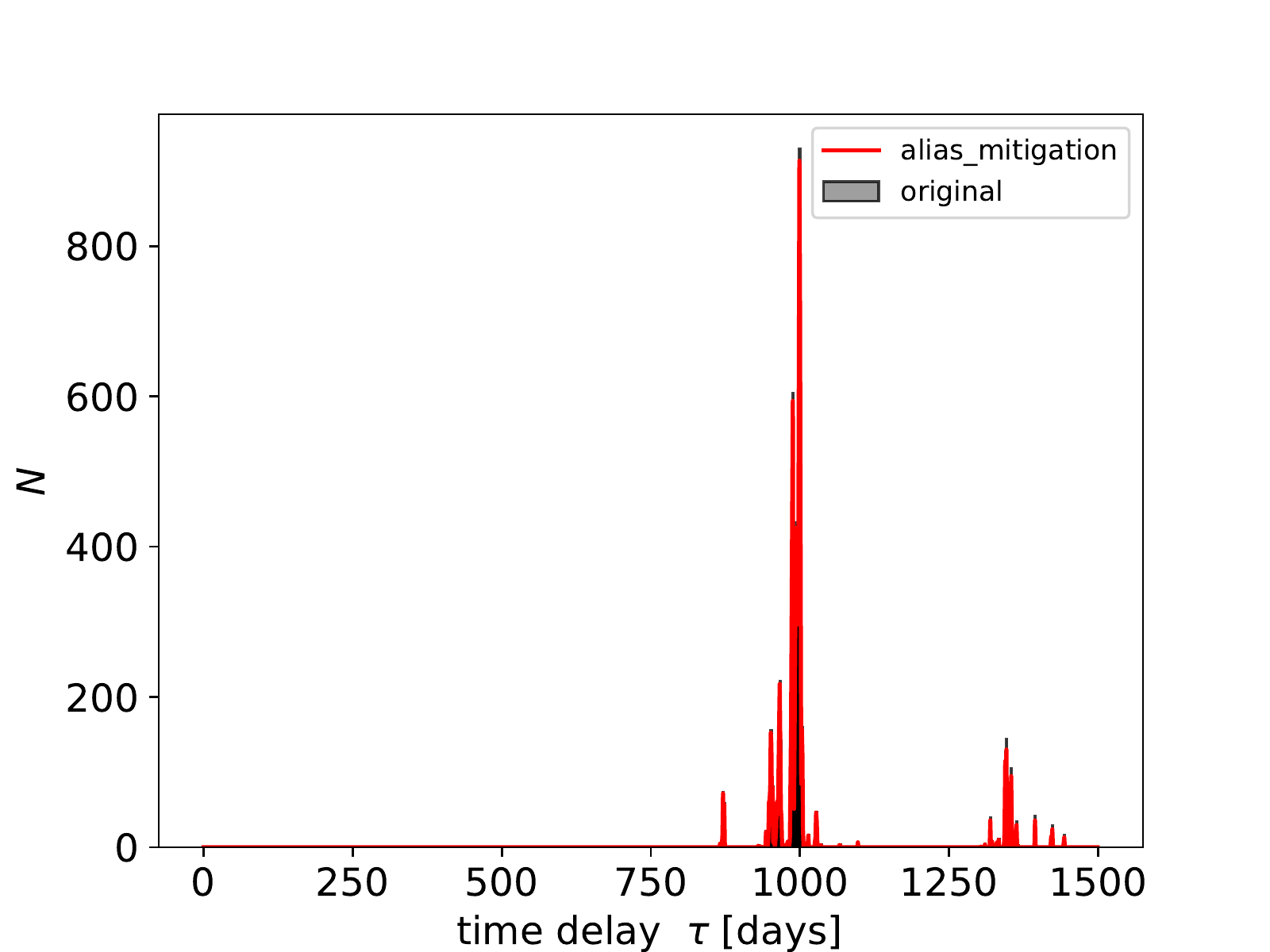}
\includegraphics[scale=0.35]{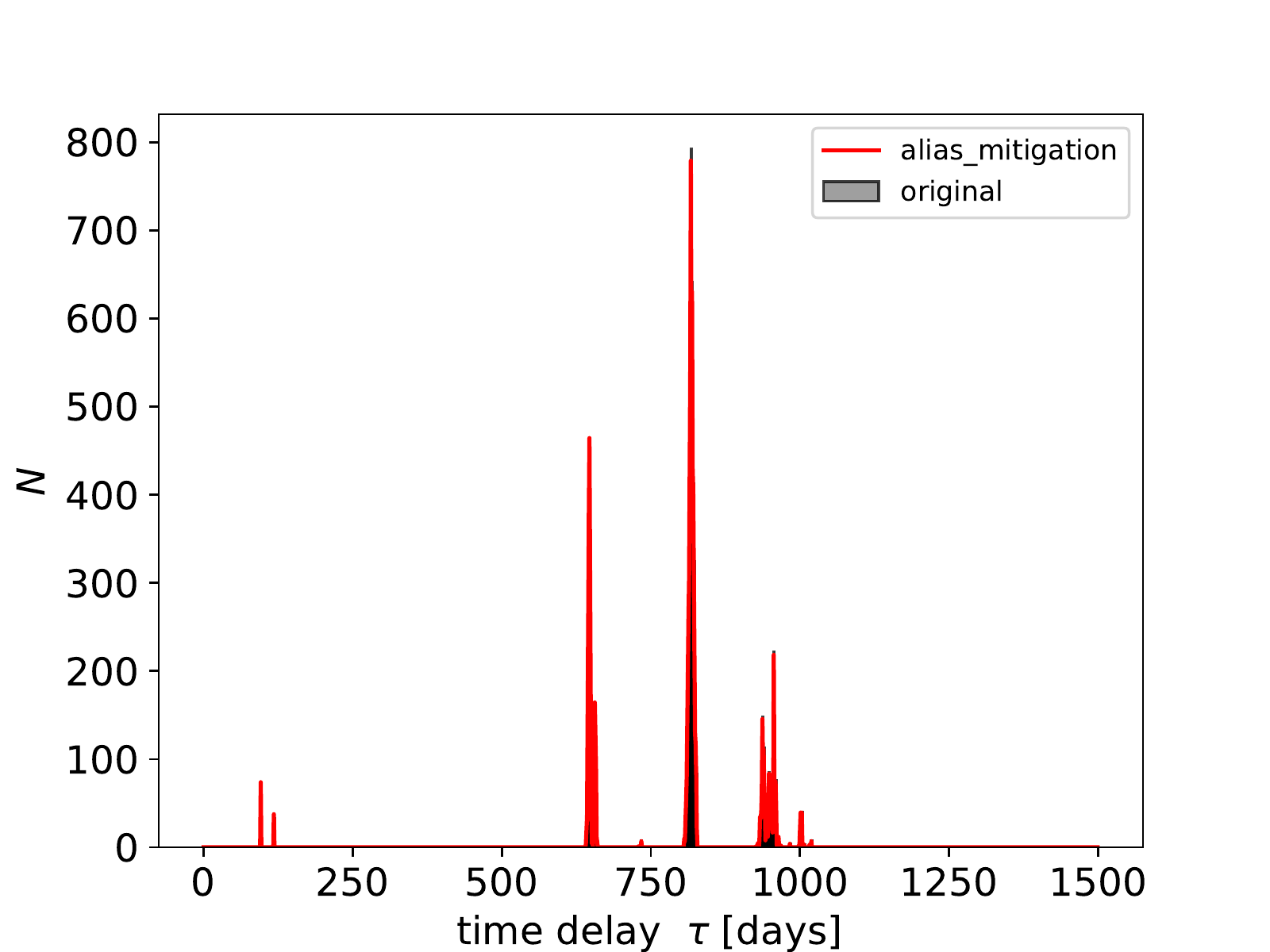}
\includegraphics[scale=0.35]{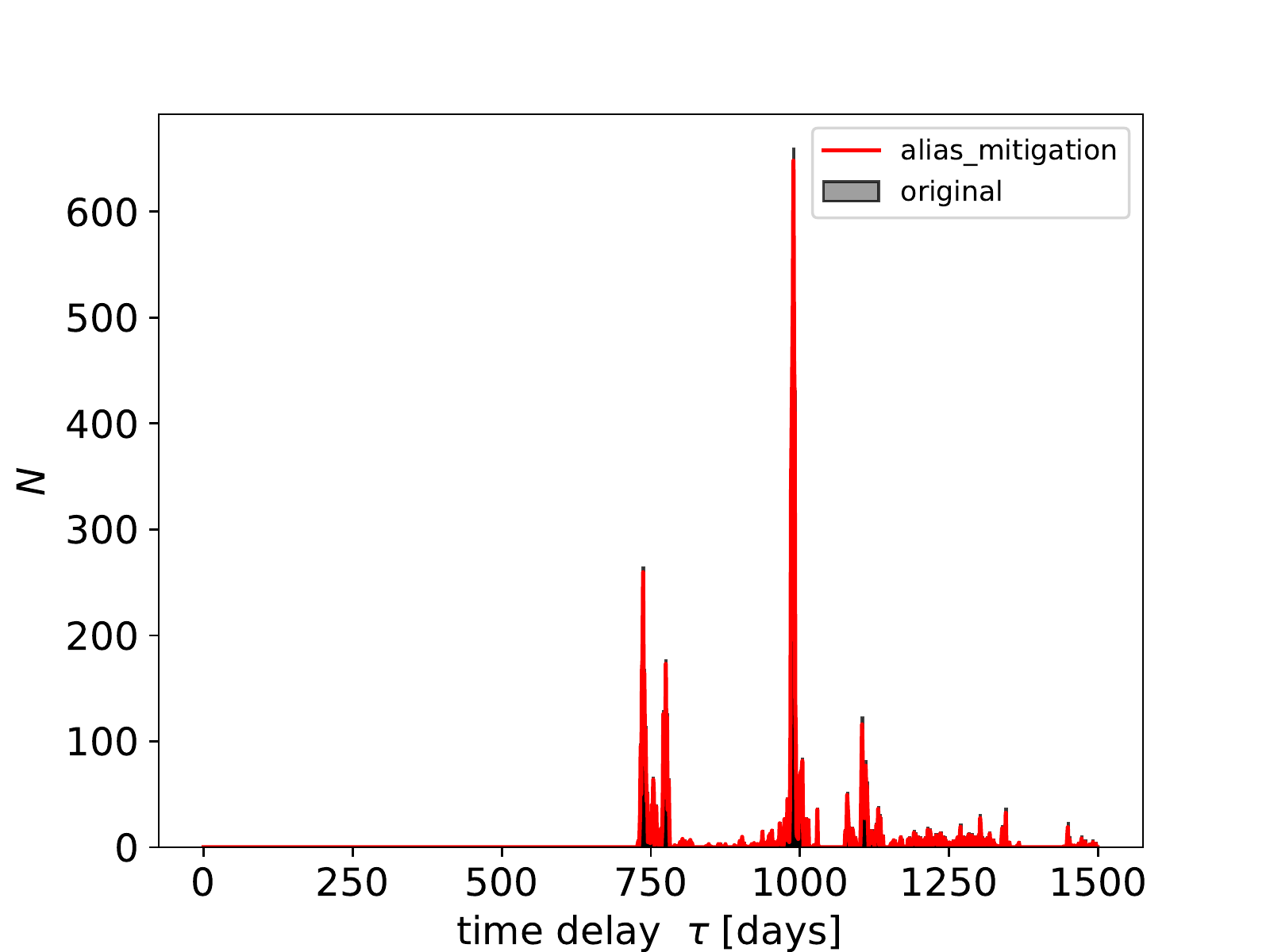}
\caption{Javelin bootstrap results from HE 0413 with 1500 realizations for all the seven curves (from left to right) along with total MgII and FeII (last 2 plots of the lower panel). The peak and  results from this are listed in Table \ref{tab:curves1_7}. We also use the alias mitigation using down-weighting by the overlapping pairs. The black histogram represents the original delay distribution and the red one is after alias mitigation.}
\label{fig:javelin_he0413}
\end{figure*}



\section{UV FeII sample}
\label{sect:UV_sample}

In Table~\ref{tab_uv_FeII_sources}, we provide the list of 4 sources, for which the significant time delay of the UV FeII line complex was measured. 

\begin{table*}[h!]
\caption{In the table, we provide (from the left to the right column) the source name, redshift, the continuum flux density at 3000\,\AA\, (in ${\rm erg\,s^{-1}\,cm^{-2}\,\AA^{-1}}$), the luminosity at 3000\,\AA\,(in ${\rm erg\,s^{-1}}$ assuming the flat $\Lambda$CDM model with $H_0=70\,{\rm km\,s^{-1}\,Mpc^{-1}}$ and $\Omega_{\rm m}=0.3$), the UV FeII rest-frame time delay in days, and the original reference.}
    \begin{tabular}{c|c|c|c|c|c}
    \hline
    \hline
    Source  & $z$ & $\log{(F_{3000}\,[{\rm erg\,s^{-1}\,cm^{-2}\,\AA^{-1}}])}$ & $\log{(L_{3000}\,[{\rm erg\,s^{-1}}])}$ & $\tau_{\rm FeII}\,[\text{days}]$ & reference \\
    \hline
    NGC 5548  & 0.0174   & $ -13.615 \pm 0.051$   & $43.696 \pm 0.051$   & $10^{+1}_{-1}$ & \citet{Maoz1993}\\  
    CTS C30.10  &  0.90052   & $ -15.060 \pm     0.026 $   & $46.023 \pm 0.026$   & $180.3^{+26.6}_{-30.0}$ & \citet{prince_CTS_2022}\\
                &    &    &    & $270.0^{+13.8}_{-25.3}$ & \citet{prince_CTS_2022}  \\
    HE 0435-4312 & 1.3764   & $ -15.179  \pm    0.036$   & $46.359 \pm 0.036$ & $284.0^{+72.9}_{-77.4}$  & this work   \\  
    HE 0413-4031 & 1.2231   & $ -14.657 \pm   0.081 $   & $46.754 \pm 0.080$ & $330.6^{+54.0}_{-87.5}$  & this work   \\              
    \hline     
    \end{tabular}
    \label{tab_uv_FeII_sources}
\end{table*}

\section{Optical FeII sample}

In Table~\ref{tab_optical_FeII_sources}, we provide the list of 20 sources, for which the significant time delay of the optical FeII line complex was measured. 

\begin{table*}[h!]
\caption{In the table, we provide (from the left to the right column) the source name, redshift, the continuum flux density at 5100\,\AA\, (in ${\rm erg\,s^{-1}\,cm^{-2}\,\AA^{-1}}$), the luminosity at 5100\,\AA\,(in ${\rm erg\,s^{-1}}$ assuming the flat $\Lambda$CDM model with $H_0=70\,{\rm km\,s^{-1}\,Mpc^{-1}}$ and $\Omega_{\rm m}=0.3$), the optical FeII rest-frame time delay in days, and the original reference.}
    \begin{tabular}{c|c|c|c|c|c}
    \hline
    \hline
    Source  & $z$ & $\log{(F_{5100}\,[{\rm erg\,s^{-1}\,cm^{-2}\,\AA^{-1}}])}$ & $\log{(L_{5100}\,[{\rm erg\,s^{-1}}])}$ & $\tau_{\rm FeII}\,[\text{days}]$ & reference \\
    \hline
    PG 1700+518  &   0.2890  &  $-14.733  \pm    0.035$  &   $45.397  \pm    0.035$  &    $209.0^{+100.0}_{-147.0}$ & \citet{Bian2010}\\
    PG 1700+518  &   0.2920  &  $-14.708  \pm    0.044$  &   $45.433  \pm    0.044$  &    $296.0^{+59.0}_{-53.0}$ & \citet{chelouche2014}\\
    NGC 4593   &   0.0090  &  $-14.096  \pm    0.049$  &   $42.868  \pm    0.049$  &    $  8.35^{+1.29}_{-1.51}$ & \citet{Barth2013}\\
    Mrk 1511   &   0.0340  &  $-14.971  \pm    0.061$  &   $43.163  \pm    0.061$  &    $  8.63^{+1.35}_{-1.31}$ & \citet{Barth2013}\\
    PG 0026+129  &   0.1420  &  $-14.595  \pm    0.034$  &   $44.845  \pm    0.034$  &    $133.0^{+16.0}_{-15.0}$ & \citet{chelouche2014}\\    
    PG 2130+099  &   0.0633  &  $-14.316  \pm    0.018$  &   $44.376  \pm    0.018$  &    $184.0^{+52.0}_{-137.0}$ & \citet{chelouche2014}\\
    PG 2130+099  &   0.0633  &  $-14.275  \pm    0.001$  &   $44.418  \pm    0.001$  &    $ 35.3^{+8.2}_{-9.9}$ & \citet{HuChen2020}\\
    PG 2130+099  &   0.0633  &  $-14.276  \pm    0.003$  &   $44.416  \pm    0.003$  &    $23.1^{+3.4}_{-5.6}$ & \citet{HuChen2020}\\     
    J113913.91+335551.1   &   0.0323  &  $-15.784  \pm   0.053$   &   $42.304   \pm   0.053$  &    $15.0^{+6.0}_{-9.0}$ & \citet{2013ApJ...773...24R}\\
    Mrk 335  &   0.0258  &  $-14.284  \pm    0.031$  &   $43.605  \pm    0.031$  &     $26.8^{+2.9}_{-2.5}$ & \citet{HuChen2015}\\
    Mrk 1044  &   0.0165  &  $-14.484  \pm    0.049$  &   $43.011  \pm    0.049$  &     $13.9^{+3.4}_{-4.7}$ &\citet{HuChen2015}\\
    IRAS 04416+1215  &   0.0889  &  $-14.570  \pm    0.045$  &   $44.433  \pm    0.045$  &     $12.6^{+16.7}_{-6.7}$ &\citet{HuChen2015}\\
    Mrk 382   &   0.0337  &  $-15.108  \pm    0.073$  &   $43.018   \pm   0.073$  &     $23.8^{+6.0}_{-6.0}$ &\citet{HuChen2015}\\
    Mrk 142  &   0.0449  &  $-14.896  \pm    0.052$  &   $43.486  \pm    0.052$  &     $ 7.6^{+1.4}_{-2.2}$ &\citet{HuChen2015}\\
    MCG +06–26–012   &   0.0328  &  $-15.468  \pm    0.094$  &   $42.634  \pm    0.094$  &     $22.4^{+9.3}_{-6.3}$ &\citet{HuChen2015}\\
    IRAS F12397+3333  &   0.0435  &  $-14.842  \pm    0.042$  &   $43.513  \pm    0.042$  &     $10.6^{+7.0}_{-1.9}$ &\citet{HuChen2015}\\
    Mrk 486    &   0.0389  &  $-14.631  \pm    0.032$  &   $43.623  \pm    0.032$  &     $17.3^{+5.8}_{-3.7}$ &\citet{HuChen2015}\\
    Mrk 493   &   0.0313  &  $-15.027  \pm    0.060$  &   $43.034  \pm    0.060$  &     $11.9^{+3.6}_{-6.5}$ &\citet{HuChen2015}\\
    3C273   &   0.1583  &  $-13.711  \pm    0.047$  &   $45.833  \pm    0.047$  &    $322.0^{+55.5}_{-57.9}$ & \citet{Zhang2019}\\
    Mrk 817 &  0.0314  &  $-14.424 \pm  0.035$ &  $43.641 \pm 0.035$  &  $51.7^{+14.9}_{-1.3}$ & \citet{2021ApJ...918...50L}\\ 
    \hline     
    \end{tabular}
    \label{tab_optical_FeII_sources}
\end{table*}

\section{UV Fe II modeling with different templates and equivalent widths}
\label{appendix_FeII_templates}
 We tested the various templates provided by the \citet{Bruhweiler2008} to model the UV Fe II emission with a fixed velocity width (km/s). The plots are shown in the left panel of Figure \ref{fig:test}. We noticed that for the quasar HE 0413-4031, except for BV9 and BV12, all the templates give very similar results. In comparison, for HE 0435-4312, all templates are a good choice. We also tested the choice of templates over various values of the velocity width (km/s), shown in the right panel of Figure \ref{fig:test}. We do not see much difference in the fitting. In addition, we also show the corresponding light curves for two templates and two velocity widths for both the quasars and the light curves are shown in Fig~\ref{fig:temp-he0413} and \ref{fig:temp-he0435}. For both the quasars, the light curves are in agreement for different templates or for different velocity widths.
\begin{figure*}
    \centering
    \includegraphics[scale=0.35]{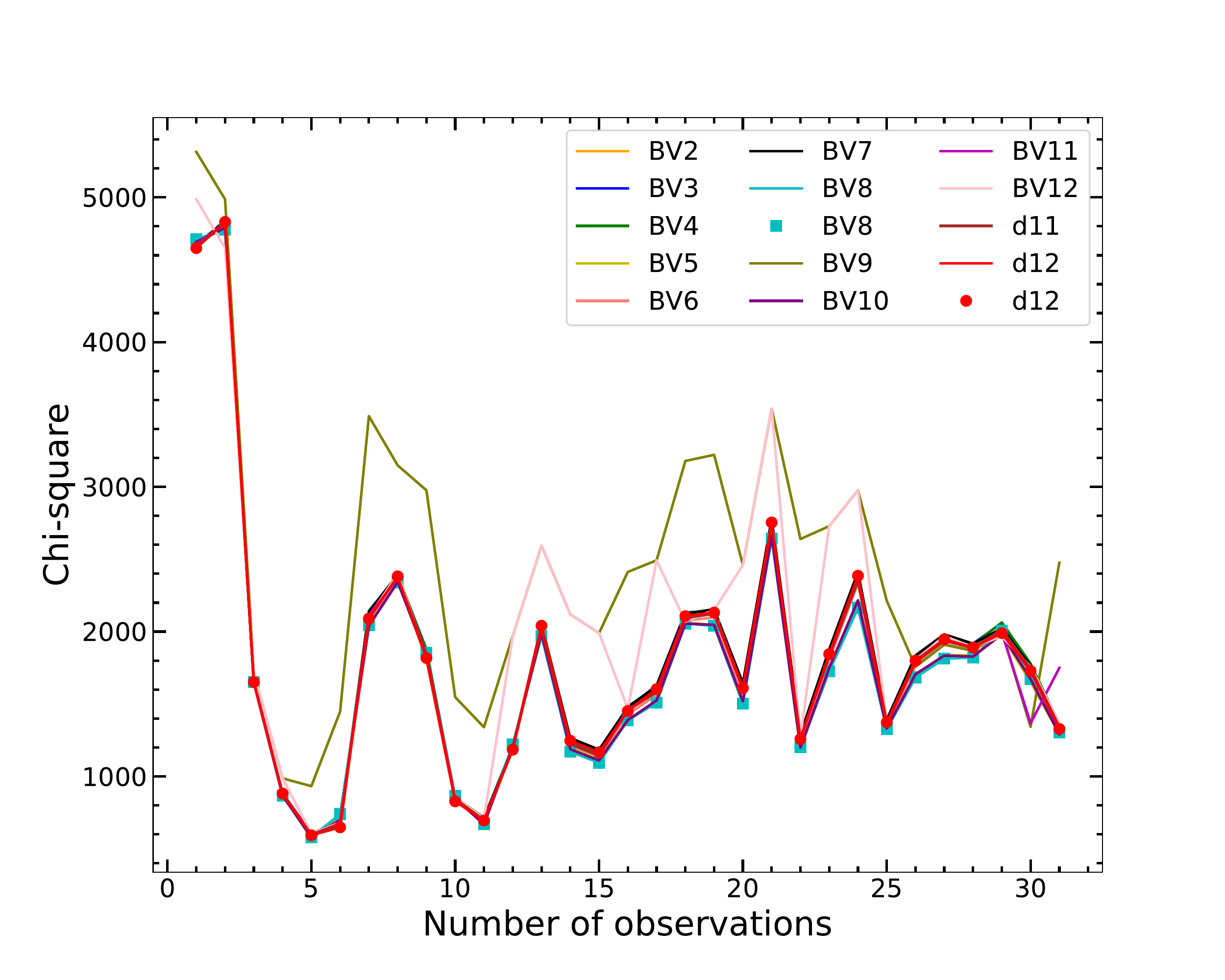}
    \includegraphics[scale=0.35]{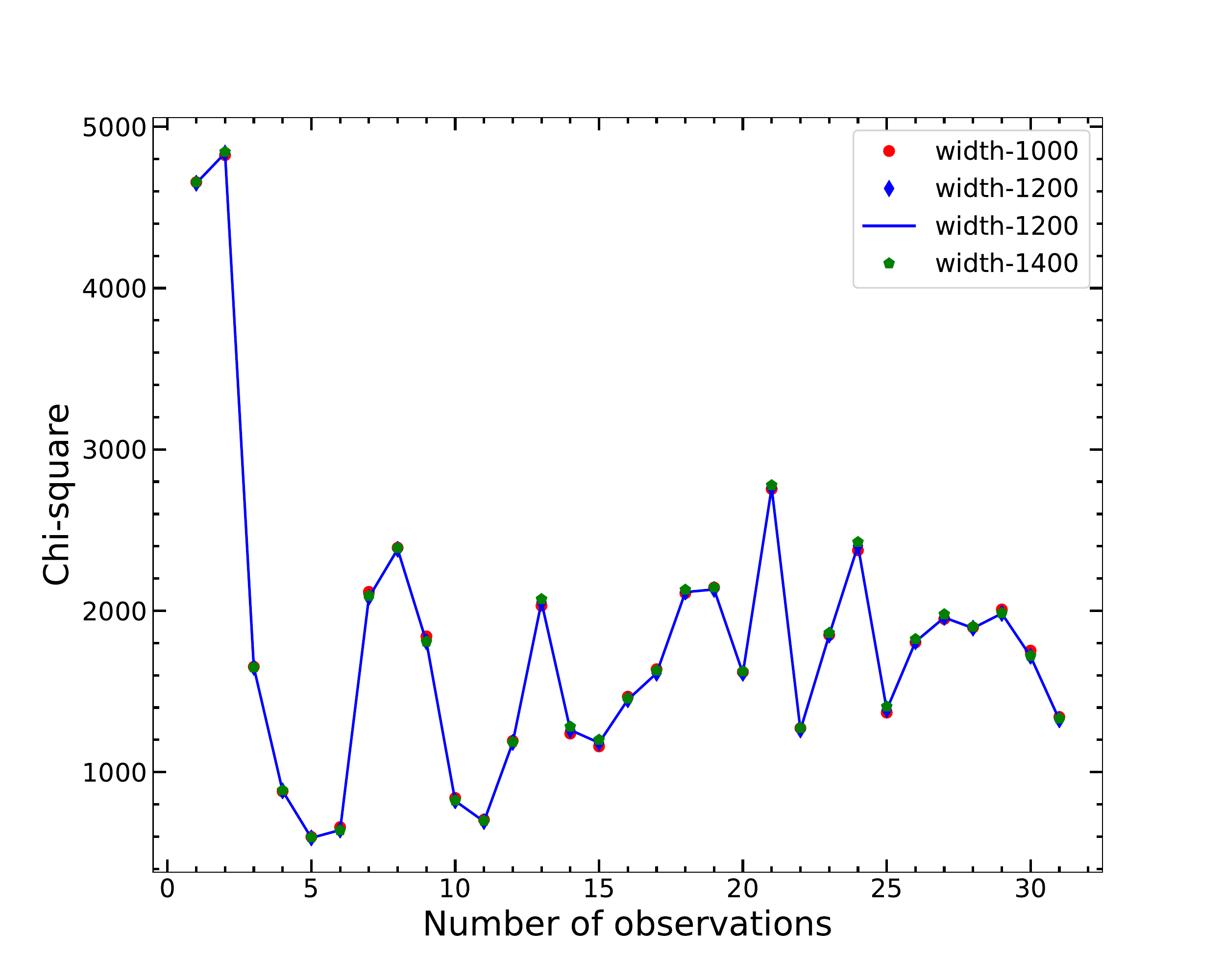}
    \includegraphics[scale=0.35]{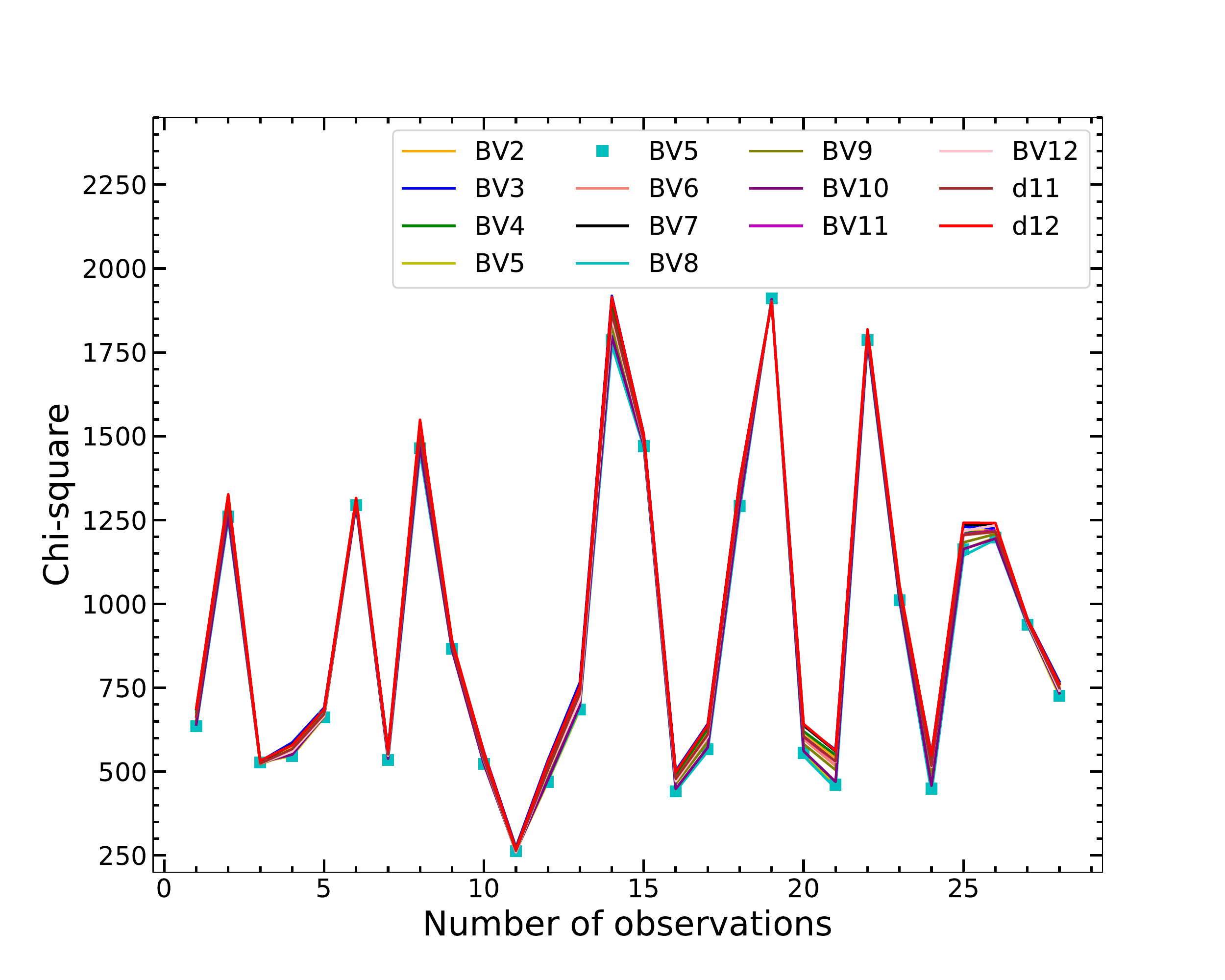}
    \includegraphics[scale=0.35]{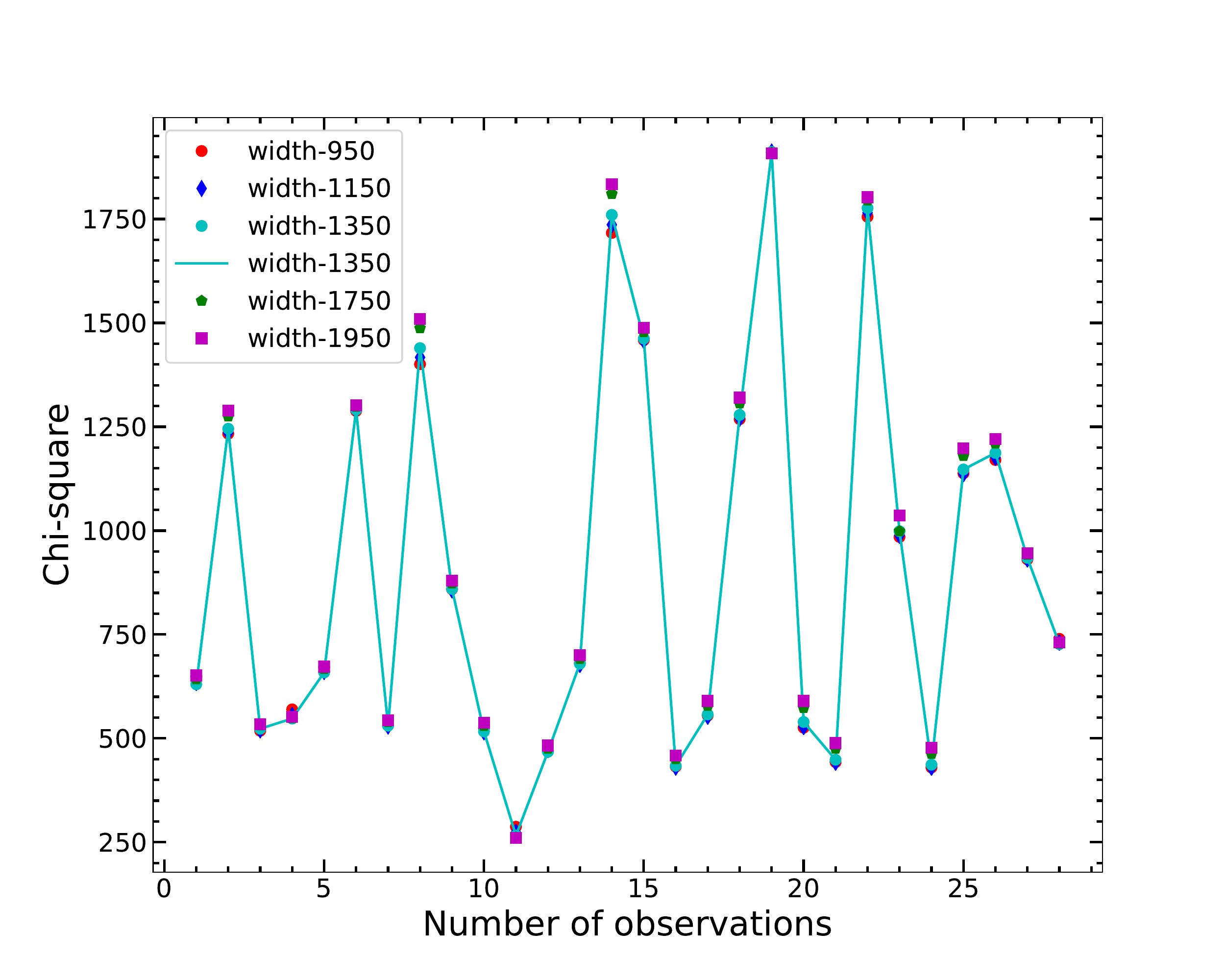}
    \caption{ The left panel represents the chi-square distribution of different templates that have been used to model the UV Fe II emission. The right panel shows the chi-square distribution when various velocity widths (km/s) were chosen to model the UV FeII. The upper panel and lower panel are for HE 0413 and HE 0435 sources. The templates name are the following: BV2: d11-5-m20-20-5.dat, BV3:  d11-m05-20-5.dat, BV4:  d11-m10-20-5.dat, BV5:  d11-m20-20.5-735.dat, BV6:  d11-m20-20-5.dat, BV7:  d11-m20-20.dat, BV8:  d11-m20-21-735.dat, BV9:  d11-m20-21.dat, BV10: d11-m30-20-5-735.dat, BV11: d11-m30-20-5.dat, BV12: d11-m50-20-5.dat, d11: d11-m20-20-5.dat, d12: d12-m20-20-5.dat. }
    \label{fig:test}
\end{figure*}

\begin{figure*}
    \centering
    \includegraphics[scale=0.35]{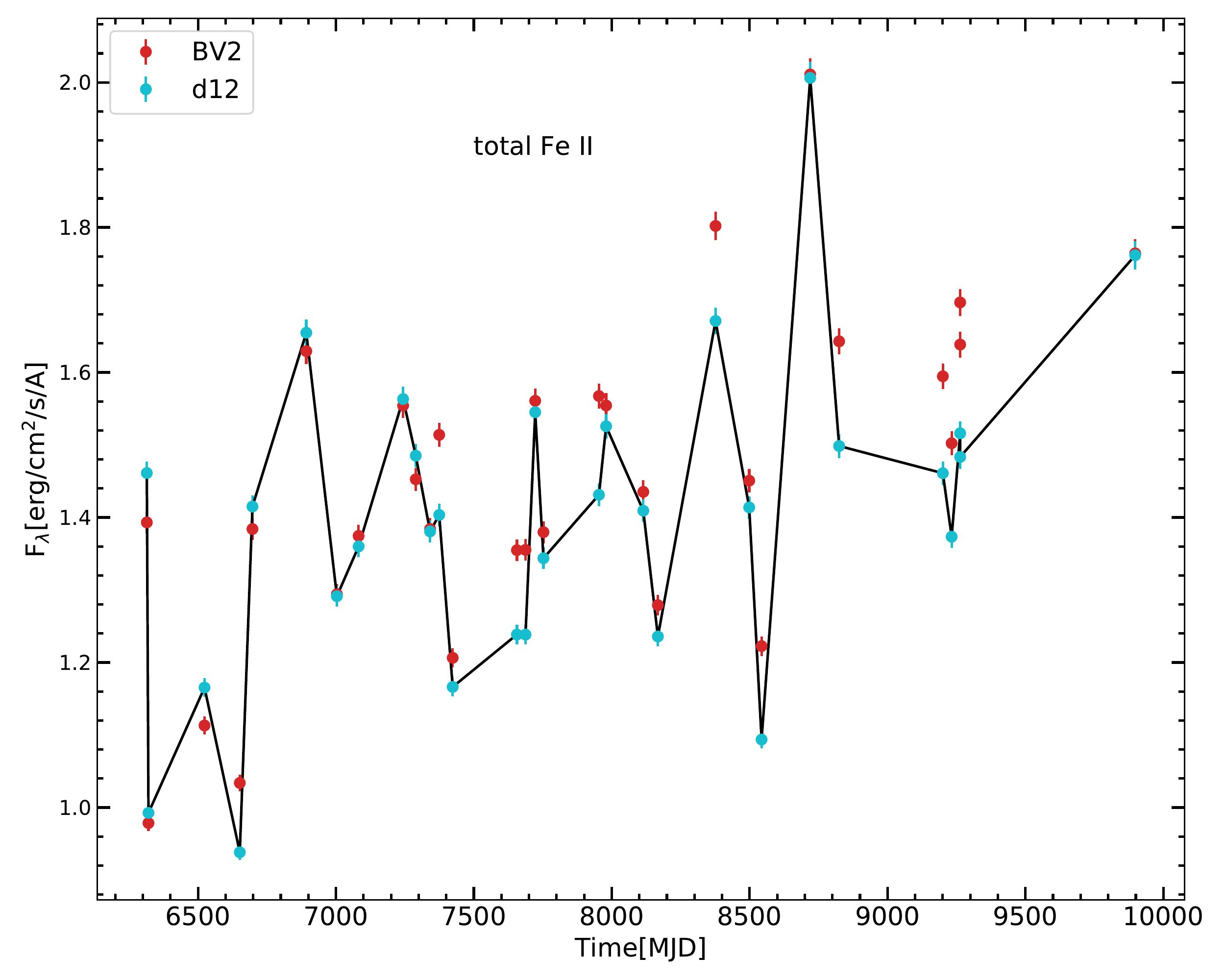}
    \includegraphics[scale=0.35]{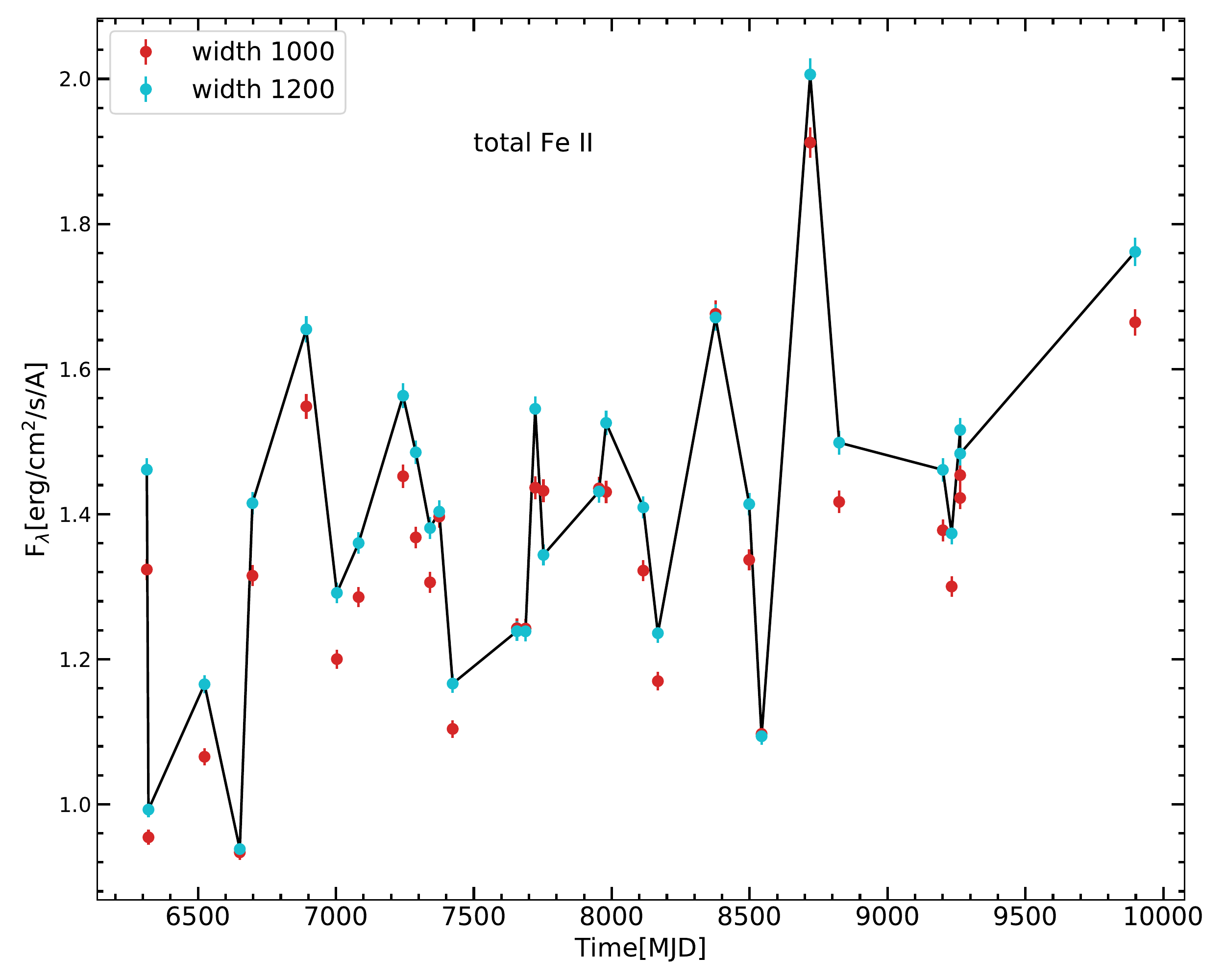}
    \caption{ The total Fe II light curve, for HE 0413, produced for the two sets of templates (left panel) and two sets of line widths (right panel). As we argued in the text the different templates and different velocity widths do not significantly affect the light curve.}
    \label{fig:temp-he0413}
\end{figure*}

\begin{figure*}
    \centering
    \includegraphics[scale=0.35]{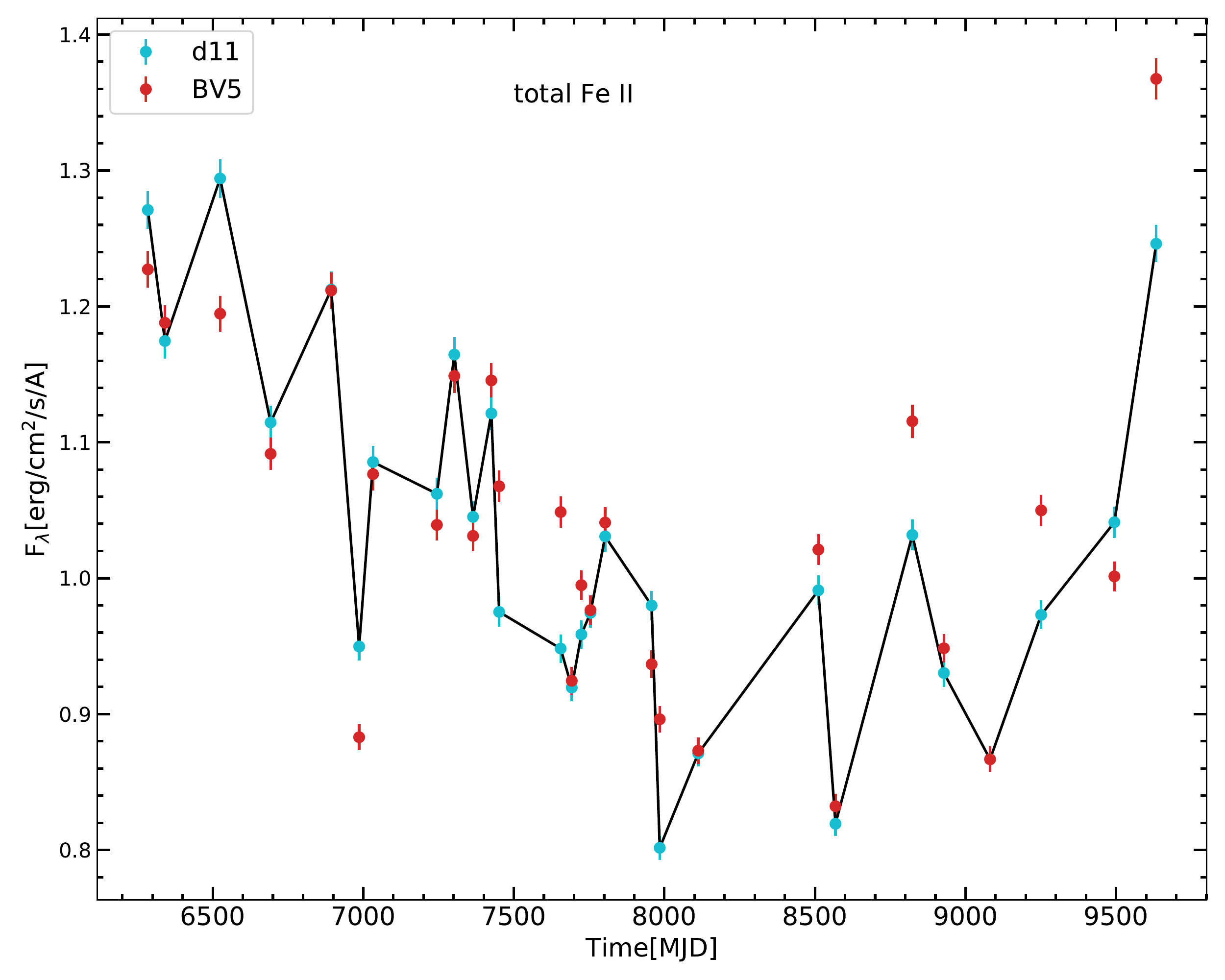}
    \includegraphics[scale=0.35]{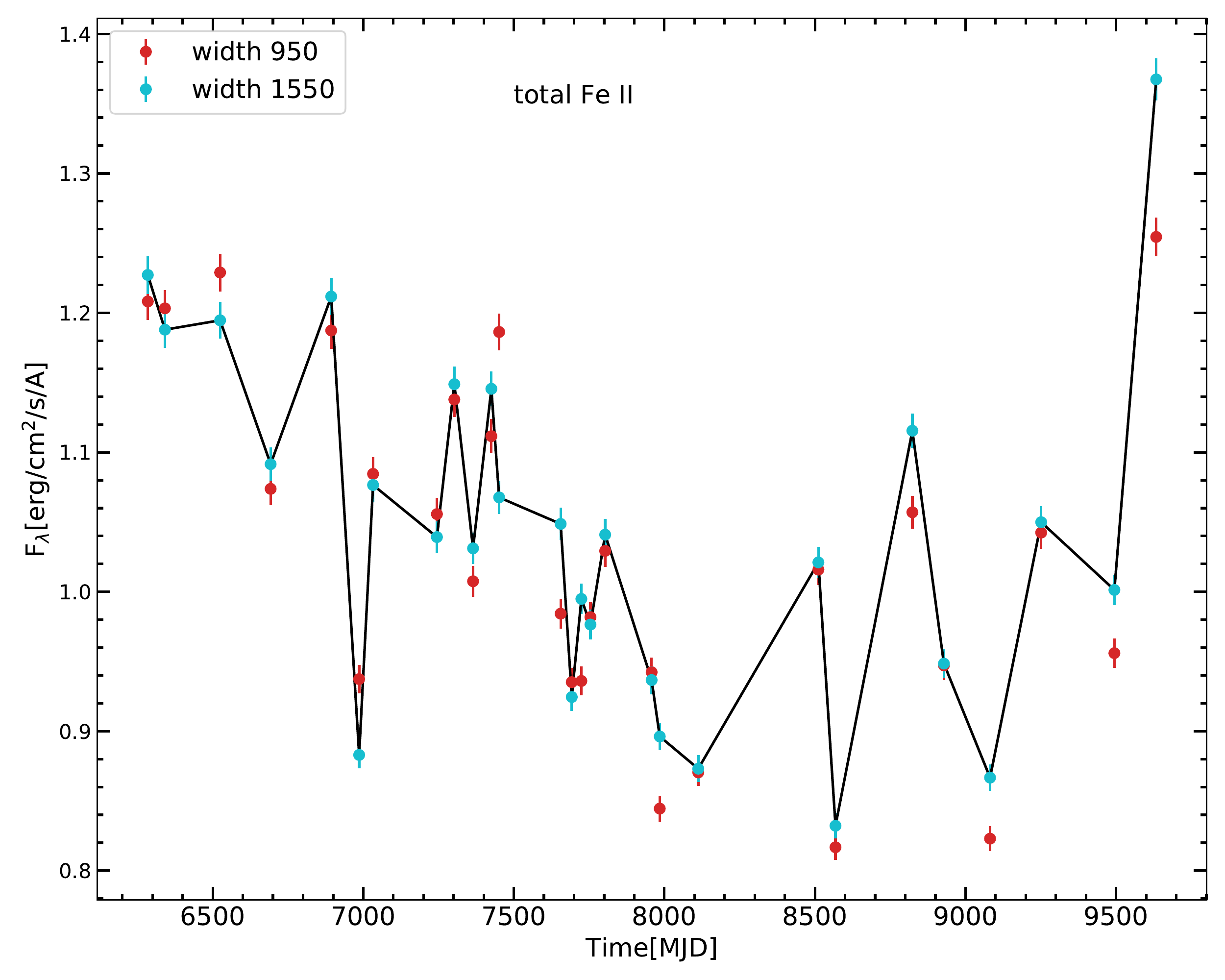}
    \caption{ The total Fe II light curve, for HE 0435, produced for the two sets of templates (left panel) and the two sets of velocity widths (right panel). As we argued in the text the different templates and different velocity widths do not significantly affect the light curve.}
    \label{fig:temp-he0435}
\end{figure*}
\end{document}